\documentclass[aip,jcp,amsmath,amssymb,preprint]{revtex4-1}
\usepackage{graphicx} % Include figure files
\usepackage{dcolumn} % Align table columns on decimal point
\usepackage{bm} % bold math
\usepackage{here}
\usepackage{color}

\begin{document}
%%%%%%%%%%%%%%%%%%%%%%%%%%%%%%%%%%%%%%%%%%%%%%%%%%%%%%%%%%%%%%%%%%%%%%%%

%%%%%%%%%%%%%%%%%%%%%%%%%%%%%%%%%%%%%%%%%%%%%%%%%%%%%%%%%%%%%%%%%%%%%%%%
\title{Self-learning path integral hybrid Monte Carlo with mixed \textit{ab initio} 
and machine learning potentials for modeling nuclear quantum effects in water}
%%%%%%%%%%%%%%%%%%%%%%%%%%%%%%%%%%%%%%%%%%%%%%%%%%%%%%%%%%%%%%%%%%%%%%%%

\author{Bo Thomsen}
\email{thomsen.bo@jaea.go.jp}
\affiliation{CCSE, Japan  Atomic Energy Agency, 178-4-4, Wakashiba, Kashiwa, Chiba, 277-0871, Japan}

\author{Yuki Nagai}
%\email{nagai.yuki@jaea.go.jp}
\affiliation{Information Technology Center, The University of Tokyo, 6-2-3 Kashiwanoha, Kashiwa, Chiba 277-0882, Japan}
%\affiliation{Mathematical Science Team, RIKEN Center for Advanced Intelligence Project (AIP), 1-4-1 Nihonbashi, Chuo-ku, Tokyo 103-0027, Japan}

\author{Keita Kobayashi}
%\email{kobayashi.keita@jaea.go.jp}
\affiliation{CCSE, Japan  Atomic Energy Agency, 178-4-4, Wakashiba, Kashiwa, Chiba, 277-0871, Japan}

\author{Ikutaro Hamada}
%\email{ihamada@prec.eng.osaka-u.ac.jp}
\affiliation{Department of Precision Engineering, Graduate School of Engineering, Osaka University, 2-1, Yamadaoka, Suita, Osaka 565-0871, Japan}

\author{Motoyuki Shiga}
\email{shiga.motoyuki@jaea.go.jp}
\affiliation{CCSE, Japan  Atomic Energy Agency, 178-4-4, Wakashiba, Kashiwa, Chiba, 277-0871, Japan}

\date{\today}

%%%%%%%%%%%%%%%%%%%%%%%%%%%%%%%%%%%%%%%%%%%%%%%%%%%%%%%%%%%%%%%%%%%%%%%%
\begin{abstract}
%%%%%%%%%%%%%%%%%%%%%%%%%%%%%%%%%%%%%%%%%%%%%%%%%%%%%%%%%%%%%%%%%%%%%%%%

The introduction of machine learned potentials (MLPs)
has greatly expanded the space available for studying
Nuclear Quantum Effects computationally
with \textit{ab initio} path integral (PI) accuracy,
with the MLPs' promise of an accuracy comparable to
that of \textit{ab initio} at a fraction of the cost.
One of the challenges in development of MLPs
is the need for a large and diverse training set
calculated by \textit{ab initio} methods.
This data set should ideally cover the entire
phase space,
while not searching this space using \textit{ab initio} methods,
as this would be counterproductive and generally intractable 
with respect to computational time.
In this paper, we present the self-learning PI
hybrid Monte Carlo Method using a mixed \textit{ab initio} and ML potential
(SL-PIHMC-MIX),
where the mixed potential allows for the study of
larger systems and the extension of the original
SL-HMC method [Nagai \textit{et al.}, Phys. Rev. B 102, 041124 (2020)] 
to PI methods and larger systems.
While the MLPs generated by this method can be directly applied to run long-time ML-PIMD simulations, we demonstrate that using PIHMC-MIX with the trained MLPs allows for an exact reproduction of the structure obtained from \textit{ab initio} PIMD. 
Specifically, we find that the PIHMC-MIX simulations require only 5,000 evaluations of the 32-bead structure, compared to the 100,000 evaluations needed for the \textit{ab initio} PIMD result.

%%%%%%%%%%%%%%%%%%%%%%%%%%%%%%%%%%%%%%%%%%%%%%%%%%%%%%%%%%%%%%%%%%%%%%%%
\end{abstract} 
%%%%%%%%%%%%%%%%%%%%%%%%%%%%%%%%%%%%%%%%%%%%%%%%%%%%%%%%%%%%%%%%%%%%%%%%

%%%%%%%%%%%%%%%%%%%%%%%%%%%%%%%%%%%%%%%%%%%%%%%%%%%%%%%%%%%%%%%%%%%%%%%%
\maketitle
%%%%%%%%%%%%%%%%%%%%%%%%%%%%%%%%%%%%%%%%%%%%%%%%%%%%%%%%%%%%%%%%%%%%%%%%

%%%%%%%%%%%%%%%%%%%%%%%%%%%%%%%%%%%%%%%%%%%%%%%%%%%%%%%%%%%%%%%%%%%%%%%%
\section{Introduction}
%%%%%%%%%%%%%%%%%%%%%%%%%%%%%%%%%%%%%%%%%%%%%%%%%%%%%%%%%%%%%%%%%%%%%%%%

Nuclear Quantum Effects (NQEs) play a large role
in determining the properties of matter containing
light atoms and, by extension, the isotope effects seen when
hydrogen (H) is exchanged for deuterium (D) or tritium (T).
One example of this is the observed differences between light (H$_2$O) 
and heavy (D$_2$O) water,\cite{ceriotti_nuclear_2016}
which has recently been investigated by a series of experiments.
\cite{zeidler_oxygen_2011, zeidler_isotope_2012, soper_radial_2013, kameda_neutron_2018}
We have also previously reported some structural and 
reactive differences between the two liquids and other 
isotopologues of water 
\cite{thomsen_nuclear_2021,Thomsen_structures_2022}
from \textit{ab initio} or first principles (FP) simulations.
Modeling of NQEs in bulk systems relies on 
path integral (PI) methods based on the Feynman 
path formulation of quantum mechanics.
\cite{feynman_statistical_1972, feynman_quantum_2010, schulman_techniques_2012}
Implementations of these methods\cite{shiga_path_2018,kapil_i-pi_2019} 
typically require the simultaneous evaluation 
of energies and gradients of $P$ copies of the 
system in each time step.
$P$ is generally considered in tens or low
hundreds for simulations at room temperature
and, thus, adds significantly to the cost of
performing FP simulations required 
for the accurate description of NQEs in materials.

%%%

In the 1990s, methods were suggested for generating 
machine learned potentials (MLPs),\cite{Blank_neural_1995, Brown_combining_1996, lorenz_representing_2004}
with accuracy close to those of FP calculations 
but at a much-reduced computational cost. 
However, MLPs were initially limited to the study of small gas phase clusters.
It was only with the introduction of high-dimensional neural network potentials
\cite{behler_generalized_2007, behler_representing_2014, behler_constructing_2015, 
behler_first_2017}
by Behler and Parrinello that the MLPs were extended to the study of bulk-phase systems. 
The development of these MLPs is continuing, with 
later generations including more physics informed 
terms, such as machine learned atomic 
charges\cite{artrith_high-dimensional_2011,morawietz_neural_2012} 
and global charge equilibration,\cite{behler_four_2021} 
for the accurate description of charge separation. 

%%%

From the first MLPs used for the simulation of liquid water,
\cite{Morawietz_how_2016} the study of bulk phase water using
MLPs has undergone a rapid development,\cite{omranpour_perspective_2024}
with the low cost of evaluation of the MLP allowing for
the molecular dynamics (MD) simulations of very large systems
both with\cite{krishnamoorthy_dielectric_2021} and without
\cite{lu_86_2021} NQEs.
Generally, fewer FP calculations are needed when training an
MLP, and one can thus explore more expensive FP methods for
describing the electronic potential in water simulations.
MB-Pol\cite{Babin_MBPol_2013, Babin_MBPol_2014, Medders_MBPol_2014} presents one
physics based model for water, which has recently\cite{Medders_MBPol_CC_2015, Zhu_MBPol_2023}
been adjusted to fit CCSD(T), i.e., the gold standard of quantum chemistry,
data for the interaction potentials in water.
This model along with other recent fitted MLPs based on FP data from CCSD(T)\cite{yu_q-aqua_2022,daru_coupled_2022,chen_data-efficient_2023}
have been shown to accurately reproduce both equilibrium and dynamic properties of water
when NQEs are considered.
The investigation of NQEs has also been undertaken by a number of 
studies due to the reduced cost of PI simulations when an MLP 
is employed,\cite{cheng_nuclear_2016, kapil_high_2016,
kapil_inexpensive_2020,yao_temperature_2020,li_static_2022,
yao_nuclear_2021}
including comparisons of isotopologues of water
\cite{ko_isotope_2019,xu_isotope_2020} and the
effect of NQEs on the behavior of the hydroxide 
and hydronium ions\cite{atsango_developing_2023} 
in the liquid phase.

%%%

Shared by all MLP models is the need for a training set made up of FP data,
which should ideally cover the entire phase space while not stemming 
from an exhaustive search using FP methods.
To efficiently carry out the search, one can use
on-the-fly learning
\cite{li_molecular_2015, jinnouchi_phase_2019, cheng_--fly_2020,young_transferable_2021,montero_de_hijes_comparing_2024}
to train a cheap potential representation, which
can be used to accelerate the search.
Several of the authors recently suggested the 
self-learning hybrid Monte Carlo (SL-HMC) method 
\cite{nagai_self-learning_2020, Kobayashi_self-learning_2021}
based on the hybrid Monte Carlo (HMC) method.
\cite{gottlieb_hybrid-molecular-dynamics_1987, duane_hybrid_1987,
mehlig_hybrid_1992, tuckerman_efficient_1993, 
shinoda_rapid_2004, nakayama_speed-up_2009}
In the SL-HMC method, a short ML-MD simulation is run between
each HMC step to allow efficient sampling of phase space
while training an MLP for the system being studied.
Extension of this method to larger systems and the 
PI domain is, however, hindered by the limitations of HMC
as the acceptance ratio scales inversely with the size of 
the system.
Here, we introduce the self-learning path integral 
hybrid Monte Carlo method using a mixed FP and ML 
potential (SL-PIHMC-MIX) 
to overcome this limitation. 
In brief, this method allows for larger discrepancies
between the FP and ML potential energies through the potential mixing, 
thus enabling larger acceptance ratios and faster sampling of the phase space of 
the mixed potential Hamiltonian. 
Thus, reweighting\cite{miao_improved_2014} and longer 
trajectories are necessary to sample the phase space of the FP Hamiltonian. 
The savings enabled by the larger acceptance rate of the potential mixing scheme are, however, great enough
that the effective length of the trajectory using potential mixing exceeds those using the pure FP 
potentials. 
The SL-PIHMC-MIX method is furthermore, 
as the SL-HMC method, fully general with respect to the FP model used and the MLP model used.

%%%

In this study, we will use SL-PIHMC-MIX to train an MLP
to model room temperature water.
After training the MLP, it will be used in a production run
using the PIHMC-MIX method,
which allows us to rapidly converge the radial distribution
functions (RDFs) and, thus, predict the structure of water
using only 5000 FP calculations 
along the bead chain, compared to the 100\,000 calculations needed in our previous FP-PIMD 
studies\cite{thomsen_nuclear_2021,Thomsen_structures_2022}
to converge the water RDFs.
The structure of water has long been a
topic of discussion,\cite{finney_structure_2024}
and FP based studies of water using density functional theory (DFT) 
have since the first report,\cite{chen_hydrogen_2003} and until the emergence of coupled cluster based MLPs, been the state of the art for studying
water, with several studies comparing the
accuracy of functionals for this purpose.
\cite{gillan_perspective_2016, villard_structure_2024}
Recent advances in algorithms for PI propagation have allowed for the study of dynamics, including NQEs using hybrid functionals,\cite{marsalek_quantum_2017}
and FP-based molecular dynamics ( FP-MD) studies have also 
been conducted at the MP2\cite{del_ben_probing_2015} and quantum 
Monte Carlo\cite{zen_ab_2015} levels of theory.
DFT and other FP based studies remain relevant in the context of solvated systems where no general high quality MLP or model is currently available.

%%%

This paper is organized as follows.
First, we will extend the SL-HMC and HMC methods to the PI formalism 
and introduce the SL-PIHMC-MIX and PIHMC-MIX methods that allow the study of 
systems containing many particles.
Reweighting of the results from PIHMC-MIX to get the structural properties
of the DFT ensemble will also be described in this section.
In Sec. III, the computational details of the simulations
used in this work are given.
In Sec. IV, the results from the PIHMC-MIX method
using an MLP that was fitted using SL-PIHMC-MIX will be compared
to the results of FP-PIMD for the RPBE-D3 functional.
The effects of the mixed potential method and the accuracy of the MLPs
produced by the SL-PIHMC-MIX method will then be discussed.
We will briefly discuss the description of heavy water (D$_2$O) using the 
PIHMC-MIX method and the MLPs produced by the SL-PIHMC-MIX method.
We will then go on to compare the results of PIHMC-MIX for  SCAN, 
rev-vdW-DF2, and optB88-vdW functionals with both experimental data 
and those from the RPBE-D3 functional.
For each of the SCAN, 
rev-vdW-DF2 and optB88-vdW functionals  a unique MLP has been fitted
using the SL-PIHMC-MIX method.
Finally, we will provide a summary of the findings of this study in Sec. V.

%%%%%%%%%%%%%%%%%%%%%%%%%%%%%%%%%%%%%%%%%%%%%%%%%%%%%%%%%%%%%%%%%%%%%%%%
\section{Theory}
%%%%%%%%%%%%%%%%%%%%%%%%%%%%%%%%%%%%%%%%%%%%%%%%%%%%%%%%%%%%%%%%%%%%%%%%

%%%%%%%%%%%%%%%%%%%%%%%%%%%%%%%%%%%%%%%%%%%%%%%%%%%%%%%%%%%%%%%%%%%%%%%%
\subsection{Self-Learning Path Integral Hybrid Monte Carlo}
%%%%%%%%%%%%%%%%%%%%%%%%%%%%%%%%%%%%%%%%%%%%%%%%%%%%%%%%%%%%%%%%%%%%%%%%

The SL-HMC method has previously been reported
by some authors.\cite{nagai_self-learning_2020, Kobayashi_self-learning_2021}
In this section this method will be extended to the PI domain, 
to the so-called SL-PIHMC method,
and then to larger system sizes in the SL-PIHMC-MIX method.
In this study, DFT with various functionals will be used as the FP method. 
However, the approach is fully general and could accommodate a wave function-based method, provided the computational time allows for thousands of evaluations across the entire bead chain.
The same holds for the low level method,
which is here an MLP denoted by ML,
and for the SL-PIHMC-MIX method any model that can be updated based on 
data from the FP data could be used.
In PIHMC, we use the path integral formulation
of quantum mechanics, 
and the atomic positions, $(\mathbf{R})$, 
are therefore expanded into $P$ imaginary time-slices 
or so-called beads, 
\textit{i.e.} $(\mathbf{R})=\left(\mathbf{R}^{(1)}_{}, 
\ldots, \mathbf{R}^{(P)}_{}\right)$.
The $j$th bead contains all the coordinates
of the $N$ atoms in the bead, 
$\left(\mathbf{R}^{j}_{}\right)=
\left(\mathbf{R}^{(j)}_{1}, 
\ldots, \mathbf{R}^{(j)}_{N}\right)$.
The equations of motion for PIMD and related 
methods are commonly derived in normal mode space.
The reason for this is to better allow
energy transfer between the modes at high
temperatures,
and to ease the derivations of the
equations of motion for the system.
\cite{tuckerman_efficient_1993, witt_applicability_2009,
shiga_path_2022}
Here, the coordinates for all the beads 
of the $I^{\mathrm{th}_{}}$ atom $\left(\mathbf{R}_{I}^{}\right)=
\left(\mathbf{R}_{I}^{(1)},\ldots ,\mathbf{R}_{I}^{(P)}\right)$ 
in the system are transformed to the normal
mode space $\mathbf{Q}_{I}^{}=
\left(\mathbf{Q}_{I}^{(1)},\ldots, \mathbf{Q}_{I}^{(P)}\right)$.
$\mathbf{Q}_{I}^{}$ and the corresponding momenta $\mathbf{P}_{I}^{}$ will in the following be assumed to be expanded as vectors.
%$\mathbf{Q}_{I}^{}$ will in the following be assumed to
%be expanded as a vector,
%as will the corresponding momenta $\mathbf{P}_{I}^{}$
%for each atom $I$ in the full physical system.

%%%

The heart of the PIHMC method is accepting or
rejecting a Monte Carlo move from the point 
in phase space $\{\mathbf{P},\mathbf{Q}\}$ 
to $\{\mathbf{P}',\mathbf{Q}'\}$
with the probability of accepting a step given as
\begin{equation}
P_{\mathrm{acc}}^{}\left(\{\mathbf{P,Q}\}\rightarrow \{\mathbf{P',Q'}\}\right) =
 \min\left(1, \exp\left(-\beta\left(H_{\mathrm{FP}}^{}\left(\{\mathbf{P',Q'}\}\right)
 -H_{\mathrm{FP}}^{}\left(\{\mathbf{P,Q}\}\right)\right)\right)\right),
\end{equation}
where $\beta = \frac{1}{Tk_{\mathrm{B}}^{}}$, $T$ is the 
temperature and $k_{\mathrm{B}}^{}$ is the Boltzmann constant.
The Hamiltonian for the whole system, where the potential energy 
is evaluated within a given model ($\mathrm{mod}$), is
\begin{equation}
H_{\mathrm{mod}}^{}(\{\mathbf{P},\mathbf{Q}\})=
\frac{1}{2}\sum_{I=1}^{N}\left(\mathbf{P}_I^{T}\boldsymbol{\mu}^{-1}_{I}\mathbf{P}_I^{}
+M_{I}^{}\omega_P^{2}\mathbf{Q}_{I}^{T}\boldsymbol{\lambda}\mathbf{Q}_{I}^{}\right)
+V_{\mathrm{av}}^{\mathrm{mod}}\left(\left\{\mathbf{Q}\right\}\right).
\end{equation}
Here, $\boldsymbol{\mu}_{I}^{}$ is a 
diagonal matrix containing the normal mode masses, 
$M_I$ is the mass of the $I$th particle in the system, 
$\boldsymbol{\lambda}$ is a diagonal matrix composed of the eigenvalues 
stemming from the normal mode transform, 
and $\omega_P=\frac{\sqrt{P}}{\beta\hbar}$.
The bead average potential in the model, $\mathrm{mod}$, is given as 
\begin{equation}
V_{\mathrm{av}}^{\mathrm{mod}}(\{\mathbf{Q}\})=\frac{1}{P}\sum_{s=1}^{P}V_{}^{\mathrm{mod}}\left(\mathbf{R}_{1}^{(s)}\left(\mathbf{Q}_{1}^{}\right),\ldots, \mathbf{R}_{N}^{(s)}\left(\mathbf{Q}_{N}^{}\right)\right)=\frac{1}{P}\sum_{s=1}^{P}V^{\mathrm{mod}}_{}\left(\mathbf{R}_{}^{(s)}\right).
\end{equation}
Here, we introduce a shorthand for the potential
energy for the $s$th bead in the system
$\left(V^{\mathrm{mod}}_{}\left(\mathbf{R}_{}^{(s)}\right)\right)$
to avoid direct reference to the normal to Cartesian coordinate transform
$\left(\mathbf{R}_{i}^{(s)}\left(\mathbf{Q}_{i}^{}\right)\right)$ and to ease the notation 
later in this manuscript.

%%%

The diagrammatic form of SL-PIHMC and PIHMC is shown in Figure 1(a).
It is assumed that one has a primitive initial guess for
the MLP.
As shown in Figure 1(b), before each Monte Carlo step, 
the system is propagated according to the MLP,
$\mathrm{mod}=\mathrm{ML}$,
for $n_{\mathrm{ML}}$ steps with the time step
$\Delta t_{\mathrm{ML}}$.
The initial momenta for the propagation 
of the ML trajectory, $\mathbf{P}$, 
are generated after each Monte Carlo step 
from a random sample of the Maxwell-Boltzmann 
distribution with the temperature $T$.
The equations of motion and details of this
propagation are widely available in the literature
\cite{parrinello_study_1984, hall_nonergodicity_1984, tuckerman_efficient_1993,
ceriotti_efficient_2010, shiga_path_2018} describing the PIMD methodology.
After $n_{\mathrm{test}}^{}$ Monte Carlo steps
$n_{\mathrm{ML}}$ can be updated,
depending on the average acceptance rate
from the previous $n_{\mathrm{test}^{}}$ steps,
$A^{}_{\mathrm{acc}}=\frac{n^{}_{\mathrm{acc}}}{n^{}_{\mathrm{test}}}$,
where $n^{}_{\mathrm{acc}}$ is the number of accepted
Monte Carlo steps out of the last $n^{}_{\mathrm{test}}$
steps.
This is in our implementation done by either doubling
$n^{}_{\mathrm{ML}}$, 
if $A^{}_{\mathrm{acc}} > P_{\mathrm{upper}}^{}$, 
to a maximum of $n_{\mathrm{ML}}^{\mathrm{max}}$ 
or halving $n^{}_{\mathrm{ML}}$,
if $A^{}_{\mathrm{acc}} < P_{\mathrm{lower}  }^{}$, 
to a minimum of $n_{\mathrm{ML}}^{\mathrm{min}}$.
All of these values, $\Delta t_{\mathrm{ML}}^{}$, $n^{}_{\mathrm{test}}$, 
$P_{\mathrm{upper}}^{}$, $P_{\mathrm{lower}}^{}$,
$n_{\mathrm{ML}}^{\mathrm{max}}$, and \textbf{$n_{\mathrm{ML}}^{\mathrm{min}}$} 
can be provided by the user on input.

%%%

A crucial feature of the SL-PIHMC method is the retraining of the MLP during 
the simulation at every $n_{\mathrm{FP}}^{}$ Monte Carlo step.
The MLP will thus implicitly depend on time in the SL-PIHMC method, 
and the potential in Eq. (3) is formally given as,
\begin{equation}
V^{\mathrm{ML}}_{n}\left(\mathbf{R}^{(s)}_{}\right), \hskip 1 cm t_n<t< t_{n+1},
\end{equation}
where $t_n^{}$ and $t_{n+1}^{}$ indicate the
simulated time span according to the collected
times propagated in the ML-PIMD trajectories.
While the practical benefits of this time dependence 
cannot be neglected,
it does not fundamentally change the working 
equations of the SL-PIHMC method.
We therefore opt to exclude the  
subscript $n$ of the MLP to simplify the notation.
As shown in the supplementary material
of Ref. \onlinecite{nagai_self-learning_2020}
the HMC method fulfills the detailed
balance requirement.
This also holds for the SL-HMC method,
as the time dependence of the MLP does 
not change the derivation given there.

%%%

In Sec. SI of the supplementary material we have derived the
following form of the acceptance probability:
\begin{equation}
P_{\mathrm{acc}}^{}\left(\{\mathbf{P},\mathbf{Q}\}\rightarrow \{\mathbf{P}',\mathbf{Q}'\}\right)\sim
\min\left(1,\exp\left(-\beta\Delta\Delta V\right)\right),
\end{equation}
under the assumption that the ML-PIMD 
trajectories conserve the energy of the 
system.
Here the difference between the FP and MLP energies
is introduced as
\begin{equation}
\Delta\Delta V\equiv \frac{1}{P}\sum_{s=1}^{P}\Delta V\left(\mathbf{R}^{(s)}_{}\right)
-\Delta V\left(\mathbf{R}'^{(s)}_{}\right),
\end{equation}
with
\begin{equation}
\Delta V\left(\mathbf{R}^{(s)}\right)\equiv V^{\mathrm{FP}}\left(\mathbf{R}^{(s)}\right)
-V^{\mathrm{ML}}\left(\mathbf{R}^{(s)}\right).
\end{equation}
It should here be stressed that while the relation
in Eq. (5) is very likely to hold,
only Eq. (1) is  used to calculate $P_{\mathrm{acc}}$
in the PIHMC method.
The relation in Eq. (5) is only introduced to illustrate what
governs the size of $P_{\mathrm{acc}}^{}$ in the PIHMC method below.

%%%

Given the relationship in Eq. (5) we see that 
if the ML-PIMD propagates to a region where the difference 
between the MLPs and FP is smaller, 
i.e., ($\Delta V \left(\mathbf{R}'^{(s)}_{}\right) 
\leq \Delta V \left(\mathbf{R}^{(s)}_{}\right)$)
for all beads, 
the step is always accepted, i.e., $P_{\mathrm{acc}}^{}=1$.
On the other hand, if the potentials of the MLPs and FP 
at the initial position are equal or very close to each other, i.e., 
$\Delta V\left(\mathbf{R}^{(s)}_{}\right) \approx 0$ 
for all beads,
$P_{\mathrm{acc}}^{}$ of the step will only depend 
on the difference between the MLPs and FP 
at the end point of the ML-PIMD propagation $\{\mathbf{P',Q'}\}$.
An example that is useful to think of here
is going from a region where the MLP
very accurately reproduces the FP energy
to a place where extrapolation error creates
an unphysical hole in the MLP.
In this case, $P_{\mathrm{acc}}$ would be greatly reduced 
since $\Delta\Delta V \approx -\frac{1}{P}\sum^P_{s=1} \Delta V(\mathbf{R'}^{(s)}_{}) \ll 0$.
Since the MLP used in the start of the SL-PIHMC 
procedure may not be well-trained across the phase space, 
it is important to avoid stepping too far into the untrained regions. 
This is essential for maintaining the high efficiency of the underlying 
PIHMC method in accurately sampling the phase space of the FP method.  

%%%%%%%%%%%%%%%%%%%%%%%%%%%%%%%%%%%%%%%%%%%%%%%%%%%%%%%%%%%%%%%%%%%%%%%%
\subsection{Self-Learning Path Integral Hybrid Monte Carlo using a Mixed FP and ML Potentials}
%%%%%%%%%%%%%%%%%%%%%%%%%%%%%%%%%%%%%%%%%%%%%%%%%%%%%%%%%%%%%%%%%%%%%%%%

While well-trained MLPs are generally believed to 
give a good approximation of the FP potential energy surface,
they will inevitably differ from the true FP potential.
In the literature, a mean absolute error (MAE) for energy per atom 
$\left(\sigma^{}_{\mathrm{at}}\right)$ in the system of
around 1 meV per atom is generally
considered a threshold for a satisfyingly
converged MLP.
Naturally, the MAE for the whole system 
$\left(\sigma^{}_{\mathrm{sys}}=N\sigma^{}_{\mathrm{at}}\right)$
will grow with the number of atoms $N$ in the system.
This is also expected to be the case if one were to 
train the MLP for a small system and then use it on a larger system.
In terms of the classical HMC, i.e., $P=1$ in PIHMC,
this means that $P^{}_{\mathrm{acc}}$ will naturally 
decrease with an increasing system size.
However, under the assumption that the two points
in phase space compared in the MC step are independent, 
the error will be dominated by the error in the MLP, which
grows larger with the system size.
For PIHMC the picture is a little more complicated
since the atomic positions in the beads are coupled.
We do, however, expect that this will lead to increasing errors
when the number of beads $P$ increases,
given that the region in the FP and ML potentials where each bead is 
located will likely have a similar error.
This will then further decrease the acceptance ratio of the 
PIHMC method over the HMC method,
especially in the case where both $N$ and $P$ are large.

%%%

The decrease in $P_{\mathrm{acc}^{}}$ directly affects the speed 
with which the phase space is sampled.
By extension, this also slows down the training of the MLP, 
which, in turn, does not allow us to reduce the errors in the
MLP by a more sampled training set.
In order to increase $P^{}_{\mathrm{acc}}$ for larger systems 
and PI simulations to maintain a reasonable acceptance, we suggest 
to modify $P^{}_{\mathrm{acc}}$ in the following way:
\begin{equation}
P_{\mathrm{acc}}\left(\{\mathbf{P},\mathbf{Q}\}\rightarrow \{\mathbf{P}',\mathbf{Q}'\}\right) =
 \min\left(1, \exp\left(-\beta\left(H_{\mathrm{MIX}}^{}\left(\{\mathbf{P}',\mathbf{Q}'\}\right)
 -H_{\mathrm{MIX}}^{}\left(\{\mathbf{P},\mathbf{Q}\}\right)\right)\right)\right)
\end{equation}
where the mixed Hamiltonian is given as
\begin{equation}
H_{\mathrm{MIX}}^{}\left(\{\mathbf{P}, \mathbf{Q}\}\right)=H_{\mathrm{FP}}\left(\{\mathbf{P}, \mathbf{Q}\}\right)-\left(1-\alpha\right)\left(V^{\mathrm{FP }}_{\mathrm{av}}\left(\{\mathbf{Q}\}\right)-V^{\mathrm{ML}}_{\mathrm{av}}\left(\{\mathbf{Q}\}\right)\right),
\end{equation}
and similarly for the phase space point
$\left\{\mathbf{P}', \mathbf{Q}'\right\}$.
$\alpha$ is a tunable parameter between 0 and 1  that effectively
 allows a bigger discrepancy between the FP and MLP.
The mixed Hamiltonian can also be seen as a special case of 
the Hamiltonian given in Eq. (2),
where the potential is given as
\begin{equation}
V^{\mathrm{MIX}}_{\mathrm{av}}\left(\left\{\mathbf{Q}\right\}\right)=
\alpha V^{\mathrm{FP }}_{\mathrm{av}}\left(\left\{\mathbf{Q}\right\}\right)
+\left(1-\alpha\right)V^{\mathrm{ML}}_{\mathrm{av}}\left(\left\{\mathbf{Q}\right\}\right).
\end{equation}
We denote this method path integral hybrid
 Monte Carlo with potential mixing (PIHCM-MIX).
Correspondingly, if we allow for MLP retraining during the
propagation, we denote the method as self-learning PIHMC-MIX (SL-PIHMC-MIX).
As shown in Figure 1(c), this method does not sample the phase 
space of the FP functional,
but rather the phase space of the  mixed potential energy 
surface, $V^{\mathrm{MIX}}_{}$.
Besides the change in the potential term, the steps in the algorithm are
the same as for the SL-PIHMC method, shown in Figure 1(a).

%%%

Choosing the value of $\alpha$ is a matter of compromise.
On the one hand, a large $\alpha$ value ensures the relevancy of the points 
sampled in the context of the phase space of the FP ensemble.
On the other hand, a small $\alpha$ value allows for faster sampling
although it is less likely that the points sampled are relevant in exploring
the phase space for the FP method.
The efficiency gain also depends on how computationally
cheap the evaluation of the MLP is.
The cheaper the evaluation, the longer one would wish to 
propagate in ML-PIMD before doing a costly  FP calculation.
Finally, it should be mentioned that $\alpha\approx 0$
can have the effect that the ML-PIMD trajectory steps
too far into regions with large extrapolation errors,
which can lead to instability in the FP calculations due to 
sampling of physically irrelevant structures of the system.
For this study, we have chosen $\alpha = 0.25$
as a compromise between the efficiency 
of the MLP and compatibility to the FP phase space.
The effects of this choice on the increase in $n_{\mathrm{ML}}^{}$ 
are shown for the SL-PIHMC-MIX method in Figure 2,
where all training sessions lead to running with $n_{\mathrm{ML}}^{}=128$
and $\left\langle A_{\mathrm{acc}}^{}\right\rangle>0.33$ relatively fast.
We do see some drops in $n_{\mathrm{ML}}^{}$ between segments,
but $A_{\mathrm{acc}}^{}$ quickly recovers and $n_{\mathrm{ML}}^{}$
is increased again.
We also observe that as the training set grows beyond the initial
1000 structures, both the acceptance rate and $n_{\mathrm{ML}}^{}$ increase.
This indicates the importance of longer trajectories to collect training 
data that represent the entire phase space of the studied system.

%%%%%%%%%%%%%%%%%%%%%%%%%%%%%%%%%%%%%%%%%%%%%%%%%%%%%%%%%%%%%%%%%%%%%%%%
\subsection{Reweighting to obtain the FP ensemble distributions of equilibrium properties}
%%%%%%%%%%%%%%%%%%%%%%%%%%%%%%%%%%%%%%%%%%%%%%%%%%%%%%%%%%%%%%%%%%%%%%%%

The PIHMC-MIX method allows us 
to accurately predict the distribution
$\rho^{\mathrm{MIX}}\left(A\right)$ of a 
structural parameter ($A$) in the phase space of 
$H_{\mathrm{MIX}}^{}\left(\{\mathbf{P}, \mathbf{Q}\}\right)$.
We do, however, wish to generate the
distributions in the phase space of 
$H_{\mathrm{FP}}^{}\left(\{\mathbf{P}, \mathbf{Q}\}\right)$,
which is guaranteed by the PIHMC method.
To that end, we employ the reweighting
scheme suggested by Miao \textit{et al.} in Ref. \onlinecite{miao_improved_2014}.
In the exact limit, the trajectory can be divided into $M$ 
equally sized bins,
and the distribution of the structural parameter $A$ of each bin 
can be reweighted in the following way to obtain the distribution
in the FP ensemble:
\begin{equation}
\rho^{\mathrm{FP}}_{}\left(A_{j}^{}\right)=
\rho^{\mathrm{MIX}}_{}\left(A_{j}^{}\right)
\frac{\left\langle \exp\left(\beta\left(\alpha-1\right)\Delta V^{\mathrm{MIX}}_{}\right)\right\rangle_j}
{\sum_{j=1}^M \left\langle \exp\left(\beta\left(\alpha-1\right)\Delta V^{\mathrm{MIX}}_{}\right)\right\rangle_{j}^{}},
\end{equation}
where the counter is the ensemble-averaged Boltzmann factor 
for the simulation frames found in the 
$j$th bin, 
and the potential difference is defined as
\begin{equation}
\Delta V^{\mathrm{MIX}} = \frac{1}{P}\sum_{s=1}^{P}
V^{\mathrm{ML}}\left(\mathbf{R}^{(s)}\right)-
V^{\mathrm{FP}}\left(\mathbf{R}^{(s)}\right).
\end{equation}

%%%

The exact reweighting is, however, difficult to converge due to the 
exponentiation of the potential differences.
To avoid this, the cumulant expansion of the average the exponential is introduced,
\begin{equation}
\left\langle\exp\left(\beta \left(\alpha-1\right)\Delta V_{}^{\mathrm{MIX}}\right)\right\rangle = \exp\left\{\sum_{k=1}^\infty\frac{\beta^k}{k!}C_k\right\},
\end{equation}
where the first cumulant is given as
\begin{equation}
C_1 = \left\langle\left(\alpha-1\right)\Delta V_{}^{\mathrm{MIX}}\right\rangle=
\left(\alpha-1\right)\left\langle\Delta V_{}^{\mathrm{MIX}}\right\rangle.
\end{equation}
The study of Miao \textit{et al.} established that considering only the first
cumulant in this expansion was sufficiently accurate to reweight
the results, 
and we will follow that procedure here.
The ensemble-averaged Boltzmann factor does in this case reduces to
\begin{equation}
 \left\langle\exp\left(\beta\left(\alpha-1\right)\Delta V^{\mathrm{MIX}}_{}\right)\right\rangle \approx 
 \exp\left(\beta C_1\right)=
 \exp\left(\beta\left\langle\left(\alpha-1\right)\Delta V^{\mathrm{MIX}}_{}\right\rangle\right).
\end{equation}
This is then inserted into Eq. (11) and forms the following expression:
\begin{equation}
\rho^{\mathrm{FP}}_{}\left(A_j\right)\approx 
\rho^{\mathrm{MIX}}_{}\left(A_j\right)
\frac{ \exp\left(\beta\left(\alpha-1\right)\left\langle\Delta V^{\mathrm{MIX}}_{}\right\rangle_j\right)}
{\sum_{j=1}^M\exp\left(\beta\left(\alpha-1\right) \left\langle \Delta V^{\mathrm{MIX}}_{}\right\rangle_j\right)}.
\end{equation}
This expression has been used to do reweighting the RDFs 
calculated from the PIHMC-MIX trajectories with a bin size 
of $M=20$.
In the weighting expression, only structures from accepted HMC steps
are considered.
This is done in order to avoid adding artificial weight to structures where
several trial ML-PIMD trajectories are needed before the MC step is accepted.
In Sec. SII of the supplementary material, we discuss the addition of higher-order terms in the cumulant expansion
and find that the resulting RDFs using the first- and second-order expansion for reweighting
PIHMC-MIX data overlap.
Furthermore, the ``anharmonicity''\cite{miao_improved_2014,Lange_proteins_2008} 
observed in the binned data suggests that binning and expansion to second order 
should be sufficient for estimating the exponential reweighting in Eq. (11).

%%%%%%%%%%%%%%%%%%%%%%%%%%%%%%%%%%%%%%%%%%%%%%%%%%%%%%%%%%%%%%%%%%%%%%%%
\section{Computational details}
%%%%%%%%%%%%%%%%%%%%%%%%%%%%%%%%%%%%%%%%%%%%%%%%%%%%%%%%%%%%%%%%%%%%%%%%

All the simulations were undertaken using the \texttt{PIMD} software package
\cite{shiga_pimd_2020}, 
which is capable of conducting PIMD, PIHMC-MIX, and SL-PIHMC-MIX simulations. 
Through an interface to the quantum chemistry software package \texttt{CP2K},\cite{kuhne_cp2k_2020}
potential energy and forces at the FP level within the 
Born-Oppenheimer approximation can be used
for HMC steps and PIMD propagation.
The \texttt{ELPA}\cite{marek_elpa_2014} and \texttt{FFTW}\cite{frigo_design_2005}
libraries were used by \texttt{CP2K} to speed up the solution of the 
electronic structure eigenvalue equations and to carry out fast 
Fourier transform, respectively.
The MLPs were trained and evaluated  using \texttt{AENET}.\cite{artrith_implementation_2016}
The parameters of the Behler-Parrinello structural fingerprint 
parameters\cite{behler_generalized_2007} used here are given in Table SI of the supplementary material.
The neural networks were all prepared with two layers with hyperbolic 
tangential activation functions and 15 nodes per layer
and a single linear combination output layer,
resulting in a total of 1290 free parameters for both the O and H atomic 
potentials. 
 
%%%
 
The RPBE,\cite{hammer_improved_1999} SCAN,\cite{sun_strongly_2015} rev-vdW-DF2,\cite{hamada_van_2014}
and optB88-vdW\cite{klimes_chemical_2009} functionals were used from their 
implementations in the \texttt{libxc} library.\cite{marques_libxc_2012,lehtola_recent_2018}
Grimme's D3 dispersion correction\cite{grimme_consistent_2010,grimme_effect_2011} 
 was employed to model the van der Waals interactions in the RPBE functional. 
The electronic structure calculations in the periodic boundary condition (PBC) 
were performed using the Gaussian and plane-wave (GWP) 
method\cite{lippert_hybrid_1997} with the plane wave cutoffs of 500 Ry for 
the RPBE functional and  800 Ry for the other functionals
to expand the charge density.
%in the study.
%
Only the $\Gamma$-point was used for the Brillouin zone sampling.
% The electronic structure calculations in the periodic boundary condition (PBC) 
% were performed for the $\Gamma$-point based on the Gaussian and plane-wave (GWP) 
% method\cite{lippert_hybrid_1997} with the plane wave cut    off at 500 Ry for 
% the RPBE functional and  800 Ry for the other functionals in the study.
%
The plane-wave basis set was combined with the TZV2P basis set \cite{vandevondele_gaussian_2007} 
 associated with the Goedecker-Teter-Hutter (GTH) pseudopotentials\cite{goedecker_separable_1996} to describe the electron-ion interactions.

%%%

All simulations were carried out in the NVT ensemble with 64 or 256 water 
molecules in a cubic box with PBC.
The volume of the cubic box was chosen to match the experimental 
density at 298.15 K (1.00 g/ml), 
i.e., the side lengths of the box at ambient conditions were
set to 12.41 and 19.71 \AA~for the systems with 64 and 256 water 
molecules respectively.
The temperature was controlled with the massive Nos{\'e}-Hoover chain (MNHC) 
thermostats\cite{hoover_canonical_1985,nose_unified_1984,martyna_nosehoover_1992} 
in all PIMD and MD simulations.
The number of imaginary time slices (the number of beads) were 
$P=1$ and $P=32$ for the classical and quantum simulations, respectively.
All simulations were conducted with a time step of $\Delta t=0.25$ fs.
ML-MD, FP-PIMD, and ML-PIMD were each propagated for 100\,000 steps, 
corresponding to a trajectory length of 25 ps for each of those trajectories,
while the AI-MD simulation was propagated for 200\,000 steps, 50 ps, to ensure convergence of the RDFs.
The error bars for the RDFs from MD and PIMD simulations were calculated by 
dividing the trajectory into four blocks and calculating the standard deviation of 
the RDFs from the blocks.
The central bold lines of the RDF plots were calculated as the 
average of the RDFs from these blocks.

%%%

The SL-HMC-MIX and SL-PIHMC-MIX simulations were initialized with an MLP trained
from around 1000 structures from short FP-MD and FP-PIMD trajectories.
These trajectories were initialized using the final structure from the previously reported RPBE-D3 FP-PIMD and FP-MD simulations for all functionals.
The self-learning process was run for 5000 steps with retraining every 
$n_{\mathrm{FP}}=500$ MC steps.
The other parameters governing the process were set to,
$n^{}_{\mathrm{test}}=50$, 
$n^{\mathrm{ML}}_{\mathrm{max}}=128$,
$n^{\mathrm{ML}}_{\mathrm{min}}=2$,
$P^{}_{\mathrm{upper}}=40 \%$, and
$P^{}_{\mathrm{lower}}=10 \%$
respectively.
The initial number of ML steps ($n^{}_{\mathrm{ML}}$) was set to 2.
For the SL-PIHMC-MIX trajectories, the 32 structures and energies of the beads
were saved in every 20th PIHMC step for use in training,
resulting in training sets containing around 9000 structures at the end
of the SL-PIHMC-MIX simulation.
While for the SL-HMC-MIX trajectory for RPBE-D3,
every structure from the HMC steps was used for training, 
resulting in a training set containing around 6000 structures.

%%%

The average acceptance rate 
$\left(\left\langle A_{\mathrm{acc}}^{}\right\rangle\right)$ 
and effective trajectory length $\left(t_{\mathrm{eff}}^{}\right)$ of SL-HMC-MIX and
SL-PIHMC-MIX trajectories used in this study are 
given in Table SII of the supplementary material.
The definition of $t_{\mathrm{eff}}^{}$ relies on dividing the
PIHMC trajectory of length $n_\mathrm{PIHMC}^{}$ into 
$O=\frac{n_\mathrm{PIHMC}^{}}{n_\mathrm{test}^{}}$ sub-trajectories.
$t_{\mathrm{eff}}^{}$ of the full PIHMC trajectory can then be calculated as
\begin{equation}
t_{\mathrm{eff}}=\sum_i^{O} n_\mathrm{acc}^{i}n_\mathrm{ML}^{i}\Delta t,
\end{equation}
where $n_\mathrm{acc}^{i}$ and $n_\mathrm{ML}^{i}$ 
are the number of accepted steps and the number of ML-PIMD 
steps taken, respectively, in the $i$th sub-trajectory.
ML-MD and ML-PIMD trajectories were run as described for the FP-PIMD 
and FP-MD simulations previously using the trained MLPs trained by 
SL-HMC-MIX and SL-PIHMC-MIX.
The HMC-MIX and PIHMC-MIX trajectories using these trained 
MLPs were all run for 5000 steps with the initial setting 
$n^{}_{\mathrm{ML}}=128$.
The resulting $t_{\mathrm{eff}}^{}$ are given in Table I and Table SIII of the supplementary material.
The SL-PIHMC-MIX trajectories served as equilibration for both the PIHMC-MIX and ML-PIMD trajectories. 
For the ML-MD and HMC-MIX trajectories, the final structure from the RPBE-D3 FP-MD simulation from our previous work was used as an equilibrated structure. 

%%%%%%%%%%%%%%%%%%%%%%%%%%%%%%%%%%%%%%%%%%%%%%%%%%%%%%%%%%%%%%%%%%%%%%%%
\section{Results}
%%%%%%%%%%%%%%%%%%%%%%%%%%%%%%%%%%%%%%%%%%%%%%%%%%%%%%%%%%%%%%%%%%%%%%%%

%%%%%%%%%%%%%%%%%%%%%%%%%%%%%%%%%%%%%%%%%%%%%%%%%%%%%%%%%%%%%%%%%%%%%%%%
\subsection{RPBE-D3 PIHMC-MIX}
%%%%%%%%%%%%%%%%%%%%%%%%%%%%%%%%%%%%%%%%%%%%%%%%%%%%%%%%%%%%%%%%%%%%%%%%

The RPBE-D3 functional has previously been used to model both room temperature 
and sub- and supercritical water in FP-MD studies by Schienbein and Marx.
\cite{schienbein_liquidvapor_2018, schienbein_supercritical_2020}
We have also used the functionals in FP-PIMD studies
of both liquid water at room temperature, and under sub-
and supercritical conditions,\cite{thomsen_ab_2021} 
and its isotopologues at room  temperature.\cite{Thomsen_structures_2022}
For Sub- and supercritical water, there are a number of differences
between the experimentally recorded structures and those
found even when including NQEs.
For room temperature, our previous works show that the RPBE-D3 
gives a good agreement with the experimental RDFs when NQEs are included, 
which can be seen in the comparisons of  FP-PIMD to 
the experimental RDFs\cite{soper_radial_2000, soper_radial_2013} 
in part (a) of Figures 3-5.
The quantitative agreement is also very good as shown by the peak positions
 and heights in Table II-IV,
where the largest discrepancies are found at the interstitial region and 
second peak of the O-O RDF, i.e., in the second hydration shell.
The height of the second peak is comparable between FP-PIMD and the 
experiment at 1.19 and 1.12, respectively,
while the position of the second peak of FP-PIMD is at 4.35 \AA, 
while for the experiment, it is at 4.53 \AA.
This indicates that the second hydration shell and disordered water 
around the first hydration shell are not well described by the RPBE-D3 
functional.
This might, however, also be a finite size effect,
as the box size is limited to 12.41 \AA\ in those studies
due to the cost of FP-PIMD for larger systems.
This claim will later be addressed by ML-PIMD studies of
larger system sizes in Subsection IV C.
We will in the following use the trajectory data from our previous studies
to confirm the ability of the PIHMC-MIX method to reproduce
the results of FP-PIMD simulations.

%%%

The PIHMC-MIX results for the RPBE-D3 functional were based on using an 
MLP trained from an SL-PIHMC-MIX trajectory with $t_\mathrm{eff}^{}=69.3$ ps
(see Table SII of the supplementary material).
One of the features of the PIHMC-MIX method is that it with the reweighting 
scheme will reproduce the results of the FP-PIMD as described in Section II C.
The only difference between the two methods is that the PIHMC-MIX method 
should be able to explore phase space more efficiently and, thus, require 
fewer FP calculations.
The production run of PIHMC-MIX after the training was completed 
had $t_{\mathrm{eff}}^{}=99.9$ ps
and maintained a high acceptance ratio of 55.5 \% while running with 128 
steps for the entire run of 5000 HMC steps.
The resulting RDFs are given in part (b) of Figures 3-5 
with the peak positions and heights given in Table
II, III and IV for O-O, O-H and H-H pairs respectively.
For the O-H and H-H pairs the FP-PIMD and PIHMC-MIX results of
the peak positions and heights as well as the other points on the curves match within $\pm 0.03$ \AA\ on $R_{\mathrm{XY}}^{}$ and $\pm 0.03$ on $G\left(R_{\mathrm{XY}}^{}\right)$, 
which we estimate to be within the error bar of the FP-PIMD 
simulation due to the length of the trajectory. 
The first peak of the O-O RDF matches similarly to the FP-PIMD result,
but for the second O-O RDF peak the maximum for the FP-PIMD
at 4.35 \AA\ shifts to 4.24 \AA\ for the PIHMC-MIX trajectory.
Part of the reason for this is believed to be the difference
in sampling of the two trajectories,
given that the PIHMC-MIX trajectory is effectively almost
four times longer than the FP-PIMD trajectory.
We conclude that the PIHMC-MIX method reproduces the
structure observed from FP-PIMD simulations,
while using an order of magnitude fewer FP calculations,
5000 vs 100\,000 for the PIHMC-MIX and FP-PIMD simulations
respectively and possibly giving a more complete sampling of
the phase space,
thus, accepting the description of the second hydration shell
calculated from the PIHMC-MIX trajectory as the correct
description within the simulations run with the RPBE-D3 functional.

%%%

The MLP trained using the SL-PIHMC-MIX trajectory 
can also be employed to conduct an HMC-MIX simulation,
i.e., a simulation without NQEs. 
The RDFs plotted in Figures 3-5 (d) show agreement between HMC-MIX and FP-MD, 
similar to that found for PIHMC-MIX and FP-PIMD. 
This is further confirmed by comparing the peak heights and positions in Tables SIV-VI of the supplementary material. 
Once again, the HMC-MIX method samples more efficiently than its FP-MD counterpart, 
achieving an effective trajectory length of 103.7 ps compared to 50 ps for the AI-MD simulation.
Additionally, only 10\,000 FP calculations were required for HMC-MIX vs 200\,000 for FP-MD.

%%%%%%%%%%%%%%%%%%%%%%%%%%%%%%%%%%%%%%%%%%%%%%%%%%%%%%%%%%%%%%%%%%%%%%%%
\subsection{Influence of $\alpha$ on the performance of PIHMC-MIX}
%%%%%%%%%%%%%%%%%%%%%%%%%%%%%%%%%%%%%%%%%%%%%%%%%%%%%%%%%%%%%%%%%%%%%%%%

Table I includes the average acceptance rates 
$\left(\left\langle A_{\mathrm{acc}}\right\rangle\right)$, 
number of ML steps ($n_{\mathrm{ML}}$), and $t_{\mathrm{eff}}^{}$ 
of the PIHMC-MIX trajectories run using the same MLP but with 
$\alpha\in\{0.25, 0.5, 0.75, 1.0\}$,
with $\alpha=1.0$ corresponding to the unmodified PIHMC method.
We find that by increasing $\alpha$ we lower both the acceptance ratio
and, more critically, $n_{ML}$,
resulting in shorter $t_{\mathrm{eff}}^{}$ for even a significantly 
larger number of HMC steps.
In Sec. SIV, the resulting RDFs for water for the  different values 
for $\alpha$ are compared.
A good qualitative and quantitative agreement for the O-H and H-H RDFs are
found for the PIHMC-MIX simulations with $\alpha\in\{0.25, 0.5, 0.75, 1.0\}$ 
with the FP-PIMD results.
As is the case for the comparison of FP-PIMD and PIHMC-MIX with $\alpha=0.25$,
however, the interstitial region and the second peak of the O-O RDFs are
not sufficiently converged for $\alpha\in\{0.5, 0.75, 1.0\}$.
This is likely due to the fact that the structure of the second hydration
shell is intrinsically harder to sample than the first hydration shell.
Given that the $t_{\mathrm{eff}}^{}$ of the PIHMC-MIX trajectory 
with $\alpha=0.25$ is 99.9 ps, 
it is assumed that this represents the most converged result reported here.
Giving enough computational time, the O-O RDFs for 
$\alpha\in\{0.5, 0.75, 1.0\}$ would converge to the same result, but the low acceptance rate might make it prohibitory expensive to extend these trajectories.

%%%%%%%%%%%%%%%%%%%%%%%%%%%%%%%%%%%%%%%%%%%%%%%%%%%%%%%%%%%%%%%%%%%%%%%%
\subsection{Accuracy of trained MLPs for water}
%%%%%%%%%%%%%%%%%%%%%%%%%%%%%%%%%%%%%%%%%%%%%%%%%%%%%%%%%%%%%%%%%%%%%%%%

The accuracy of the MLPs generated by the SL-PIHMC-MIX method
is generally found to be comparable to those trained in other studies,
which bodes well for their use in ML-MD and ML-PIMD studies.
It is, however, important to stress that no matter how poorly trained the MLP is, 
PIHMC-MIX will still be able to reproduce the FP-PIMD result through the 
reweighting of the property distributions,
given that $t_{\mathrm{eff}}^{}$ is long enough.
A simple way of checking the quality of the MLP is the instantaneous acceptance 
rate and $n_{\mathrm{ML}}^{}$ in PIHMC-MIX,
which both in the case of a well-trained MLP should be high.
In this section, we will look more carefully at the trained MLPs and the accuracy
of ML-PIMD based on the trained MLPs compared to FP-PIMD and PIHMC-MIX.

%%%

In part (c) of Figs. 3-5 the O-O, O-H and H-H RDFs, respectively, 
are displayed for ML-PIMD simulations with water
systems containing 64 and 256 water molecules using
the MLP fitted during the SL-PIHMC-MIX training process 
with the RPBE-D3 functional.
The quantitative agreement with FP-PIMD is found to be slightly 
worse than the case for PIHMC-MIX, 
as seen from the peak positions and heights given in 
Table SIV-SVI of the supplementary material.
The larger water systems are included in an effort to examine finite size 
effects on the RFDs and test the behavior of the MLP under NPT-like conditions
for the first and second hydration shells.
In the comparison between the systems containing 64 and 256 water molecules,
we find no significant finite size effects and, thus, 
conclude that the fitted MLP is extendable to larger water system sizes.
Furthermore, the size of the systems studied using FP-PIMD and PIHMC-MIX are
sufficiently large for studies of the first and second hydration shell
structure of water.

%%%

Validation of the MLPs themselves is done in Section SVII of the supplementary material,
where the energies and forces obtained from FP and ML 
calculations of the same structures are compared.
The results are in line with those of previous studies training MLPs
for water systems.
The MAE for energy per atom $\left(\sigma_{\mathrm{E}}^{\mathrm{at}}\right)$ is 0.36 meV/atom and the MEA for force  
$\left(\sigma_{\mathrm{F}}^{\mathrm{at}}\right)$ is 79.0 meV/\AA,
these quantities are described in Eq. (S12) and (S14).
We also tested the transferability of an MLP trained using SL-HMC-MIX, 
i.e., the MLP is constructed without considering NQEs when
creating the FP training data.
Here, we find that both $\sigma_{\mathrm{E}}^{\mathrm{at}}$ and $\sigma_{\mathrm{F}}^{\mathrm{at}}$
are more significant at 3.58 meV/atom and 199.9 meV/\AA\ respectively.
This indicates that the MLP trained without including data reflecting NQEs 
in their training sets, while transferable, 
will not give accurate modeling of the NQEs of the system studied.
In Figures  3-5 (d), 
we have plotted the RDFs of ML-MD simulations using the MLP trained by the SL-PIHMC-MIX method, 
which reproduces the FP-MD results with good qualitative agreement. 
The quantitative agreement of the peak positions, 
as shown in Tables SIV-VI of the supplementary material, 
is also found to be acceptable. 
In Sec. SVIII of the supplementary material we find that the transferability of 
the MLP trained by SL-PIHMC-MIX is generally greater than that trained using 
SL-HMC-MIX when considering the RDFs calculated by either method using ML-MD or ML-PIMD.

%%%

The transfer of the MLP trained using SL-PIHMC-MIX to be used 
in an HMC-MIX production run is, however, found to be smoother.
Here, we find $t_{\mathrm{eff}}^{}$ to be around 88 ps 
and $\sigma_{\mathrm{E}}^{\mathrm{at}}$ and $\sigma_{\mathrm{F}}^{\mathrm{at}}$
at 0.70 meV/atom and 0.20 eV/\AA~, respectively.
This agreement can, however, stem from
the selection of training data in the SL-PIHMC-MIX method,
where the proximity of the 32 beads from each HMC step
could be argued to form a training set similar to that
suggested by Cooper \textit{et al.} in Ref. \onlinecite{cooper_efficient_2020} to approximate the inclusion
of gradients in the fitting of the MLP.
In that study, the FP data-set was augmented with slightly distorted 
structures where the energy was calculated by Taylor expansion using
the FP energy and gradients of a known structure.
Here, we do not extrapolate, 
rather we calculate the FP energies of several 
distorted points directly, 
but this might lead to an increase in the accuracy of the
MLP, as shown in Ref \onlinecite{cooper_efficient_2020}. 

%%%%%%%%%%%%%%%%%%%%%%%%%%%%%%%%%%%%%%%%%%%%%%%%%%%%%%%%%%%%%%%%%%%%%%%%
\subsection{Simulations of heavy water (D$_2$O)}
%%%%%%%%%%%%%%%%%%%%%%%%%%%%%%%%%%%%%%%%%%%%%%%%%%%%%%%%%%%%%%%%%%%%%%%%

Another way of examining the transfer-ability of the MLPs and FP models is the 
comparison of NQEs in both H$_{2}^{}$O and D$_{2}^{}$O.
As the structure of both liquids at room temperature is known 
experimentally\cite{soper_radial_2000,soper_quantum_2008,soper_radial_2013}
and shows significant differences,
these differences are large enough to not have overlapping error bars
in theoretical studies.
It is however rarely done due to the added cost of running two 
separate PIMD simulations in place of one.
When two simulations are run, they might reveal problems with the 
underlying potential model.
For example the GGA functional BLYP-D2 was found to not reproduce
the correct order of the O-O peak heights of H$_{2}^{}$O and D$_{2}^{}$O 
at room temperature,\cite{machida_nuclear_2017}
this difference was ascribed to the description of dispersion in the
functional and the D2 correction.
This failure to reproduce the correct ordering is, however, not present for 
RPBE-D3.\cite{Thomsen_structures_2022}
Furthermore, previous studies using MLPs trained on FP data from the 
PBE0-TS hybrid functional\cite{ko_isotope_2019} and SCAN 
meta-GGA functional,\cite{xu_isotope_2020}
also show the correct isotopic ordering while being overall over structured
compared to the experimental results for both H$_{2}^{}$O and D$_{2}^{}$O.
This exemplifies the delicate balance in the description
of the intermolecular potentials needed to model NQEs correctly 
in both isotopologues of water.

%%%

In Fig. 6, the RDFs for all pairs in D$_{2}^{}$O are presented for FP-PIMD, 
\cite{Thomsen_structures_2022} PIHMC-MIX,
and ML-PIMD based on the MLP fitted by SL-PIHMC-MIX for H$_{2}^{}$O.
The peak heights and positions of these RDFs are given in Table SXI of the 
supplementary material.
For the PIHMC-MIX trajectory, we obtain a result that agrees 
with the FP-PIMD reference data for RPBE-D3.
For the ML-PIMD trajectory, minor deviations from the FP-PIMD results are found
for the O-D and D-D RDFs.
The second peak of the O-O RDF shows similar deviations as those
discussed for H$_{2}^{}$O between the three models.

%%%

The first peak and the interstitial region of the O-O RDFs do occur at 
similar distances for all trajectories,
and the heights for these two extrema (h$_{1}^{\mathrm{OO}}$ 
and h$_{\mathrm{min}}^{\mathrm{OO}}$) are, however, quite different.
These heights for FP-PIMD are found to have the values (2.65, 0.87) for D$_{2}^{}$O,
whereas they are (2.47, 0.83) in the case of H$_{2}^{}$O.
These results are in line with those we have calculated here by PIHMC-MIX, 
(2.61, 0.73) and (2.53, 0.77) for D$_{2}^{}$O and H$_{2}^{}$O, respectively.
Both FP-PIMD and PIHMC-MIX results match well with the experimental values 
for D$_{2}^{}$O, (2.62, 0.79),\cite{soper_quantum_2008} 
and H$_{2}^{}$O, (2.50, 0.78).\cite{soper_radial_2013}
For ML-PIMD, these heights are (2.55, 0.78) and (2.42,0.87) for D$_{2}^{}$O and
H$_{2}^{}$O, respectively.
This gives the impression that the MLP on its own does not fully reproduce 
the FP-PIMD results, 
especially in the case of D$_2$O.

%%%

In order to improve the agreement between ML-PIMD and FP-PIMD, some D$_{2}^{}$O
structures and energies were added to the FP training set of the MLP by 
running an additional 2000 step SL-PIHMC-MIX simulation for D$_{2}^{}$O after 
the initial 5000 steps for H$_2$O.
The resulting MLP is then used for ML-PIMD, the RDFs are given in Fig. S5 
of the supplementary material, and the peak positions and heights are given 
in Tables SXI and SXII of the supplementary material for D$_{2}^{}$O 
and H$_{2}^{}$O, respectively.
A better agreement for D$_2$O is observed with the heights of the two first O-O RDF
extrema given as (2.62, 0.75),
while for H$_{2}^{}$O, these are found to be (2.40, 0.87), which
is slightly worse than before.
It is therefore not certain if it is possible to make
a balanced MLP able to reproduce the O-O RDFs calculated by the FP-PIMD
result for both H$_2$O and D$_2$O simultaneously.
The cheap cost of running the PIHMC method does however make it feasible
to simply run a simulation for both H$_{2}^{}$O and D$_{2}^{}$O
to confirm the values of equilibrium properties.
If one needed the MLPs for studying the dynamics of the liquid,
it would be recommended to run a separate SL-PIHMC-MIX training
for D$_{2}^{}$O,
where the trained MLP for H$_{2}^{}$O could be used to speed
up the sampling of phase space significantly.

%%%%%%%%%%%%%%%%%%%%%%%%%%%%%%%%%%%%%%%%%%%%%%%%%%%%%%%%%%%%%%%%%%%%%%%%
\subsection{SCAN, rev-vdW-DF2 and optB88-vdW results}
%%%%%%%%%%%%%%%%%%%%%%%%%%%%%%%%%%%%%%%%%%%%%%%%%%%%%%%%%%%%%%%%%%%%%%%%

Given the efficiency gains demonstrated for the RPBE-D3 functional,
we are able to extend the study of the effect of NQEs in DFT
functionals to the SCAN, rev-vdW-DF2, and optB88-vdW functionals
using a limited computational effort compared to that required
to run FP-PIMD simulations for each functional.
While there are no FP-PIMD data available for all of these functionals,
the PIHMC-MIX method has been shown to reproduce the FP-PIMD
results in 5000 HMC steps, given a high $n_{\mathrm{ML}}^{}$ and
$\left\langle A_{\mathrm{acc}}^{}\right\rangle$.
$\left\langle A_{\mathrm{acc}}^{}\right\rangle$ and
$t_{\mathrm{eff}}^{}$ for the three functionals are given in Table II and Table SII of the 
supplementary material for the PIHMC-MIX and HMC-MIX trajectories, respectively.
We find that while the acceptance rates are smaller
than they were for RPBE-D3, they are still high enough
for the SCAN and rev-vdW-DF2 functionals to be able to
run PIHMC-MIX with $t_{\mathrm{eff}}^{}$ of
96.8 and 88.7 ps, respectively.
The performance for the optB88-vdW functional is, however, less promising, 
with an average acceptance rate of 36.8 \% and $t_{\mathrm{eff}}^{}$
drops to 59.4 ps.

%%%

The accuracy of the MLPs from the SL-PIHMC-MIX trajectories is 
analyzed in Figs. S3(b) and S3(c) of the supplementary material 
for the SCAN functional and rev-vdW-DF2 functional, respectively.
In those figures, the energies and forces calculated by FP and ML from the same
structure taken from the PIHMC-MIX trajectories are compared.
$\sigma_{\mathrm{E}}^{\mathrm{at}}$ and $\sigma_{\mathrm{F}}^{\mathrm{at}}$ for the 
SCAN and rev-vdW-DF2 functionals are 0.44 meV/atom, 61.3 meV/\AA\ and 0.59 meV/atom, 60.2 meV/\AA\ respectively.
The MLP trained using the optB88-vdW functional has larger errors
when comparing to FP results at 2.51 meV/atom and 109.2 meV/\AA\ 
for $\sigma_{\mathrm{E}}^{\mathrm{at}}$ and $\sigma_{\mathrm{F}}^{\mathrm{at}}$, respectively,
and the distributions of energies and forces in Fig. S3 (d) of the supplementary information
are also more spread out than for MLPs constructed using FP data from 
the other functionals.
The quality of the underlying MLP for a given functional
should not change the results of PIHMC-MIX;
it should only affect $t_{\mathrm{eff}}^{}$ through
low $\left\langle A_{\mathrm{acc}}^{}\right\rangle$ 
and $n_{\mathrm{ML}}^{}$.
The 59.4 ps trajectory for optB88-vdW should in this context still
be sufficient to converge the RDFs of the two first hydration
shells of water.

%%%

For the RDFs calculated using ML-PIMD based on FP data from 
SCAN and rev-vdW-DF2, the agreement to the PIHMC-MIX results
from the same functional is found to be similar
to that of the RPBE-D3 functional discussed above for 
the O-H and H-H RDFs.
For the O-O RDFs, the MLPs for SCAN and rev-vdW-DF2 even
seems better at reproducing the PIHMC-MIX results than
for the RPBE-D3 functional.
This might be due to the descriptor chosen to be better at describing
the more structured O-O RDFs found for SCAN and rev-vdW-DF2.
In the case of optB88-vdW, we find significant discrepancies in all RDFs,
especially the first O-O peak and the secondary O-H and H-H peaks.
This indicates that the description of the H-bond is not the same in the FP 
and ML potentials.
The results from PIHMC-MIX should, however, be correct for this
and should be indicative of the true performance of the optB88-vdW
functional for modeling water.

%%%

In the Sec. \ref{scansub}, \ref{revsub} and \ref{optsub} 
below we will analyze the calculated RDFs
of the SCAN, rev-vdW-DF2 and optB88-vdW functionals.
The effects of NQEs on the RDFs will also be discussed by comparing 
the PIHMC-MIX results to those obtained from HMC-MIX,
in both cases using the MLPs trained by SL-PIHMC-MIX. 
The RDFs including NQEs calculated using PIHMC-MIX 
are give in Figures 7-9, 
with the peak positions and heights are given in Table II, III, and IV
for O-O, O-H and H-H RDFs respectively.
For the RDFs calculated using HMC-MIX see Figure S6 of the SM and
the peak positions and heights are given in Table SXIII.

The inclusion of NQEs does naturally soften the intramolecular O-H bonds
and H-O-H angles the most due to the low mass of the hydrogen atoms 
and high zero point energies of the intramolecular degrees of freedom.
We thus find that for all functionals studied here that the
O-H and H-H RDFs in general and in particular the first peaks of these, 
are softened from the values obtained by HMC-MIX
by the inclusion of NQEs in the PIHMC-MIX simulations.
We will therefore focus on the softening of the O-O RDFs
when comparing classical and quantum results in the following sections,
as these are more sensitive to the intermolecular interactions and 
thus are more challenging to reproduce accurately.

%%%%%%%%%%%%%%%%%%%%%%%%%%%%%%%%%%%%%%%%%%%%%%%%%%%%%%%%%%%%%%%%%%%%%%%%
\subsubsection{The Scan Functional}\label{scansub}
%%%%%%%%%%%%%%%%%%%%%%%%%%%%%%%%%%%%%%%%%%%%%%%%%%%%%%%%%%%%%%%%%%%%%%%%

With the SCAN functional, the O-O RDF from PIHMC-MIX
in figure 7(b) is over-structured compared to the 
experimental RDFs.
Looking to the second peaks of the O-H and
H-H RDFs, we find that the hydrogen positions
for the hydrogens participating in the hydrogen
bond are more localized than in the experimental
RDFs.
The inclusion of NQEs in the PIHMC-MIX simulation
softens the liquid structure somewhat,
with the O-O RDFs first peak height changing from
3.41 to 3.24 in the HMC-MIX and PIHMC-MIX simulations
respectively.
Furthermore, softening is also observed for the
second peaks of the O-H and H-H RDF with
the inclusion of NQEs.
The changes in heights are comparable to those
observed for the RPBE-D3 functional,
but given that the O-O RDF without NQEs is much
more structured for the SCAN functional,
this softening with the inclusion of NQEs
is not enough to reproduce the experimental
structure.

%%%

The SCAN functional has been studied using both
FP-MD\cite{chen_ab_2017},
 FP-PIMD\cite{li_static_2022, herrero_connection_2022},
and MLP based methods\cite{yao_temperature_2020} 
using MLPs trained on FP data from the SCAN functional.
These  studies have also shown a tendency to over structure 
the liquid in the NVT ensemble,
even when including NQEs at room temperature.
The over-structuring when using classical MD simulations
have been attributed to the lack of NQEs,
which lead to the practice of simulating water
at 330 K in an effort to emulate the effects of NQEs
\cite{chen_ab_2017}.
However, a study by Yao and Kanai\cite{yao_temperature_2020} found this practice
problematic due to a fortuitous cancellation of errors in the underlying potential
energy surface, which allowed for an accurate reproduction of the O-O RDFs and other 
properties of the room temperature liquid.
In this study, we similarly examined the local structure of the hydrogen bond and compared to the experimental work of Modig, Pfrommer and Halle\cite{modig_temperature-dependent_2003} in section SXII.
Our results show that the inclusion of NQEs widens the hydrogen bond angle ($\beta\left(\mathrm{O\cdots O-H}\right)$) and contracts the hydrogen bond donor distance ($R_\mathrm{H\cdots O}^{}$), consistent with Yao and Kanai's conclusions for the SCAN functional at room temperature.
We note that Li, Peasani and Voth\cite{li_static_2022} also explored this issue, 
finding that the dynamical properties from classical simulations at 330 K
do not match the effect of NQEs at room temperature across several model potentials and MLPs including one based on the SCAN functional.
Given that the PIHMC-MIX result presented here
stems from a simulation with a long $t_{\mathrm{eff}}^{}$,
we can also conclude that the room temperature
structure of SCAN water is significantly
different from the experimental structure
even when NQEs are included.

%%%%%%%%%%%%%%%%%%%%%%%%%%%%%%%%%%%%%%%%%%%%%%%%%%%%%%%%%%%%%%%%%%%%%%%%
\subsubsection{The rev-vdW-DF2 Functional}\label{revsub}
%%%%%%%%%%%%%%%%%%%%%%%%%%%%%%%%%%%%%%%%%%%%%%%%%%%%%%%%%%%%%%%%%%%%%%%%

The rev-vdW-DF2 functional shares a similar accuracy with that of the SCAN functional, while being less expensive to execute and at the same time including van der Waals forces directly in the functional.
Compared to the experiment we find that using rev-vdW-DF2 the first hydration
shell is over-structured when considering all pair RDFs
as seen in figures 7(c), 8(c) and 9(c).
This over structuring is similar in size to that of the SCAN functional,
giving a much more structured liquid phase than for the RPBE-D3 functional.
The intramolecular peaks of the O-H and H-H RDFs are also significantly
different from the experimental ones,
indicating that the differences in the hydrogen bond structure
stems from a small difference in molecular structure.
Removing the NQEs by using HMC-MIX as shown in the SM leads to
a less structured liquid in terms of the O-O RDFs.
This indicates that the rev-vdW-DF2 functional is not capable of 
reproducing the delicate balance in the hydrogen bonds which generally soften 
the liquid structure as the NQEs are introduced.
In figure S9 (e) and (f) it is observed that $\beta\left(\mathrm{O\cdots O-H}\right)$ 
does not widen to the same degree as were the case for the SCAN functional.
Additionally, $R_{\mathrm{H\cdots O}}$ contracts more significantly, 
suggesting a much stronger hydrogen bond when NQEs are combined with the rev-vdW-DF2 functional, which 
might explain the larger degree of structure found in PIHMC-MIX compared to HMC-MIX.
The description of room temperature liquid water using
the rev-vdW-DF2 functional is thus considered to be worse than that
of both the SCAN and the RPBE-D3 functionals.

%%%%%%%%%%%%%%%%%%%%%%%%%%%%%%%%%%%%%%%%%%%%%%%%%%%%%%%%%%%%%%%%%%%%%%%%
\subsubsection{The optB88-vdW Functional}\label{optsub}
%%%%%%%%%%%%%%%%%%%%%%%%%%%%%%%%%%%%%%%%%%%%%%%%%%%%%%%%%%%%%%%%%%%%%%%%

The MLP constructed by the SL-PIHMC-MIX method with optB88-vdW is the 
least accurate in reproducing the results from FP 
calculations among the four functionals studied here.
The results from PIHMC-MIX are not improved upon the poor 
performance of ML-PIMD with respect to the experimental RDFs.
It leads to further over structuring of the RDFs
as shown in figures 7(d), 8(d) and 9(d).
As in the case for rev-vdW-DF2, RDFs are not only over structured,
but also the inclusion of NQEs do not have the effect
of softening the O-O RDFs.
The hydrogen bond structures reported in figure S9 (g) and (h) are much tighter than in any of the other functionals studied here, this along with a trend similar to that observed for rev-vdW-DF2 are likely the cause for the poor performance of the optB88-vdW functional in this study.
It should be noted that previous studies which report the 
structure of water using the optB88-vdW
functional\cite{del_ben_probing_2015,herrero_connection_2022}
find a better agreement with the experimental RDFs.
The improvements do however not change the fact that the 
water is over-structured when described by this functional,
to an extent that suggests that the inclusion of NQEs
should not lead to a better agreement than what is found
for the RPBE-D3 functional.
However, we cannot rule out the possibility that the current computational setup is a part of the reason for the poor performance of optB88-vdW shown here.

%%%%%%%%%%%%%%%%%%%%%%%%%%%%%%%%%%%%%%%%%%%%%%%%%%%%%%%%%%%%%%%%%%%%%%%%
\section{Conclusions}
%%%%%%%%%%%%%%%%%%%%%%%%%%%%%%%%%%%%%%%%%%%%%%%%%%%%%%%%%%%%%%%%%%%%%%%%

The PIHMC-MIX method has been shown to reproduce the accuracy 
of FP-PIMD simulations,
while requiring an order of magnitude fewer FP calculations.
This speedup does however require training of an MLP,
which we have shown can be done on the fly through the SL-PIHMC-MIX method.
The cost of fitting the MLP is however not prohibitory expensive,
and the computational cost of the method is much smaller than 
that of the FP-PIMD while allowing for the study of longer
$t_{\mathrm{eff}}^{}$ and thus more efficient sampling of
the phase space.
The mixing of FP and MLPs through the $\alpha$ parameter 
in the PIHMC-MIX method is essential in the context of both
PI methods and larger systems, such as the case for the water 
systems studied here.
This is shown for the RPBE-D3 functional,
where setting $\alpha = 1$,
\textit{i.e.,} using the PIHMC method,
results in low acceptance rates in
the HMC step.
The PIHMC method would thus require prohibitively 
long trajectories and extensive number of FP calculations
for convergence of the RDFs.
We have also tested the extend-ability of the method to other states of water,
namely ice I$^{}_\mathrm{h}$ in Section SXI of the supplementary information.
Here it was found that the PIHMC-MIX method using the trained MLP 
for liquid water were able to converge the RDF within 5,000 steps with acceptance ratio and effective trajectory length slightly smaller and shorter than were the case for the PIHMC-MIX simulations of water.
The PIHMC-MIX model thus shows promise for extending the study of water across its complex phase diagram.
This along with studies of more complex systems will be the subject of future studies.

%%%

The MLPs trained by the SL-PIHMC-MIX method were also found 
to reproduce the FP-PIMD results for all cases studied,
except the MLP trained on optB88-vdW data.
This gives the perspective of further computational savings by 
running ML-PIMD simulations instead of the more expensive 
PIHMC-MIX simulations.
Furthermore, it would be possible to study dynamical properties 
using the MLPs in  methods such as ring polymer molecular dynamics (RPMD)
\cite{craig_quantum_2004, braams_short-time_2006}, 
centroid molecular dynamics (CMD)
\cite{cao_formulation_1994, liu_path_2016}, 
or the recently proposed Brownian chain molecular
dynamics (BCMD) method\cite{shiga_path_2022}.
From our results we do however find reasons to caution direct transfer of
an MLP from H$_{2}^{}$O to other isotopologues of water,
\textit{i.e.} D$_{2}^{}$O,
and more extremely an MLP trained on only data without NQEs
being transferred to a system where NQEs are considered.
The MLP is not guarantied to accurately model
the differences caused by NQEs,
unless they are specifically trained for them,
or that PIHMC-MIX is used to guarantee convergence to the
FP-PIMD results.
It should be noted that for pure water using either MB-Pol\cite{Zhu_MBPol_2023} another MLPs\cite{yu_q-aqua_2022,daru_coupled_2022,chen_data-efficient_2023} trained on CCSD(T) data
would produce more accurate results than what is found here.
DFT based FP-PIMD or ML-PIMD trained on DFT data will however still be necessary to
study more complex systems,
leaving a wide field of applications of SL-PIHMC-MIX for training MLPs and PIHMC-MIX for
studying static properties at the DFT level of theory.

%%%

Finally, we have been able to provide a survey of the effects 
of NQEs in the simulations of H$_{2}^{}$O with the RPBE-D3, 
SCAN, rev-vdW-DF2 and optB88-vdW functionals.
We find an increased structuring of O-O RDFs for the 
rev-vdW-DF2 and optB88-vdW functions when NQEs are considered.
From the analysis of the shift in hydrogen bond parameters as NQEs 
are included, this behavior can be explained as the NQEs for these 
two functionals are found to tighten the hydrogen bonds. 
For the SCAN functional a slight softening,
especially in the hydrogen bond angle is found, 
leading to a loosening of the structure with the inclusion of NQEs. 
However, for RPBE-D3 the softening of the hydrogen bond parameters 
are more subtle and the averages are further from the experiments\cite{modig_temperature-dependent_2003} 
than the other functionals as seen in figure S9 (a) and (b). 
However, it seems that the trend of having longer hydrogen bond donor distance 
($R_{\mathrm{H\cdots O}}^{}$) in the distribution are key to the good performance of the functional. 
The conclusion is that among the four functionals studied here, 
the RPBE-D3 performs the best for studying the structure of water 
at room temperature in the NVT ensemble. 
The situation might change for the NPT ensemble, 
and for higher temperatures and pressures,
where non hydrogen bonded contacts between the water molecules 
become more important.

%%%%%%%%%%%%%%%%%%%%%%%%%%%%%%%%%%%%%%%%%%%%%%%%%%%%%%%%%%%%%%%%%%%%%%%%
\section{Supplementary Material}
%%%%%%%%%%%%%%%%%%%%%%%%%%%%%%%%%%%%%%%%%%%%%%%%%%%%%%%%%%%%%%%%%%%%%%%%

The supplementary material (SM) contains the derivation of Eq. (5),
details on the Behler-Parrinello structure fingerprint used in the MLPs,
and additional analysis and data on the simulations and RDFs presented in the main text.

%%%%%%%%%%%%%%%%%%%%%%%%%%%%%%%%%%%%%%%%%%%%%%%%%%%%%%%%%%%%%%%%%%%%%%%%
\section{Acknowledgements}
%%%%%%%%%%%%%%%%%%%%%%%%%%%%%%%%%%%%%%%%%%%%%%%%%%%%%%%%%%%%%%%%%%%%%%%%

We thank the JSPS Grant-in-Aid for Scientific Research (24K01145, 23K04670, 21H01603, 23H01273, 18H05519) for financial support.
The calculations were conducted using the supercomputer facilities at Japan Atomic Energy Agency (JAEA).

%%%%%%%%%%%%%%%%%%%%%%%%%%%%%%%%%%%%%%%%%%%%%%%%%%%%%%%%%%%%%%%%%%%%%%%%
\section{Data Availability}
%%%%%%%%%%%%%%%%%%%%%%%%%%%%%%%%%%%%%%%%%%%%%%%%%%%%%%%%%%%%%%%%%%%%%%%%

The data that support the findings of this study are available within
 the article.

%%%%%%%%%%%%%%%%%%%%%%%%%%%%%%%%%%%%%%%%%%%%%%%%%%%%%%%%%%%%%%%%%%%%%%%%
\bibliography{SuperCritWater.bib}
%%%%%%%%%%%%%%%%%%%%%%%%%%%%%%%%%%%%%%%%%%%%%%%%%%%%%%%%%%%%%%%%%%%%%%%%

%%%%%%%%%%%%%%%%%%%%%%%%%%%%%%%%%%%%%%%%%%%%%%%%%%%%%%%%%%%%%%%%%%%%%%%%
\clearpage
\begin{center}
{\bf FIGURE CAPTIONS}
\end{center}
%%%%%%%%%%%%%%%%%%%%%%%%%%%%%%%%%%%%%%%%%%%%%%%%%%%%%%%%%%%%%%%%%%%%%%%%

Figure 1: 
%Diagrams explaining the SL-PIHMC method.
%
(a) Diagram describing the flow of an SL-PIHMC simulation.
The parts shown in black are the core of the PIHMC method.
The orange part updates $n_\mathrm{ML}$ every $n_{\mathrm{test}}^{}$ steps 
and is not strictly required in the PIHMC method.
The green part denotes the SL part of the SL-PIHMC method, 
and is not active during a pure PIHMC simulation.
(b) Diagram of a single step in the PIHMC method.
(c) Diagram of a single step of PIHMC-MIX method.
Here the acceptance is judged based on the 
Hamiltonian using the $V^\mathrm{MIX}$, rather than $V^\mathrm{DFT}$ (see the text).
The numbers on parts (b) and (c) refer to the step in the diagram given in (a).

\ \\ \\
Figure 2: (Top) The evolution of the number of ML steps 
($n_{\mathrm{ML}}^{}$) between HMC steps during the 
training process for the RPBE-D3 functional 
using SL-PIHMC-MIX (blue),
the RPBE-D3 functional using SL-HMC-MIX (lightblue),
the SCAN functional using SL-PIHMC-MIX (green),
the rev-vdW-DF2 functional using SL-PIHMC-MIX (orange),
and the optB88-vdW functional using SL-PIHMC-MIX (red).
(Center) The evolution of the instantaneous acceptance rate
$\left(A_{\mathrm{acc}}^{}\right)$
for every $n_{\mathrm{test}}^{}$ steps. 
(Bottom) The evolution of the accumulated average acceptance rate
$\left(\left\langle A_{\mathrm{acc}}^{}\right\rangle\right)$
over the SL-PIHMC-MIX simulations.
\ \\ \\
Figure 3: O-O RDFs calculated using the
RPBE-D3 functional and  MLPs trained 
on FP data from the same functional
using SL-PIHMC-MIX.
(a) Comparison of FP-MD (green) and FP-PIMD (blue)
from our previous works\cite{thomsen_ab_2021,Thomsen_structures_2022}, 
with that of experiment 
\cite{soper_radial_2013} (black).
(b) Comparison of the FP-PIMD (blue) and
PIHMC-MIX (purple) with that of experiment
\cite{soper_radial_2013} (black).
(c) Comparison of FP-PIMD (blue) with
the results of ML-PIMD for a system containing
64 water molecules (yellow) or 256 water molecules
(orange).
(d)
Comparison of FP-MD (green) with HMC-MIX (light blue)
and ML-MD for systems containing 64 water molecules (red)
and 256 water molecules (pink).
Note that these results are different from the results trained using
SL-HMC-MIX given in Section SVII.
\ \\ \\
Figure 4: Same as Figure 3 (a-d) for the O-H RDFs. 
The first O-H RDF peaks are shown in the insets.
The experimental data for the first RDF peak is taken from Ref. \onlinecite{soper_radial_2000} and is given as a dashed black line.
\ \\ \\
Figure 5: Same as Figure 3 (a-d) for the H-H RDFs. 
The first H-H RDF peaks for the simulations without NQEs in figure (a) and  (d) are shown in the insets.
The experimental data for the first RDF peak is taken from Ref. \onlinecite{soper_radial_2000}, and is given as a dashed black line.
\ \\ \\
Figure 6: (a) O-O, (b) O-D and (c) D-D RDFs for D$_2$O.
In all figures, the experimental data
\cite{soper_quantum_2008} (black) are given as a reference.
The result from our previously published FP-PIMD
\cite{Thomsen_structures_2022} result are given in red,
in green are the results from ML-PIMD,
and in blue are the results from PIHMC-MIX.
The MLP used in ML-PIMD and PIHMC-MIX stems 
from the RPBE-D3 data from the SL-PIHMC-MIX trajectory 
of H$_2$O.
The peak heights and positions are given in Table SXII.
In Figure S5, the ML-PIMD results are compared with
results for an MLP partially trained using data
from SL-PIHMC-MIX for D$_{2}^{}$O,
the heights and positions of which are also given in Table SXII.
\ \\ \\
Figure 7: O-O RDFs calculated using various functions
and MLPs trained on FP data from said functions.
In all figures, the experimental data
\cite{soper_radial_2013} (black) are given as a reference.
(a) Comparison of the results from PIHMC-MIX (blue) and
ML-PIMD (light blue) using the RPBE-D3 functional.
(b) Comparison of the results from PIHMC-MIX (green) and
ML-PIMD (light green) using the SCAN functional.
(c) Comparison of the results from PIHMC-MIX (orange) and
ML-PIMD (yellow) using the rev-vdW-DF2 functional.
(d) Comparison of the results from PIHMC-MIX (red) and
ML-PIMD (pink) using the optB88-vdW functional.
\ \\ \\
Figure 8: Same as for figure 7 (a-d) for the O-H RDFs.
Note that all figures contain a subplot of the first
O-H RDF peaks, as this goes out of scale when compared to
the secondary and tertiary peaks.
Furthermore, the experimental data for the first
RDF peak is taken from Ref. \onlinecite{soper_radial_2000},
and is given as a dashed black line.
\ \\ \\
Figure 9: Same as for figure 7 (a-d) for the H-H RDFs.
Note that the experimental data for the first
RDF peak is taken from Ref. \onlinecite{soper_radial_2000},
and is given as a dashed black line.

%%%%%%%%%%%%%%%%%%%%%%%%%%%%%%%%%%%%%%%%%%%%%%%%%%%%%%%%%%%%%%%%%%%%%%%%
% Figure 1
%%%%%%%%%%%%%%%%%%%%%%%%%%%%%%%%%%%%%%%%%%%%%%%%%%%%%%%%%%%%%%%%%%%%%%%%

\clearpage
\begin{figure}[H] % [htbp]
\centering
\includegraphics[width=0.99\linewidth]{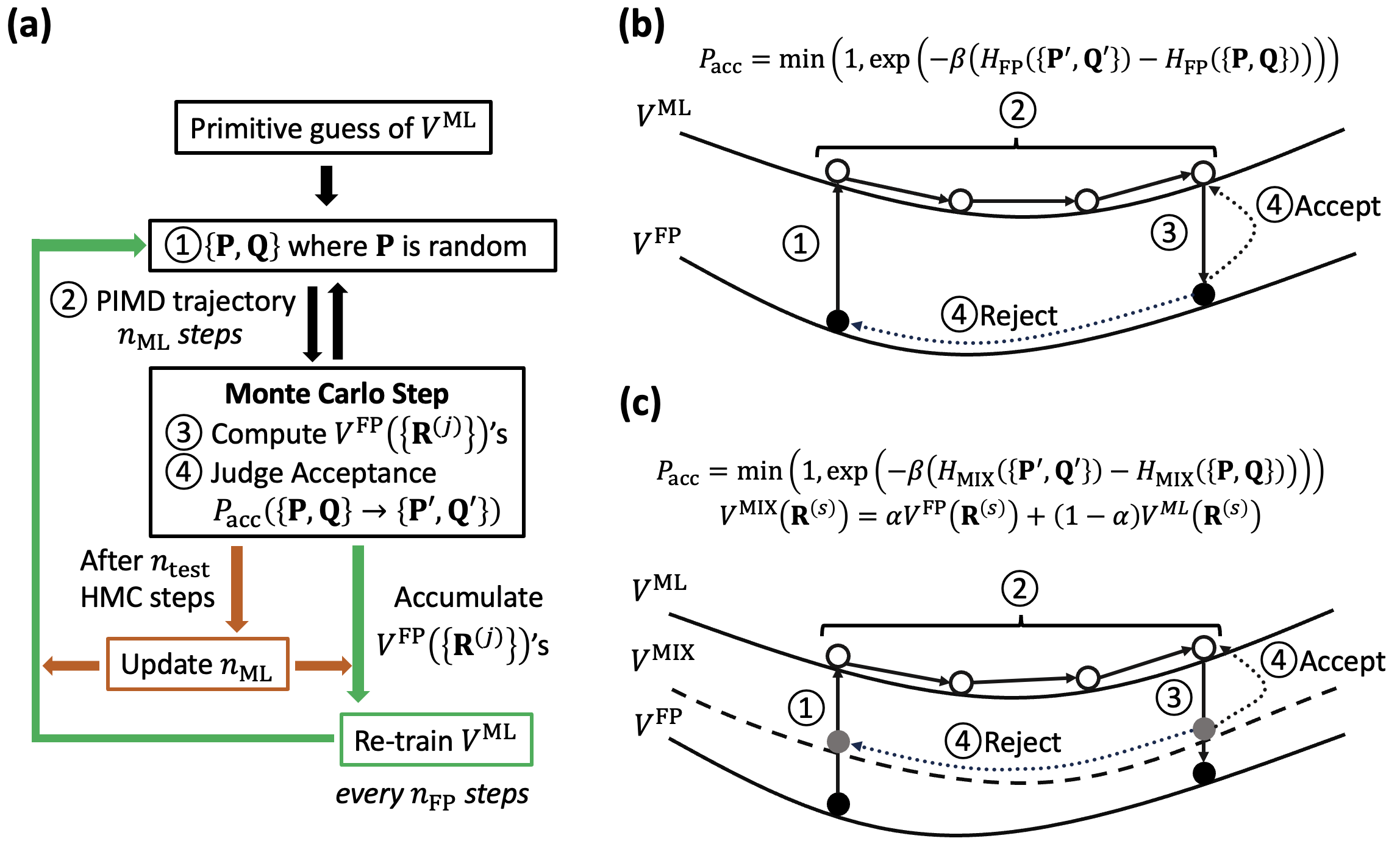}
\end{figure}
Figure 1, B. Thomsen, Y. Nagai, K. Kobayashi, I. Hamada and M. Shiga, submitted to J. Chem. Phys.

%%%%%%%%%%%%%%%%%%%%%%%%%%%%%%%%%%%%%%%%%%%%%%%%%%%%%%%%%%%%%%%%%%%%%%%%
% Figure 2
%%%%%%%%%%%%%%%%%%%%%%%%%%%%%%%%%%%%%%%%%%%%%%%%%%%%%%%%%%%%%%%%%%%%%%%%

\clearpage
\begin{figure}[H] % [htbp]
\centering
\includegraphics[width=0.50\linewidth]{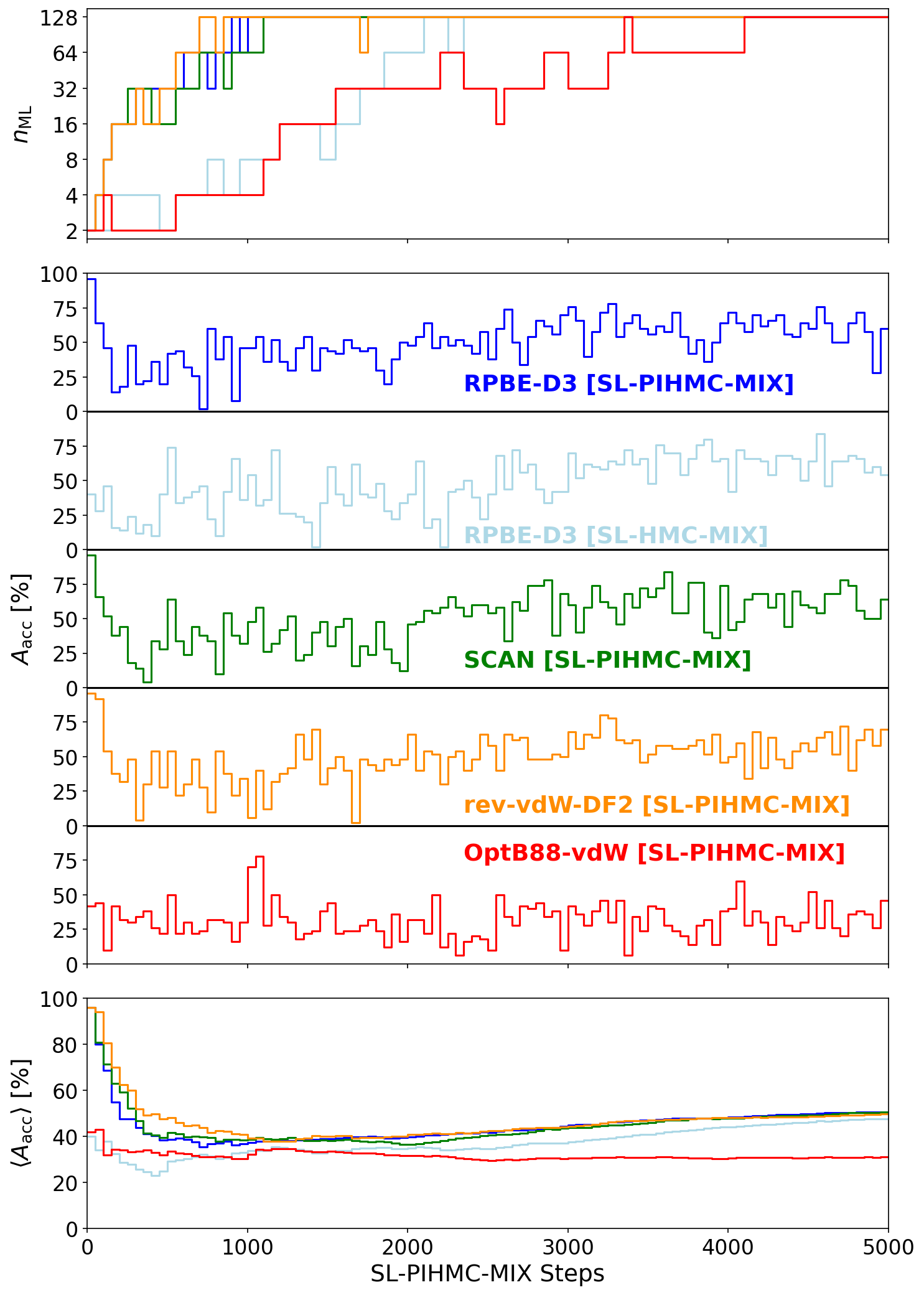}
\end{figure}
Figure 2, B. Thomsen, Y. Nagai, K. Kobayashi, I. Hamada and M. Shiga, submitted to J. Chem. Phys.

%%%%%%%%%%%%%%%%%%%%%%%%%%%%%%%%%%%%%%%%%%%%%%%%%%%%%%%%%%%%%%%%%%%%%%%%
% Figure 3
%%%%%%%%%%%%%%%%%%%%%%%%%%%%%%%%%%%%%%%%%%%%%%%%%%%%%%%%%%%%%%%%%%%%%%%%

\clearpage
\begin{figure}[H] % [htbp]
\centering
\includegraphics[width=0.99\linewidth]{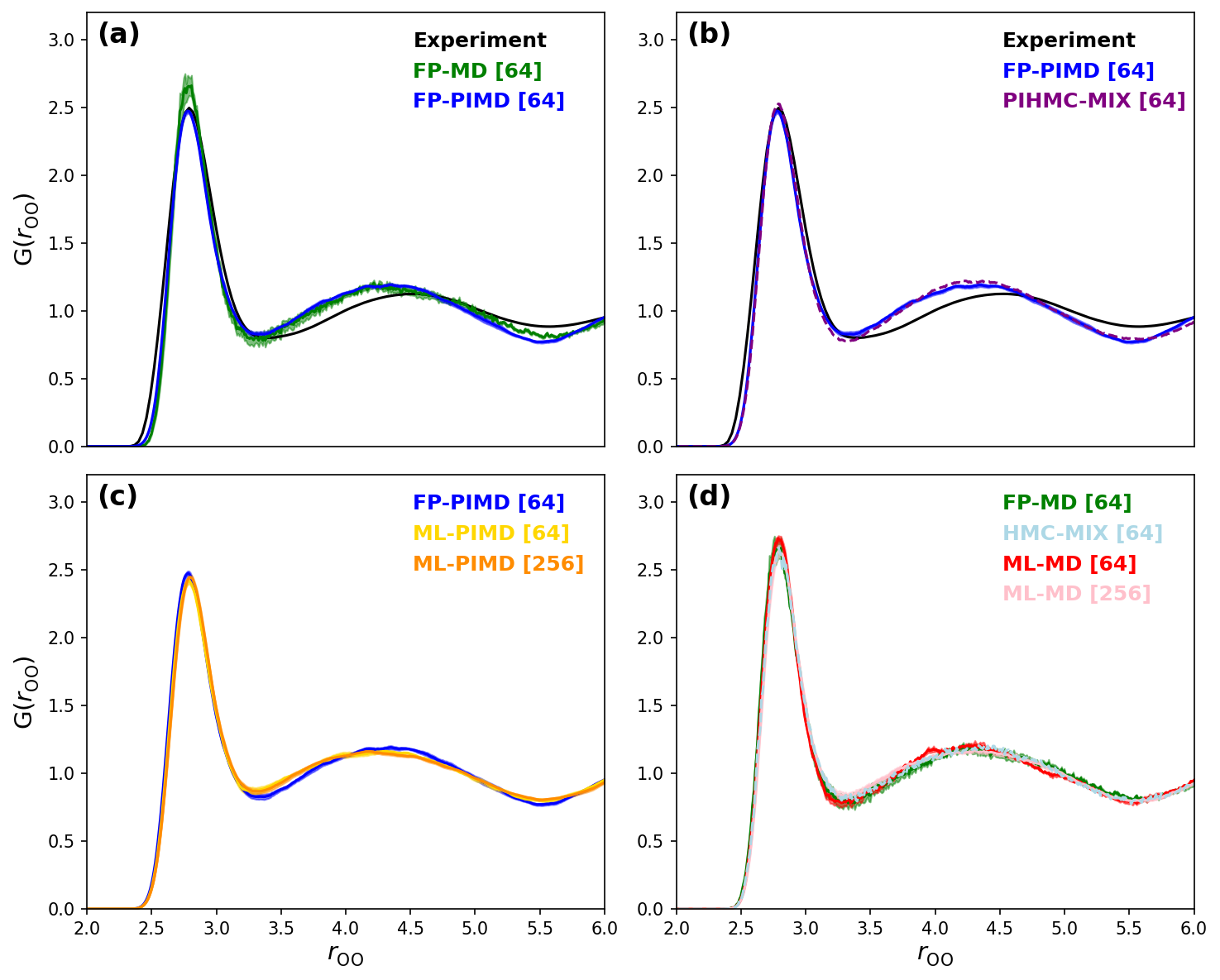}
\end{figure}
Figure 3, B. Thomsen, Y. Nagai, K. Kobayashi, I. Hamada and M. Shiga, submitted to J. Chem. Phys.

%%%%%%%%%%%%%%%%%%%%%%%%%%%%%%%%%%%%%%%%%%%%%%%%%%%%%%%%%%%%%%%%%%%%%%%%
% Figure 4
%%%%%%%%%%%%%%%%%%%%%%%%%%%%%%%%%%%%%%%%%%%%%%%%%%%%%%%%%%%%%%%%%%%%%%%%

\clearpage
\begin{figure}[H] % [htbp]
\centering
\includegraphics[width=0.99\linewidth]{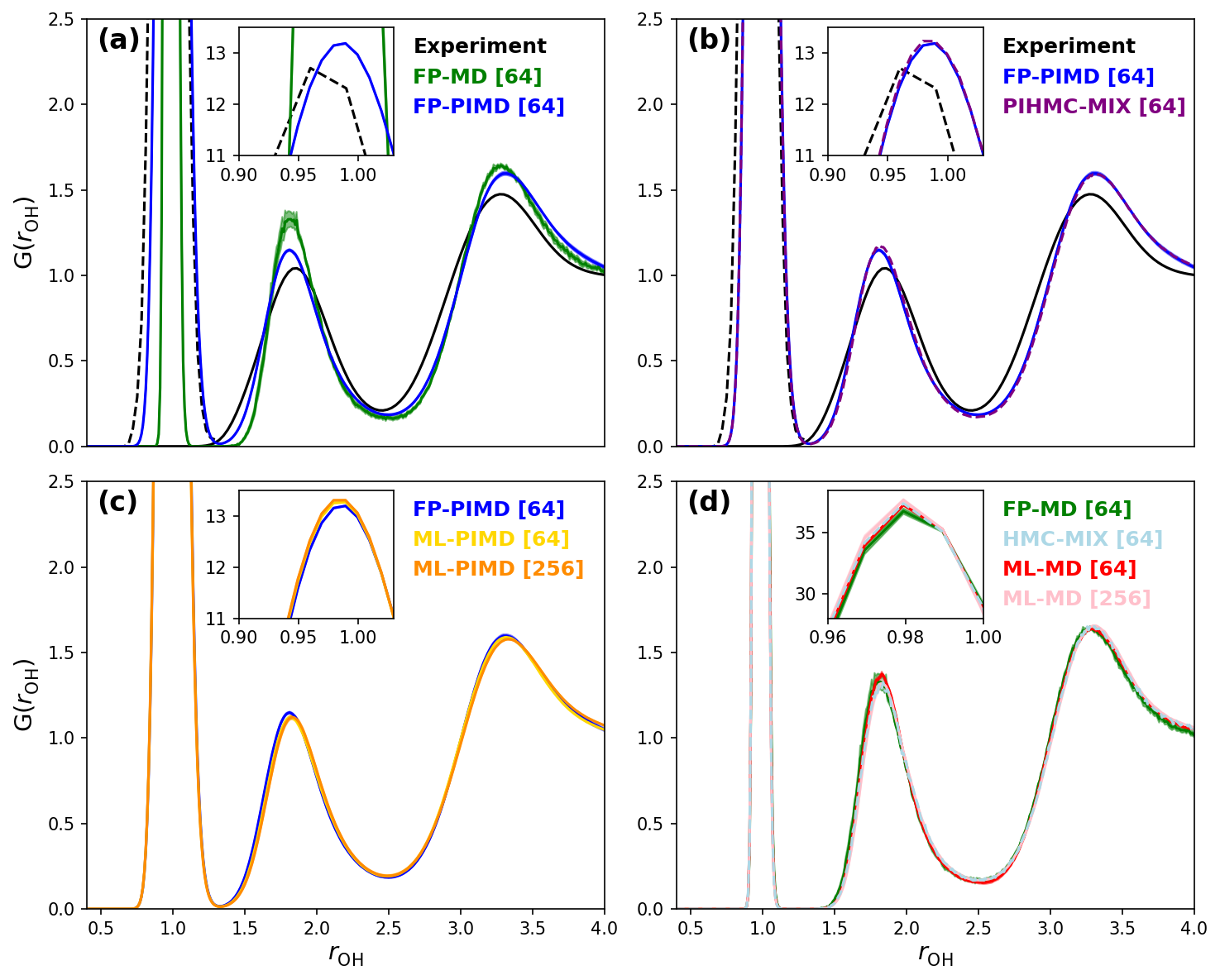}
\end{figure}
Figure 4, B. Thomsen, Y. Nagai, K. Kobayashi, I. Hamada and M. Shiga, submitted to J. Chem. Phys.

%%%%%%%%%%%%%%%%%%%%%%%%%%%%%%%%%%%%%%%%%%%%%%%%%%%%%%%%%%%%%%%%%%%%%%%%
% Figure 5
%%%%%%%%%%%%%%%%%%%%%%%%%%%%%%%%%%%%%%%%%%%%%%%%%%%%%%%%%%%%%%%%%%%%%%%%

\clearpage
\begin{figure}[H] % [htbp]
\centering
\includegraphics[width=0.99\linewidth]{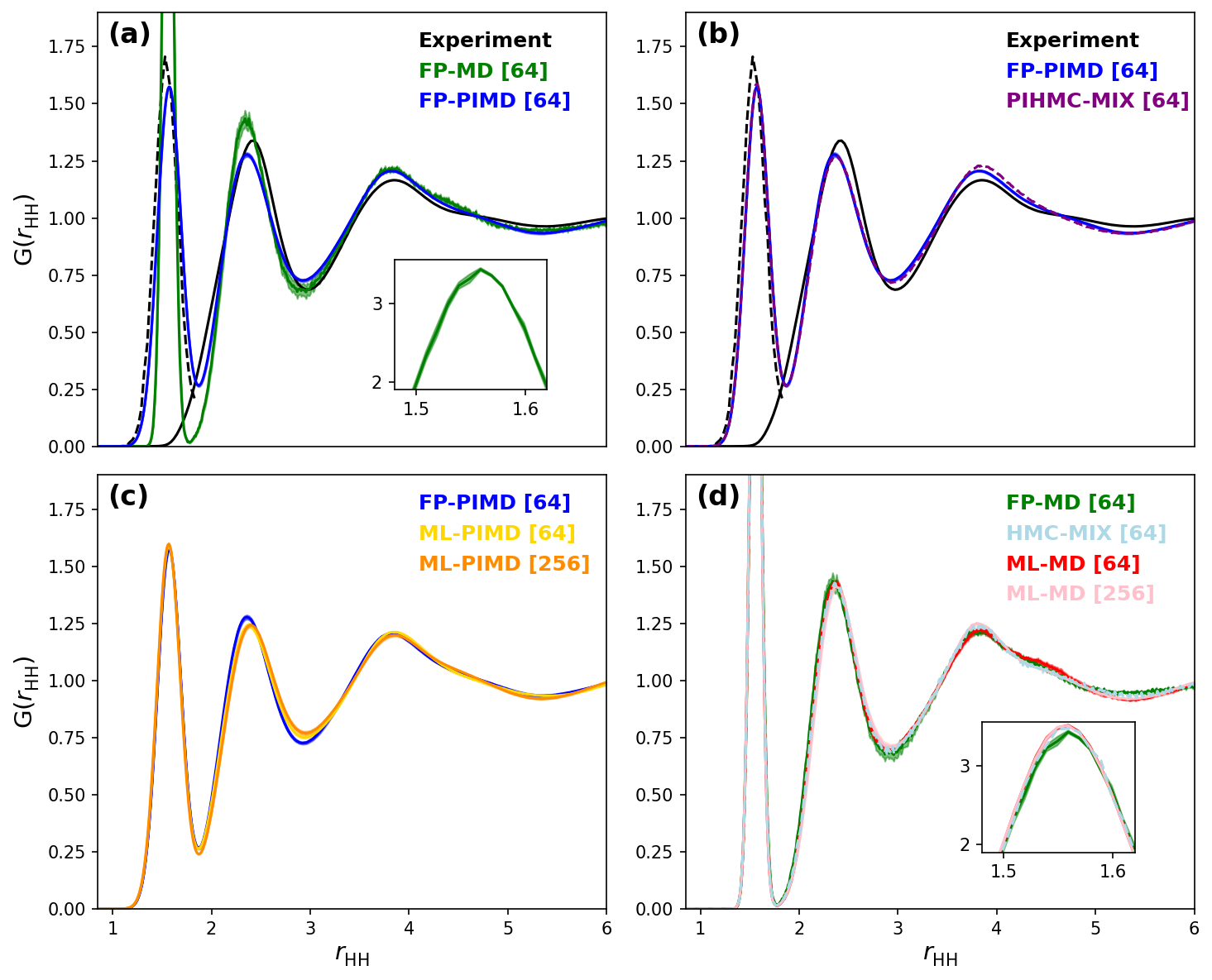}
\end{figure}
Figure 5, B. Thomsen, Y. Nagai, K. Kobayashi, I. Hamada and M. Shiga, submitted to J. Chem. Phys.

%%%%%%%%%%%%%%%%%%%%%%%%%%%%%%%%%%%%%%%%%%%%%%%%%%%%%%%%%%%%%%%%%%%%%%%%
% Figure 6
%%%%%%%%%%%%%%%%%%%%%%%%%%%%%%%%%%%%%%%%%%%%%%%%%%%%%%%%%%%%%%%%%%%%%%%%

\clearpage
\begin{figure}[H] % [htbp]
\centering
\includegraphics[width=0.99\linewidth]{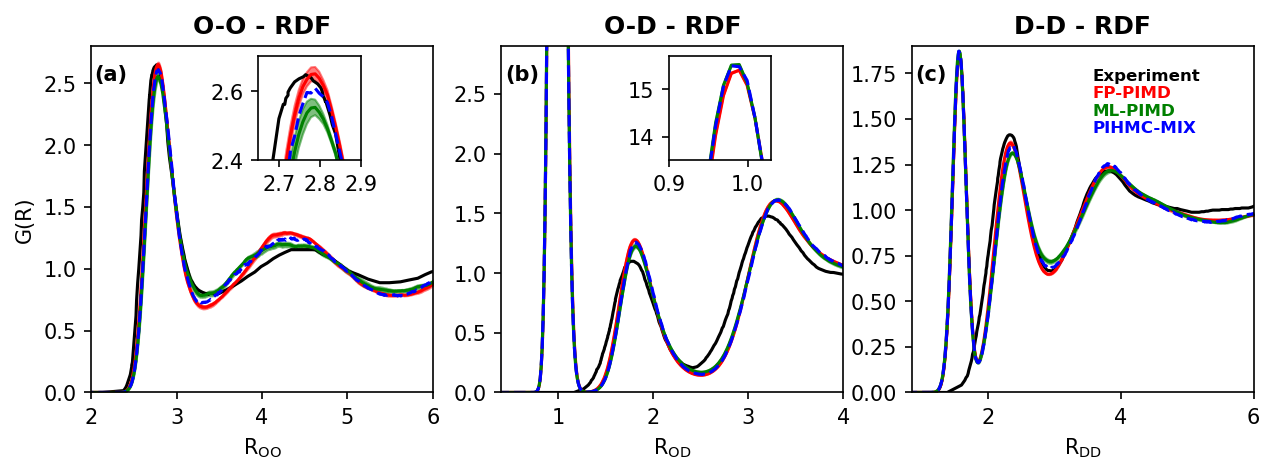}
\end{figure}
Figure 6, B. Thomsen, Y. Nagai, K. Kobayashi, I. Hamada and M. Shiga, submitted to J. Chem. Phys.

%%%%%%%%%%%%%%%%%%%%%%%%%%%%%%%%%%%%%%%%%%%%%%%%%%%%%%%%%%%%%%%%%%%%%%%%
% Figure 7
%%%%%%%%%%%%%%%%%%%%%%%%%%%%%%%%%%%%%%%%%%%%%%%%%%%%%%%%%%%%%%%%%%%%%%%%

\clearpage
\begin{figure}[H] % [htbp]
\centering
\includegraphics[width=0.99\linewidth]{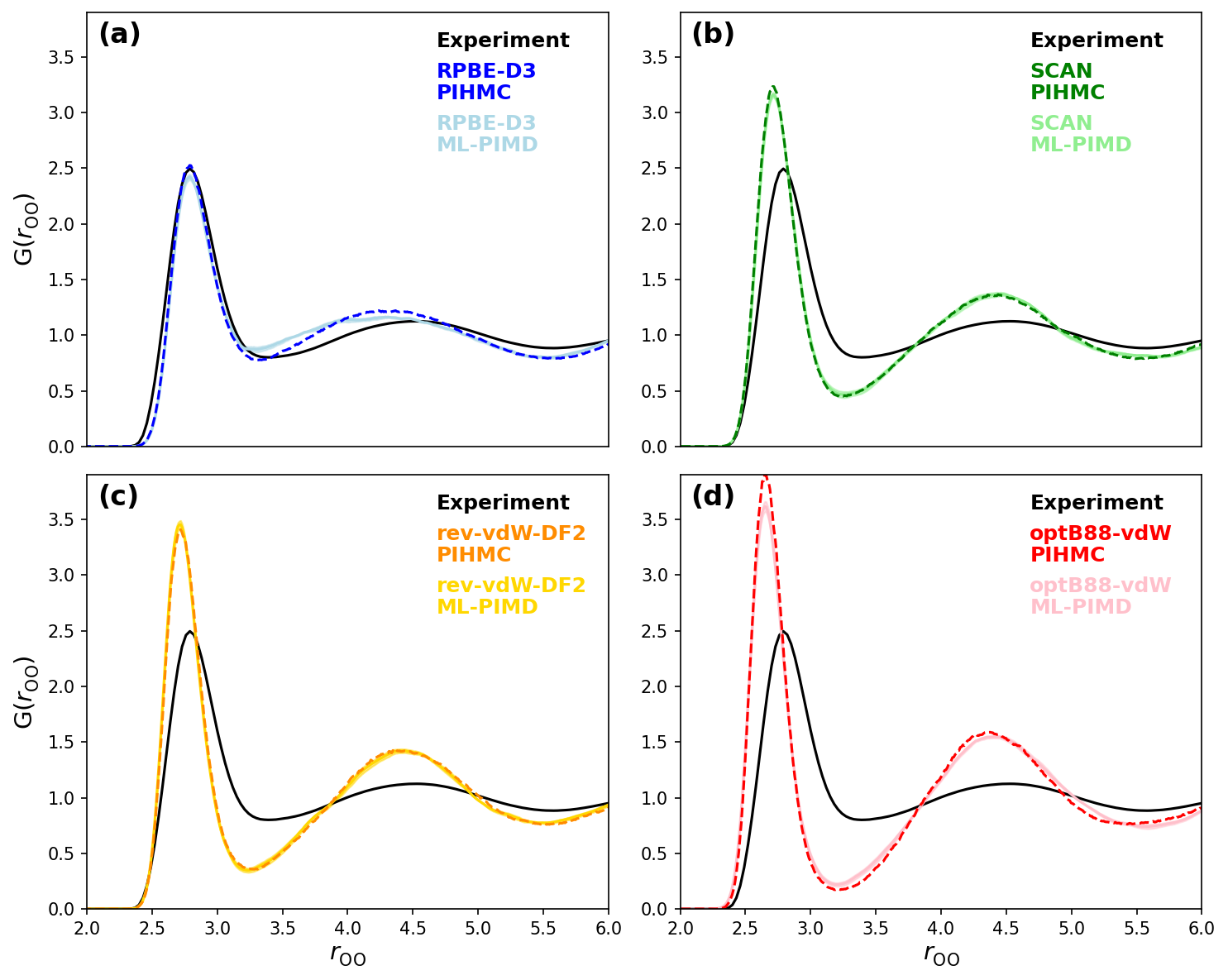}
\end{figure}
Figure 7, B. Thomsen, Y. Nagai, K. Kobayashi, I. Hamada and M. Shiga, submitted to J. Chem. Phys.

%%%%%%%%%%%%%%%%%%%%%%%%%%%%%%%%%%%%%%%%%%%%%%%%%%%%%%%%%%%%%%%%%%%%%%%%
% Figure 8
%%%%%%%%%%%%%%%%%%%%%%%%%%%%%%%%%%%%%%%%%%%%%%%%%%%%%%%%%%%%%%%%%%%%%%%%

\clearpage
\begin{figure}[H] % [htbp]
\centering
\includegraphics[width=0.99\linewidth]{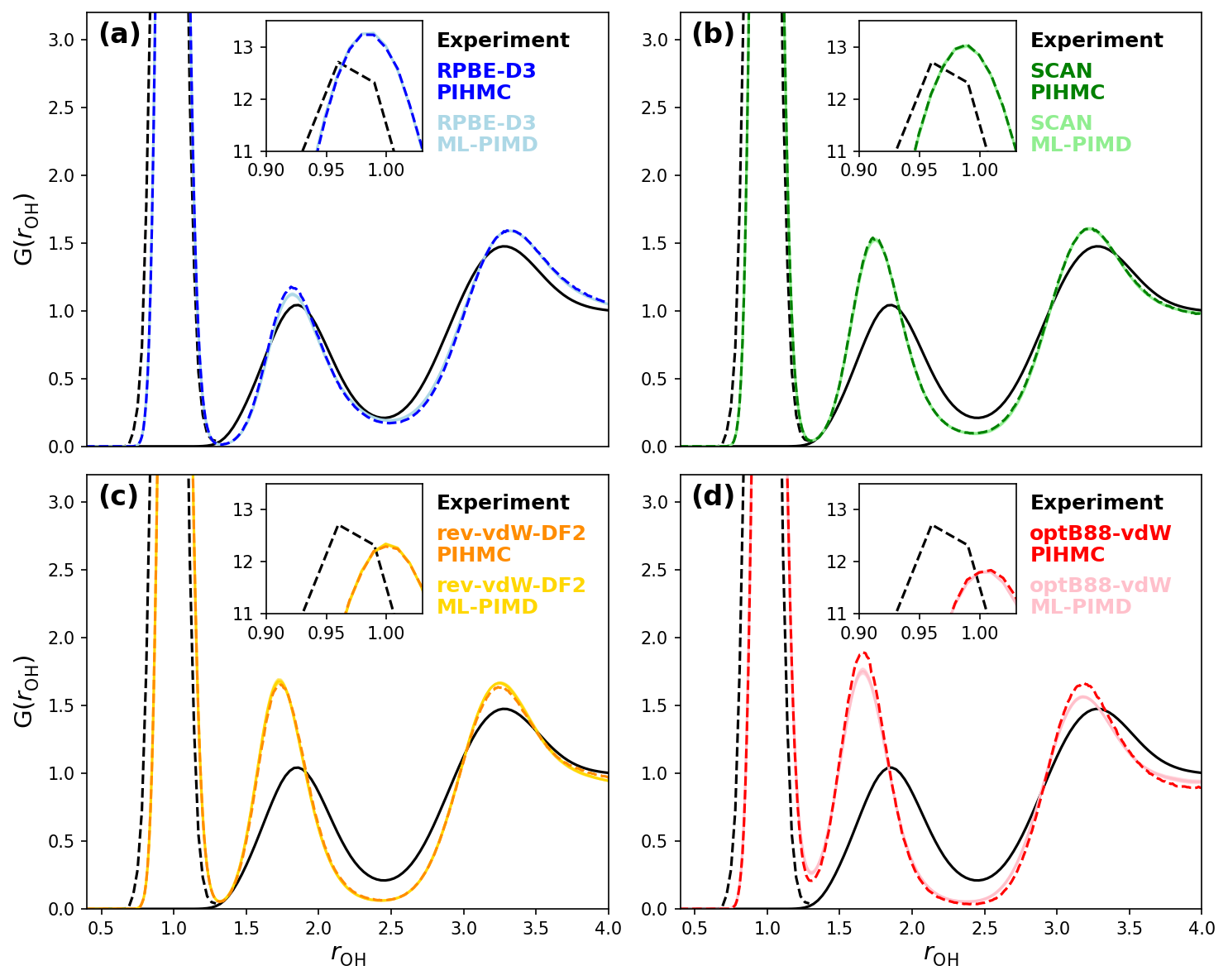}
\end{figure}
Figure 8, B. Thomsen, Y. Nagai, K. Kobayashi, I. Hamada and M. Shiga, submitted to J. Chem. Phys.

%%%%%%%%%%%%%%%%%%%%%%%%%%%%%%%%%%%%%%%%%%%%%%%%%%%%%%%%%%%%%%%%%%%%%%%%
% Figure 9
%%%%%%%%%%%%%%%%%%%%%%%%%%%%%%%%%%%%%%%%%%%%%%%%%%%%%%%%%%%%%%%%%%%%%%%%

\clearpage
\begin{figure}[H] % [htbp]
\centering
\includegraphics[width=0.99\linewidth]{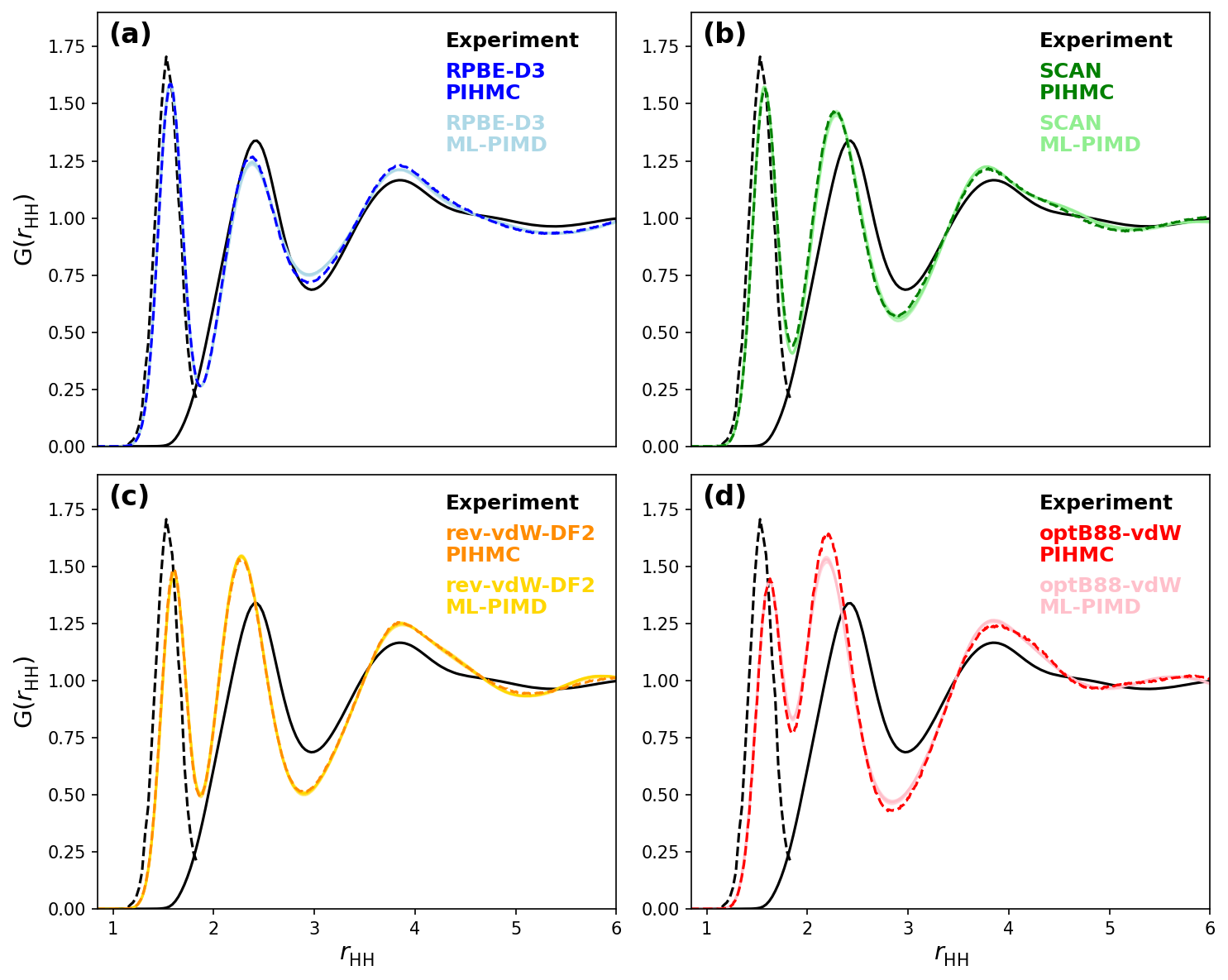}
\end{figure}
Figure 9, B. Thomsen, Y. Nagai, K. Kobayashi, I. Hamada and M. Shiga, submitted to J. Chem. Phys.

%%%%%%%%%%%%%%%%%%%%%%%%%%%%%%%%%%%%%%%%%%%%%%%%%%%%%%%%%%%%%%%%%%%%%%%%
% Tables
%%%%%%%%%%%%%%%%%%%%%%%%%%%%%%%%%%%%%%%%%%%%%%%%%%%%%%%%%%%%%%%%%%%%%%%%

\clearpage
\begin{table*}[htp]
\caption{ The functionals used for the FP calculations, 
  the method used for trajectory propagation,
  the number of steps, HMC or MD depending on the method, 
  $\left( N_{\mathrm{steps}^{}}\right)$,
  the $\alpha$ value used for PIHMC-MIX,
  the average acceptance ratio $\left(\left\langle A_{\mathrm{test}}^{}\right\rangle\right)$,
  the number of steps in ML-PIMD for the PIHMC-MIX method $\left( n_{\mathrm{ML}}^{}\right)$,
  and the effective trajectory length $\left(t_{\mathrm{eff}}^{}\right)$ in picoseconds
  for all simulations presented in the main text.
  See Table SII in the SM for the SL-PIHMC-MIX trajectories run to train the MLPs,
  and Table SIII for the additional PIHMC-MIX trajectories 
  only used in the SM.
  }
\begin{ruledtabular}
\begin{tabular}{ccccccc}
Functional & Method & $N_{\mathrm{steps}}$ & 
$\alpha$ & $\left\langle A_{\mathrm{acc}}\right\rangle$ (\%) & 
$n_{\mathrm{ML}}$ & $t_{\mathrm{eff}}^{}$ (ps) \\
\hline
RPBE-D3$^a$ & FP-MD     & 200,000 & -    & -    & -      & 50.0\\
RPBE-D3     & HMC-MIX   & 10,000  & 0.25 & 55.3 & 128    & 103.7\\
RPBE-D3$^a$ & FP-PIMD   & 100,000 & -    & -    & -      & 25.0\\
RPBE-D3     & PIHMC     & 15,000  & 1.0  & 24.8 & 8-128  & 17.9\\
RPBE-D3     & PIHMC-MIX & 10,000  & 0.75 & 25.9 & 8-128  & 22.3\\
RPBE-D3     & PIHMC-MIX & 7,000   & 0.5  & 31.5 & 16-128 & 62.4\\
RPBE-D3     & PIHMC-MIX & 5,000   & 0.25 & 55.5 & 128    & 99.9\\
SCAN        & PIHMC-MIX & 5,000   & 0.25 & 54.4 & 128    & 96.8 \\
rev-vdW-DF2 & PIHMC-MIX & 5,000   & 0.25 & 51.7 & 64-128 & 88.7\\
optB88-vdW  & PIHMC-MIX & 5,000   & 0.25 & 36.8 & 128    & 59.0\\
\end{tabular}
\end{ruledtabular}
\parbox{\linewidth}{\raggedright $^a$ These trajectories are from Refs. \onlinecite{thomsen_ab_2021, Thomsen_structures_2022}}

\end{table*}

%%%

\clearpage
\begin{table*}[htp]
\caption{
% Table II.
The positions and heights of 
the  peaks in the O-O RDFs presented in Figure 7. 
%
% All peak positions are given in \AA.
%  
The data is denoted either with a $r^{\mathrm{OO}}_i$ or $h^{\mathrm{OO}}_i$ 
referring to the peak position and heights respectively.
$r_{\mathrm{min}}^{\mathrm{OO}}$ and
$h_{\mathrm{min}}^{\mathrm{OO}}$ refer to the height and position
of the minimum of RDF found in the first interstitial region.
The experimental results stem from 
Ref. \onlinecite{soper_radial_2013}.}
\begin{ruledtabular}
\begin{tabular}{cccccccc}
DFT Functional & Model & $r_1^{\mathrm{OO}}$ & $h_1^{\mathrm{OO}}$ &
  $r_{\mathrm{min}}^{\mathrm{OO}}$ & $h_{\mathrm{min}}^{\mathrm{OO}}$ &
  $r_2^{\mathrm{OO}}$  & $h_2^{\mathrm{OO}}$\\
  & & (\AA) & & (\AA) & & (\AA) &\\
\hline
RPBE-D3 & FP-PIMD & 2.78 & 2.47 & 3.33 & 0.83 & 4.35 & 1.19 \\
RPBE-D3 & PIHMC & 2.79 & 2.53 & 3.33 & 0.77 & 4.24 & 1.22 \\
SCAN & PIHMC & 2.72 & 3.24 & 3.23 & 0.44 & 4.36 & 1.36 \\
rev-vdw-DF2 & PIHMC & 2.72 & 3.43 & 3.23 & 0.36 & 4.46 & 1.43 \\
optB88-vdW & PIHMC & 2.65 & 3.88 & 3.20 & 0.17 & 4.36 & 1.58 \\
\hline
&Experiment & 2.79 & 2.50 & 3.36 & 0.78 & 4.53 & 1.12
\end{tabular}
\end{ruledtabular}
\end{table*}

%%%

\clearpage
\begin{table*}[htp]
\caption{
The positions and heights of the  peaks in the O-H RDFs 
presented in figure 8. 
All peak positions are given in \AA.  
The data is denoted either with a $r^{\mathrm{OH}}_i$ or $h^{\mathrm{OH}}_i$ 
referring to the peak position and heights respectively. 
The experimental results stem from Ref. \onlinecite{soper_radial_2013}, 
except those marked by $^*$ which are from Ref. \onlinecite{soper_radial_2000}.
 } 
\begin{ruledtabular}
\begin{tabular}{cccccccc}
DFT Functional & Model & $r_1^{\mathrm{OH}}$ (Å) & $h_1^{\mathrm{OH}}$ & 
 $r_2^{\mathrm{OH}}$ (\AA) & $h_2^{\mathrm{OH}}$ & 
 $r_3^{\mathrm{OH}}$ (\AA) & $h_3^{\mathrm{OH}}$\\
\hline
RPBE-D3 & FP-PIMD & 0.99 & 13.19 & 1.81 & 1.15 & 3.32 & 1.60 \\
RPBE-D3 & PIHMC & 0.99 & 13.22 & 1.81 & 1.18 & 3.32 & 1.59 \\
SCAN & PIHMC & 0.99 & 13.04 & 1.75 & 1.54 & 3.24 & 1.61 \\
rev-vdw-DF2 & PIHMC & 1.00 & 12.33 & 1.74 & 1.66 & 3.26 & 1.63 \\
optB88-vdW & PIHMC & 1.01 & 11.82 & 1.66 & 1.89 & 3.19 & 1.66 \\
\hline
& Experiment &0.96$^*$ & 12.71$^*$ & 1.86 & 1.04 & 3.27 & 1.48\\
\end{tabular}
\end{ruledtabular}
\end{table*}

%%%

\clearpage
\begin{table*}[htp]
\caption{ 
 The positions and heights of the  peaks in 
the H-H RDFs presented in figure 9.
All peak positions are given in \AA.
The data is denoted either with a r$^{\mathrm{HH}}_i$ or h$^{\mathrm{HH}}_i$ 
referring to the peak position and heights respectively.
The experimental results stem from Ref. \onlinecite{soper_radial_2013}, 
except those marked by $^*$ which are from Ref. \onlinecite{soper_radial_2000}.
}
\begin{ruledtabular}
\begin{tabular}{cccccccc}
DFT Functional & Model & $r_1^{\mathrm{HH}}$ (Å) & $h_1^{\mathrm{HH}}$ &
  $r_2^{\mathrm{HH}}$ (\AA) & $h_2^{\mathrm{HH}}$ & 
  $r_3^{\mathrm{HH}}$ (\AA) & $h_3^{\mathrm{HH}}$\\
\hline
RPBE-D3 & FP-PIMD & 1.57 & 1.57 & 2.36 & 1.28 & 3.83 & 1.21 \\
RPBE-D3 & PIHMC & 1.57 & 1.58 & 2.36 & 1.27 & 3.84 & 1.23 \\
SCAN & PIHMC & 1.57 & 1.56 & 2.28 & 1.47 & 3.80 & 1.21 \\
rev-vdW-DF2 & PIHMC & 1.60 & 1.49 & 2.25 & 1.53 & 3.85 & 1.25 \\
optB88-vdW & PIHMC & 1.63 & 1.44 & 2.21 & 1.64 & 3.89 & 1.24 \\
\hline
&Experiment & 1.53$^*$ & 1.71$^*$ & 2.43 & 1.34 & 3.84 & 1.17\\
\end{tabular}
\end{ruledtabular}
\end{table*}

\clearpage

\section*{Supplemental Materials}
\setcounter{section}{0}
\renewcommand{\thesection}{S\arabic{section}}
\renewcommand{\thefigure}{S\arabic{figure}}
\setcounter{figure}{0}
\renewcommand{\thetable}{S\arabic{table}}
\setcounter{table}{0}

\section{Derivation of equation (5)}
%%%%%%%%%%%%%%%%%%%%%%%%%%%%%%%%%%%%%%%%%%%%%%%%%%%%%%%%%%%%%%%%%%%%%%%%
Here we use the vector and scalars introduced in the theory 
section of the main text.
To ease the notion we introduce the following shorthand 
for the kinetic energy of the system
\begin{equation}
E_{\mathrm{kin}}^{\mathrm{MOD}}\left(\mathbf{P}\right)=
\frac{1}{2}\sum_{I=1}^{N}
\mathbf{P}^{T}_{}\boldsymbol{\mu}^{-1}_{I}\mathbf{P}^{}_{}
\end{equation}
and for the effective potential term
\begin{equation}
V^{\mathrm{MOD}}_{}\left(\mathbf{Q}\right)=
V^{\mathrm{MOD}}_{\mathrm{av}}\left(\mathbf{Q}\right)+
\frac{1}{2}\sum_{I=1}^{N}M_{I}^{}\omega_{p}^{2}\mathbf{Q}^{T}_{}
\boldsymbol{\lambda}\mathbf{Q}^{}_{}.
\end{equation}
Taking the difference between the Hamiltonians
for the initial and trial point in the PIHMC
acceptance criteria from Eq. (1) 
and introducing Eq. (S1-2) we obtain,
\begin{equation}
H_{\mathrm{FP}}(\{\mathbf{P},\mathbf{Q}\})
-H_{\mathrm{FP}}(\{\mathbf{P}',\mathbf{Q}'\})=
E_{\mathrm{kin}}^{\mathrm{FP}}\left(\mathbf{P}\right)-
E_{\mathrm{kin}}^{\mathrm{FP}}\left(\mathbf{P}'\right)+
V^{\mathrm{FP}}_{}\left(\mathbf{Q}\right)-
V^{\mathrm{FP}}_{}\left(\mathbf{Q}'\right).
\end{equation}
Since the momenta from the final step of the 
trial ML-PIMD trajectory are used when determining 
the acceptance, we can replace $E_{\mathrm{kin}}^{\mathrm{FP}}$ with 
$E_{\mathrm{kin}}^{\mathrm{ML}}$ in the equation above.
Furthermore, assuming conservation of 
energy in the ML-PIMD trajectory, the following relation holds
\begin{equation}
E_{\mathrm{kin}}^{\mathrm{ML}}\left(\mathbf{P}\right)-
E_{\mathrm{kin}}^{\mathrm{ML}}\left(\mathbf{P}'\right)=
-\left(V^{\mathrm{ML}}_{}\left(\mathbf{Q}\right)
-V^{\mathrm{ML}}_{}\left(\mathbf{Q}'\right)\right).
\end{equation}
Inserting into Eq. (S3) and rearranging we get the following,
\begin{equation}
H_{\mathrm{FP}}(\{\mathbf{P},\mathbf{Q}\})
-H_{\mathrm{FP}}(\{\mathbf{P}',\mathbf{Q}'\})=
V^{\mathrm{ FP }}_{}\left(\mathbf{Q}\right)-
V^{\mathrm{ML}}_{}\left(\mathbf{Q}\right)+
V^{\mathrm{ML}}_{}\left(\mathbf{Q}'\right)-
V^{\mathrm{ FP }}_{}\left(\mathbf{Q}'\right).
\end{equation}
Since the second term in Eq. (S2) only depends
on the coordinates,
which are the same for the two differences in
the above equation,
this term will cancel out, and we are left with
\begin{equation}
H_{\mathrm{ FP }}(\{\mathbf{P},\mathbf{Q}\})
-H_{\mathrm{ FP }}(\{\mathbf{P}',\mathbf{Q}'\})=
V^{\mathrm{ FP }}_{\mathrm{av}}\left(\mathbf{Q}\right)-
V^{\mathrm{ML}}_{\mathrm{av}}\left(\mathbf{Q}\right)+
V^{\mathrm{ML}}_{\mathrm{av}}\left(\mathbf{Q}'\right)-
V^{\mathrm{ FP }}_{\mathrm{av}}\left(\mathbf{Q}'\right).
\end{equation}
Which by rearrangement and introduction of the terms
given in Eq. (6-7) can be seen to correspond to
$\Delta\Delta V$ of Eq. (5).

%%%%%%%%%%%%%%%%%%%%%%%%%%%%%%%%%%%%%%%%%%%%%%%%%%%%%%%%%%%%%%%%%%%%%%%%
\section{Reweighting procedure}
%%%%%%%%%%%%%%%%%%%%%%%%%%%%%%%%%%%%%%%%%%%%%%%%%%%%%%%%%%%%%%%%%%%%%%%%

\begin{figure}
    \centering
    \includegraphics[width=0.99\linewidth]{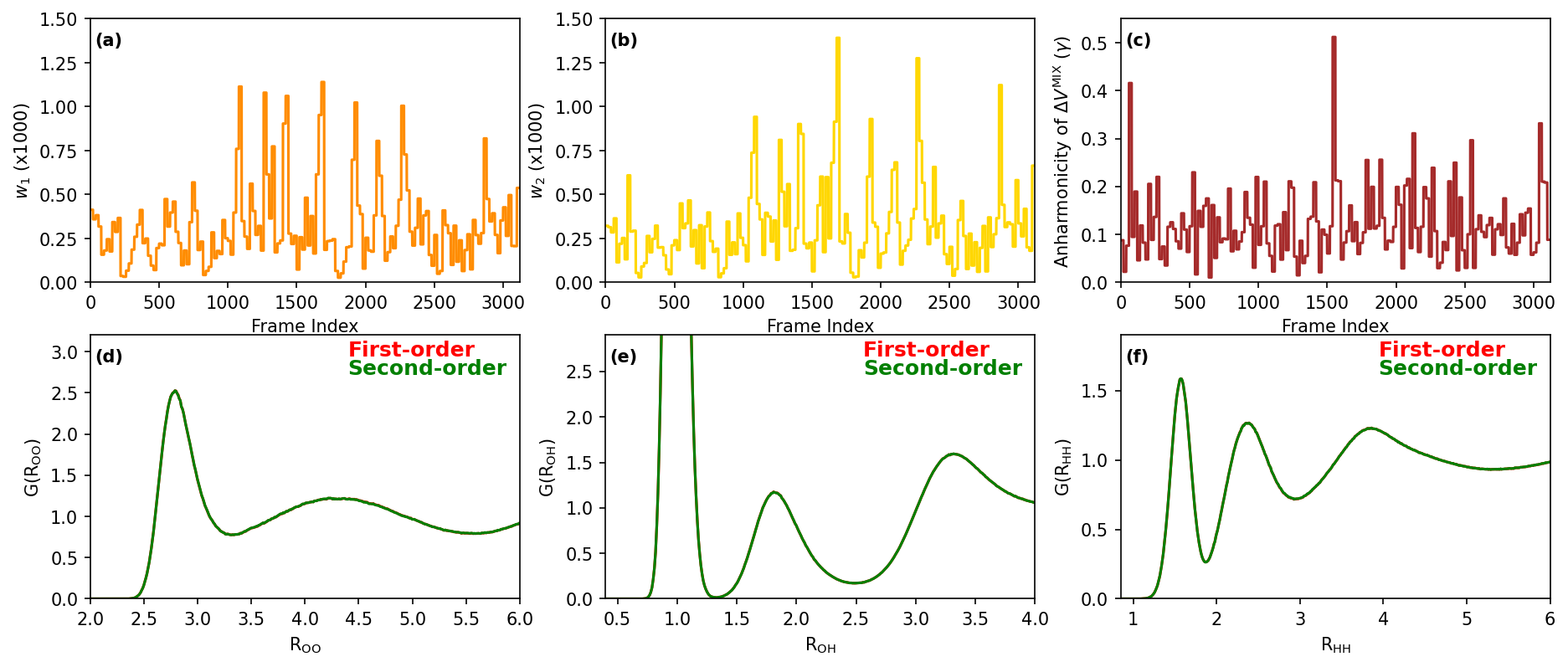}
    \caption{The weights for the reweighting procedure for the first- (a) 
    and second-order (b) as shown in Eq. (16) and Eq. (S8) respectively.
    In figure (c) the ``anharmonicity'' as defined in Eq. (S11) is shown.
    The RDFs for PIHMC-MIX for the RPBE-D3 functionals are plotted in
    figures (d), (e) and (f) for O-O, O-H and H-H respectively.
    The ones reweighted with the first-order expression, Eq. (16), are plotted in red
    and those from the second-order expression, Eq. (S8), are plotted in green.}
\end{figure}
In section 2C of the main text, we outline the reweighting procedure by Miao \textit{et al.}\cite{miao_improved_2014} to the first order in the cumulant expansion. 
In this section, we will briefly discuss the second-order expansion of the exponential
reweighting and the ``anharmonicity''\cite{miao_improved_2014,Lange_proteins_2008} of the
binned data.
We will in the following use the results from PIHMC-MIX for RPBE-D3 as an example,
but the findings are found to be general across the simulations done in this study.
The second-order term of the cumulant expansion is given as 
\begin{equation}
C_2 = \left(1-\alpha\right)^2\left(\left\langle  
\left(\Delta V^{\mathrm{MIX}}_{}\right)^2\right\rangle - \left\langle  
\Delta V^{\mathrm{MIX}}_{}\right\rangle^2\right)=
\left(1-\alpha\right)^2\sigma^2_{}\left(V^{\mathrm{MIX}}_{}\right)
\end{equation}
where $\sigma^{2}_{}\left(\Delta V^{\Delta\mathrm{MIX}}_{}\right)$ is the standard deviation 
of the potential difference.
The reweighting expression to the second order is then defined as
\begin{equation}
\rho^{\mathrm{FP}}_{}\left(A\right)\approx
\rho^{\mathrm{MIX}}_{}\left(A\right)
\frac{\exp\left(
\beta\left(1-\alpha\right)
\left(\left\langle \Delta V^{\mathrm{MIX}}_{}\right\rangle_j+
\frac{\beta}{2}\left(1-\alpha\right)
\sigma^{2}_{j}\left(\Delta V_{}^{\mathrm{MIX}}\right)
\right)\right)}
{\sum_j^{M}\exp \left(\beta\left(1-\alpha\right)
\left(\left\langle \Delta V^{\mathrm{MIX}}_{}\right\rangle_j+\frac{\beta}{2}\left(1-\alpha\right)\sigma^{2}_{j}\left(\Delta V_{}^{\mathrm{MIX}}\right)\right)\right)}
\end{equation}
We have plotted the weights for reweighting using the first and second order expansion
in figure S1 (a) and (b) respectively.
The weights while different express no systematic shifts when using the second order expansion.

%%%

The second order expansion is important, since if the data in each bin were drawn from a
normal distribution, 
then the second order cumulant expansion is exact. 
To determine if the data in the bins are normally distributed we consider their differential entropy defined as
\begin{equation}
S_{\Delta V}=-\int_0^{\infty} p\left(\Delta V^{\mathrm{MIX}}_{}\right)\ln\left(p\left(\Delta V^{\mathrm{MIX}}_{}\right)\right) \mathrm{d}\Delta V^{\mathrm{MIX}}_{}.
\end{equation}
Where $p\left(\Delta V^{\mathrm{MIX}}_{}\right)$ is the probability distribution.
Assuming a normal distribution of $\Delta V_{}^{\mathrm{MIX}}$ the maximum entropy is given by
\begin{equation}
S^{}_{\max}=\frac{1}{2}\ln\left(2\pi e\sigma^2\left(\Delta V^{\mathrm{MIX}}_{}\right)\right).
\end{equation}
The "anharmonicty"\cite{Lange_proteins_2008} ($\gamma$) can then be defined 
as a difference between these two entropies,
\begin{equation}
\gamma = S^{}_{\max} - S_{\Delta V} = \frac{1}{2}\ln\left(2\pi e\sigma^2\right)+\int_0^{\infty} p\left(\Delta V^{\mathrm{MIX}}_{}\right)\ln\left(p\left(\Delta V^{\mathrm{MIX}}_{}\right)\right) \mathrm{d}\Delta V^{\mathrm{MIX}}_{},
\end{equation}
which, if the data were truly normally distributed, would be zero and will always be positive.
$\gamma$ can thus serve as an indicator of the accuracy of the cumulant expansion of second-order.
The ``anharmonicity'' for the RPBE-D3 PIHMC-MIX data were calculated using SciPy\cite{2020SciPy-NMeth} version 1.13.1 and are plotted in figure S1 (c).
The anharmonicity is generally low when compared to the results of Miao \textit{et al.},
and we do not find any correlation between the size of the weights and the anharmonicity 
of the data in each bin.
With similar results obtained for the other simulations carried out in this study,
we conclude that binning the data into bins with $M=20$ according to the simulation steps
and using the second-order cumulant expansion is sufficiently accurate.

%%%

We now turn to examining the differences between the first- and second-order cumulant expansions,
given in Eq. (16) and (S8) respectively.
The resulting O-O, O-H and H-H RDFs are plotted in figures Y (d), (e) and (f) respectively.
By visual inspection, the plots for the RDFs with the first and second order cumulant expansion
are identical.
The sum of absolute differences between the G($R$)'s of the RDFs are 0.79, 0.43 and 0.41
for O-O, O-H and H-H respectively across the 800 bins used to construct the plots.
The above findings allow us to conclude that expansion to the first-order does not deviate significantly from the second-order expansion.
Thus, the choice of using the first-order expansion only in this study is justified.

%%%%%%%%%%%%%%%%%%%%%%%%%%%%%%%%%%%%%%%%%%%%%%%%%%%%%%%%%%%%%%%%%%%%%%%%
\section{Parameters for the descriptors for the water MLPs}
%%%%%%%%%%%%%%%%%%%%%%%%%%%%%%%%%%%%%%%%%%%%%%%%%%%%%%%%%%%%%%%%%%%%%%%%

The Behler-Parrinello structural fingerprint parameters
\cite{behler_generalized_2007} used for all MLPs considered 
in this study are given in Table SI.
The brackets are used to indicate that the terms should be
expanded by taking the direct product of the vectors
to generate the full set of descriptors.
\textit{E.g.} in the first three index
descriptors both the terms with 
$(\lambda, \eta)=(-1,1)$ and 
$(\lambda, \eta)=(1,2)$ are included in
the total descriptor, as well as 4 other
combinations of $(\lambda, \eta)$.
For the atom types Type2 and Type3 (Y,Z) 
we consider all possible atom types
, (Y)=\{(O), (H)\}, for the two body interactions,
and for the three body interactions we consider
all non-redundant pairs, \textit{i.e.}
(Y,Z)=\{(H,H), (H,O), (O,O)\}.
For the Type1 (X) atom type, we consider 
oxygen and hydrogen separately,
depending on which the descriptor
aims to describe. 
This results in a descriptor vector for 
both atomic species with a length of 70.

\begin{table*}[htp]
\caption{The parameters of the radial and angular descriptors 
used to model water in this study.
For the $\lambda$ and $\zeta$ parameters, the numbers in brackets 
should be taken as a direct product to form the full set of 
descriptors used in the angular space. 
X, Y, Z correspond to the atom type,
\textit{i.e.} either H or O. }
\begin{ruledtabular}
\begin{tabular}{cccccccc}
Type1 & Type2 & R$_s$ & R$_c$ & $\eta$  \\
 \hline
X & Y & 0 & 6.5 & 0.003214 \\
X & Y & 0 & 6.5 & 0.035711 \\
X & Y & 0 & 6.5 & 0.071421 \\
X & Y & 0 & 6.5 & 0.124987 \\
X & Y & 0 & 6.5 & 0.214264 \\
X & Y & 0 & 6.5 & 0.357106 \\
X & Y & 0 & 6.5 & 0.714213  \\
X & Y & 0 & 6.5 & 1.428426 \\
\hline\hline
Type1 & Type2 & Type3 & R$_c$& $\eta$ & $\lambda$ & $\zeta$ \\
\hline
X & Y & Z & 6.5 & 0.000357 &  $\{-1, 1\}$ & $\{1,2,4\}$ \\
X & Y & Z & 6.5 & 0.0028569 & $\{-1, 1\}$ & $\{1,2,4\}$ \\
X & Y & Z & 6.5 & 0.089277 & $\{-1, 1\}$ & $\{1,2,4\}$\\
\end{tabular}
\end{ruledtabular}
\end{table*}

%%%%%%%%%%%%%%%%%%%%%%%%%%%%%%%%%%%%%%%%%%%%%%%%%%%%%%%%%%%%%%%%%%%%%%%%
\section{Additional information about (SL-)PIHMC-MIX and (SL-)HMC-MIX simulations}
%%%%%%%%%%%%%%%%%%%%%%%%%%%%%%%%%%%%%%%%%%%%%%%%%%%%%%%%%%%%%%%%%%%%%%%%

In this section we present the number of HMC steps 
$\left(N_{\mathrm{steps}^{}}\right)$, 
the average acceptance ratios 
$\left(\left\langle A_{\mathrm{test}}\right\rangle\right)$, 
and the effective trajectory length 
$\left(t_{\mathrm{eff}}^{}\right)$ for the SL-PIHMC-MIX and 
SL-HMC-MIX trajectories used to generate the MLPs that were used 
for the PIHMC-MIX and HMC-MIX trajectories.
Note that for all the training trajectories, the number of ML-PIMD
steps in each trial trajectory ($n_{\mathrm{ML}}^{}$)
is initialized at 2 and eventually grows to 128.
We furthermore list $N_{\mathrm{steps}^{}}$, 
$\left\langle A_{\mathrm{test}}\right\rangle$,
$n_{\mathrm{ML}}^{}$, and $t_{\mathrm{eff}}^{}$
for the D$_2$O trajectories and classical (HMC-MIX) trajectories 
for SCAN, rev-vdW-DF2 and optB88-vdW in Table SIII.
Common for all trajectories in table SII and SIII are that they are
run with $\alpha=0.25$.

\begin{table*}[htp]
\caption{
 Number of HMC steps $\left(N_{\mathrm{steps}^{}}\right)$, 
average acceptance ratios 
$\left(\left\langle A_{\mathrm{test}}\right\rangle\right)$ 
and effective trajectory length $\left(t_{\mathrm{eff}}^{}\right)$ in picoseconds for the SL-PIHMC-MIX 
and SL-HMC-MIX trajectories used to train the MLPs used in this study. 
$t_{\mathrm{eff}}^{}$ were calculated
based on taking 2-128 ML-PIMD steps in between each
HMC step.}
\begin{ruledtabular}
 \begin{tabular}{ccccc}
Functional & Method & $N_{\mathrm{steps}}$ & 
$\left\langle A_{\mathrm{test}}\right\rangle$ (\%) & 
$t_{\mathrm{eff}}^{}$ (ps) \\
\hline
RPBE-D3 & SL-HMC-MIX & 5,000 & 45.7 & 52.9\\
RPBE-D3 & SL-PIMC-MIX & 5,000 & 48.5 & 69.3\\
RPBE-D3$^a$ & SL-PIMC-MIX & 2,000 & 63.4 & 40.6\\
SCAN & SL-PIHMC-MIX & 5,000 & 48.4 & 67.5\\
rev-vdW-DF2 & SL-PIHMC-MIX & 5,000 & 47.8 & 67.6\\
Opt88-vdW & SL-PIHMC-MIX & 5,000 & 29.0 & 17.3\\
\end{tabular}
\end{ruledtabular}
\parbox{\linewidth}{\raggedright $^a$This is the additional training done for the D$_2$O MLP, in this case $n_{\mathrm{ML}}=128$ for the entire trajectory.}
\end{table*}

\begin{table*}[htp]
\caption{ Number of HMC steps $\left(N_{\mathrm{steps}^{}}\right)$, 
average acceptance ratios 
$\left(\left\langle A_{\mathrm{test}}\right\rangle\right)$,
the number of ML-PIMD steps in each trial trajectory 
($n_{\mathrm{ML}}^{}$), and effective trajectory length $\left(t_{\mathrm{eff}}^{}\right)$
in picoseconds for additional trajectories in this study 
not covered in Table I.} 
\begin{ruledtabular}
\begin{tabular}{cccccc}

Functional & Method & $N_{\mathrm{steps}}$ & $\left\langle A_{\mathrm{test}}\right\rangle$ (\%) & 
$n_{\mathrm{ML}}$ & $t_{\mathrm{eff}}^{}$ (ps) \\
\hline
RPBE-D3$^{a}$ & HMC-MIX & 5,000 & 64.8 & 128 & 103.7\\
RPBE-D3$^{a}$ & PIHMC-MIX & 5,000 & 25.7 & 8-128 & 5.1\\
RPBE-D3$^{b}$ & PIHMC-MIX & 5,000 & 60.5 & 128 & 96.8\\
SCAN$^{c}$ & HMC-MIX & 10,000 & 59.6 & 128 &  96.8\\
rev-vdW-DF2$^{c}$ & HMC-MIX & 10,000 & 49.4 & 128 & 88.7\\
Opt88-vdW$^{c}$ & HMC-MIX & 10,000 & 31.5 & 32-128 & 59.0\\

\end{tabular}
\end{ruledtabular}
\parbox{\linewidth}{\raggedright $^a$ These trajectories were made with the MLP 
generated by SL-HMC-MIX, which are discussed in Sections IV(A) and SIV.}
\parbox{\linewidth}{\raggedright $^b$ This is the PIHMC-MIX trajectory for D$_2$O which is presented in Section IV(D).}
\parbox{\linewidth}{\raggedright $^c$ The RDFs for these simulations are presented in Sections IV(E) and SIX.}
\end{table*}

%%%%%%%%%%%%%%%%%%%%%%%%%%%%%%%%%%%%%%%%%%%%%%%%%%%%%%%%%%%%%%%%%%%%%%%%
\section{Peak positions for H$_2$O with the RPBE-D3 Functional}
%%%%%%%%%%%%%%%%%%%%%%%%%%%%%%%%%%%%%%%%%%%%%%%%%%%%%%%%%%%%%%%%%%%%%%%%

In this section we report the peak positions 
of the RDFs for RPBE-D3 shown in Figures 2, 3 and 4 
in Table SIV, SV and SVI respectively.
These peak positions and heights are used in the discussions 
in Sections IV(A-C).

\begin{table}[H]
\caption{The positions and heights of the  peaks 
in the O-O RDFs for H$_2$O  presented in Figure 3. 
All peak positions are given in Å.
The data is denoted either with a r$^{\mathrm{OO}}_i$ 
or h$^{\mathrm{OO}}_i$ referring to the peak position 
and heights respectively.
The position (h$^{\mathrm{OO}}_{\mathrm{min}}$) and 
height (r$^{\mathrm{OO}}_{\mathrm{min}}$) of the minimum 
of the first interstitial region. 
The experimental reference stem from Ref. 
\onlinecite{soper_radial_2013}.}
\begin{ruledtabular}
\begin{tabular}{ccccccccc}
Potential & Method & nwat & r$_1^{\mathrm{OO}}$ (Å) & h$_1^{\mathrm{OO}}$ &
r$_{\mathrm{min}}^{\mathrm{OO}}$ (Å) & h$_{\mathrm{min}}^{\mathrm{OO}}$ &
  r$_2^{\mathrm{OO}}$ (Å) & h$_2^{\mathrm{OO}}$\\
\hline
 FP & MD & 64 & 2.81 & 2.66 & 
3.33 & 0.78 & 4.22 & 1.20 \\
ML & HMC-MIX & 64 & 2.79 & 2.66
& 3.40 & 0.79 & 4.39 & 1.21 \\ 
ML & MD & 64 & 2.79 & 2.73 & 3.28 & 0.78 & 
4.31 & 1.21 \\
ML & MD & 256 & 2.79 & 2.64 & 3.32 & 0.83 & 4.30 & 1.17 \\
 FP  & PIMD & 64 & 2.78 & 2.47 & 3.33 & 0.83 & 4.35 & 1.19 \\
ML & PIHMC-MIX & 64 & 2.79 & 2.53 & 3.33 & 0.77 & 4.24 & 1.22 \\
ML & PIMD & 64 & 2.79 & 2.42 & 3.30 & 0.87 & 4.29 & 1.16 \\
ML & PIMD & 256 & 2.80 & 2.45 & 3.32 & 0.87 & 4.19 & 1.16 \\
\hline
& Experiment & & 2.79 & 2.50 & 3.36 & 0.78 & 4.53 & 1.12\\
\end{tabular}
\end{ruledtabular}
\end{table} 

\begin{table}[H]
\caption{The positions and heights of the peaks in 
the O-H RDFs of H$_2$O presented in Figure 4. 
All peak positions are given in Å.
The data is denoted either with a r$^{\mathrm{OH}}_i$ 
or H$^{\mathrm{OH}}_i$ referring to the peak position 
and heights respectively. 
The experimental reference stem from Ref. 
\onlinecite{soper_radial_2013}, 
except those marked by $^*$ which are from 
Ref. \onlinecite{soper_radial_2000}.}
\begin{ruledtabular}
\begin{tabular}{ccccccccc}
Potential & Method & nwat & r$_1^{\mathrm{OH}}$ (Å) & h$_1^{\mathrm{OH}}$ & 
 r$_2^{\mathrm{OH}}$ (Å) & h$_2^{\mathrm{OH}}$ & 
 r$_3^{\mathrm{OH}}$ (Å) & h$_3^{\mathrm{OH}}$\\
\hline
 FP & MD & 64 & 0.98 & 36.78 & 
1.82 & 1.33 & 3.28 & 1.64 \\
ML & HMC-MIX & 64 & 
0.98 & 37.42 & 1.84 & 
1.33 & 3.28 & 1.66 \\ 
ML & MD & 64 & 0.98 & 37.28 & 1.83 & 1.37 & 
3.29 & 1.64 \\
ML & MD & 256 & 0.98 & 37.69 & 1.83 & 1.32 & 3.30 & 1.66 \\
 FP  & PIMD & 64 & 0.99 & 13.19 & 1.81 & 1.15 & 3.32 & 1.60 \\
ML & PIHMC-MIX & 64 & 0.99 & 13.22 & 1.81 & 1.18 & 3.32 & 1.59 \\
ML & PIMD & 64 & 0.99 & 13.27 & 1.82 & 1.12 & 3.32 & 1.59 \\
ML & PIMD & 256 & 0.99 & 13.30 & 1.83 & 1.12 & 3.33 & 1.58 \\
\hline
& Experiment & & 0.96$^*$ & 12.71$^*$ & 1.86 & 1.04 & 3.27 & 1.48\\
\end{tabular}
\end{ruledtabular}
\end{table}

\begin{table}[H]
\caption{The positions and heights of the  peaks 
in the H-H RDFs of H$_2$O presented in Figure 5. 
All peak positions are given in Å.
The data is denoted either with r$^{\mathrm{HH}}_i$ 
or h$^{\mathrm{HH}}_i$ referring to the peak position 
and heights respectively.
The experimental reference stem from Ref. 
\onlinecite{soper_radial_2013}, 
except those marked by $^*$ which are from Ref. 
\onlinecite{soper_radial_2000}.}
\begin{ruledtabular}
\begin{tabular}{ccccccccc}
Potential & Method & nwat & r$_1^{\mathrm{HH}}$ (Å) & h$_1^{\mathrm{HH}}$ &
  r$_2^{\mathrm{HH}}$ (Å) & h$_2^{\mathrm{HH}}$ & 
  r$_3^{\mathrm{HH}}$ (Å) & h$_3^{\mathrm{HH}}$\\
\hline
 FP & MD & 64 & 1.56 & 3.43 & 
2.34 & 1.44 & 3.88 & 1.22 \\
 ML & HMC-MIX & 64 & 1.56 & 3.47 & 2.36 & 1.43 & 3.80 & 1.24 \\
ML & MD & 64 & 1.56 & 3.51 & 2.38 & 1.43 & 3.83 & 1.22 \\
ML & MD & 256 & 1.56 & 3.51 & 2.37 & 1.41 & 3.80 & 1.25 \\
 FP  & PIMD & 64 & 1.57 & 1.57 & 2.36 & 1.28 & 3.83 & 1.21 \\
ML & PIHMC-MIX & 64 & 1.57 & 1.58 & 2.36 & 1.27 & 3.84 & 1.23 \\
ML & PIMD & 64 & 1.57 & 1.59 & 2.38 & 1.24 & 3.83 & 1.21 \\
ML & PIMD & 256 & 1.57 & 1.60 & 2.40 & 1.24 & 3.85 & 1.20 \\
\hline
& Experiment & & 1.53$^*$ & 1.71$^*$ & 2.43 & 1.34 & 3.84 & 1.17\\
\end{tabular}
\end{ruledtabular}
\end{table}

%%%%%%%%%%%%%%%%%%%%%%%%%%%%%%%%%%%%%%%%%%%%%%%%%%%%%%%%%%%%%%%%%%%%%%%%
\section{Results for different values of $\alpha$ in PIHMC-MIX}
%%%%%%%%%%%%%%%%%%%%%%%%%%%%%%%%%%%%%%%%%%%%%%%%%%%%%%%%%%%%%%%%%%%%%%%%

In this section, we will provide the structural data for 
different values of $\alpha$ in the PIHMC-MIX procedure.
In Figure S1, we compare the structures of 
$\alpha = \{1.0, 0.75, 0.5\}$ with the value used throughout
the main text, $\alpha = 0.25$.
The peak positions and heights are given in Table SVII.
These results are discussed in Section 4(A) of the main text.

\begin{figure}
    \centering
    \includegraphics[width=0.99\linewidth]{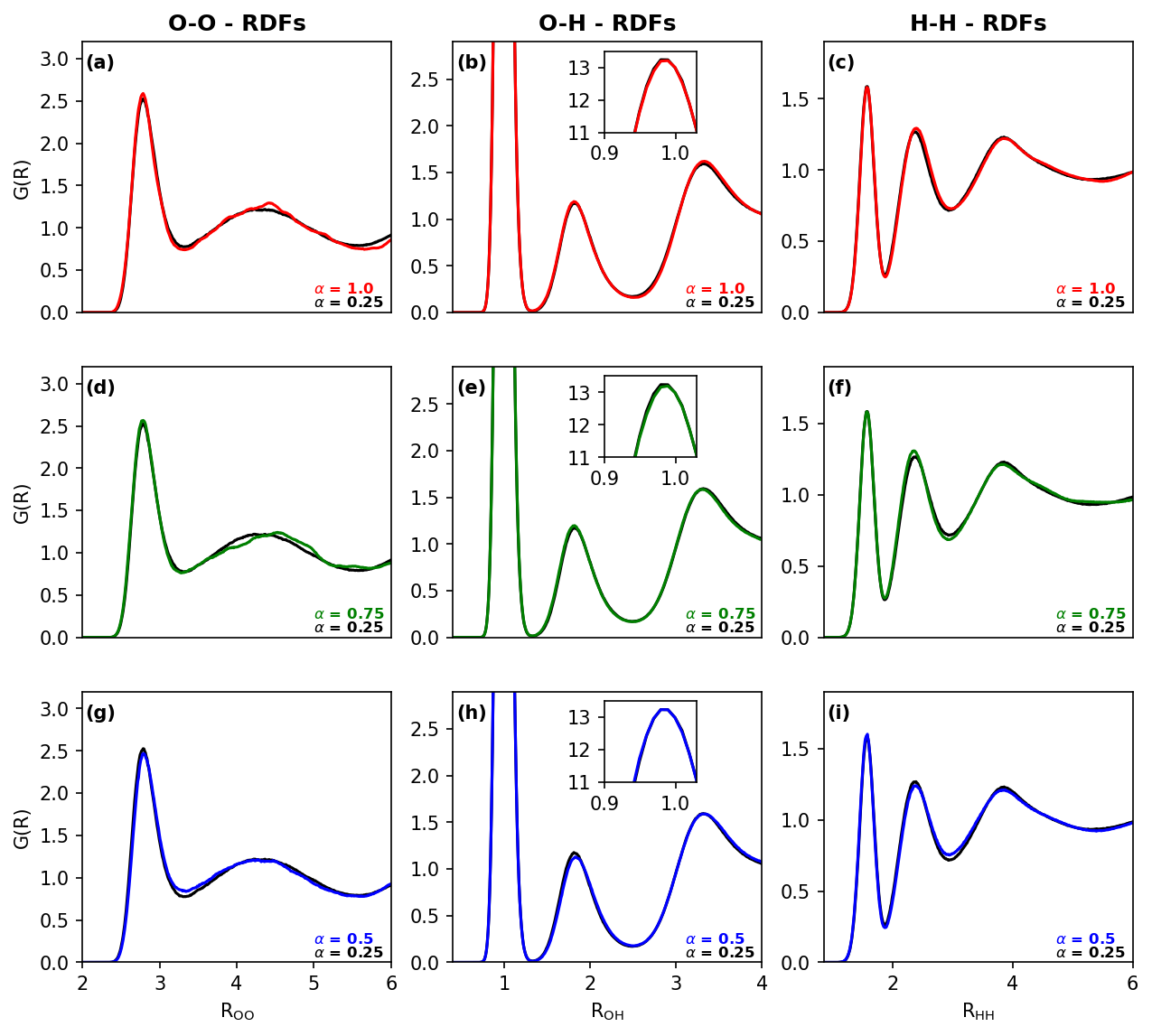}
    \caption{The O-O (a,d,g), O-H (b,e,h) and H-H (c,f,i) RDFs 
    calculated from PIHMC-MIX with $\alpha=\{1.0, 0.75, 0.5\}$ 
    plotted in red (a-c), green (d-f) and blue (g-i) respectively. 
    The results for $\alpha=0.25$ are given as reference
    in black in all figures.}
    \label{fig:enter-label}
\end{figure}

\begin{table}[H]
\caption{The positions and heights of the peaks 
in the O-H RDFs of H$_2$O presented in Figure S1.
All peak positions are given in Å. 
The data is denoted either with r$^{\mathrm{XY}}_i$ 
or h$^{\mathrm{XY}}_i$ referring to the peak
position and heights respectively for the pair XY
$\in\{\mathrm{OO}, \mathrm{OH},\mathrm{HH}\}$.
For the OO pair, we also give the position 
(r$_{\mathrm{min}}^{\mathrm{OO}}$) and height 
(h$_{\mathrm{min}}^{\mathrm{OO}}$) of the minima 
of the first interstitial region.}
\begin{ruledtabular}
\begin{tabular}{ccccccc}
$\alpha$ & r$_1^{\mathrm{OO}}$ (Å) & h$_1^{\mathrm{OO}}$ &
 r$_{\mathrm{min}}^{\mathrm{OO}}$ (Å) & h$_{\mathrm{min}}^{\mathrm{OO}}$ &
  r$_2^{\mathrm{OO}}$ (Å) & h$_2^{\mathrm{OO}}$\\
\hline
1.00 & 2.79 & 2.59 & 3.33 & 0.74 & 4.42 & 1.29 \\
0.75 & 2.78 & 2.57 & 3.28 & 0.76 & 4.53 & 1.24 \\
0.50 & 2.79 & 2.46 & 3.35 & 0.84 & 4.28 & 1.21 \\
0.25 & 2.79 & 2.53 & 3.33 & 0.77 & 4.24 & 1.22 \\
\hline\hline
$\alpha$ & r$_1^{\mathrm{OH}}$ (Å) & h$_1^{\mathrm{OH}}$ & 
 r$_2^{\mathrm{OH}}$ (Å) & h$_2^{\mathrm{OH}}$ & 
 r$_3^{\mathrm{OH}}$ (Å) & h$_3^{\mathrm{OH}}$\\
\hline
1.00 & 0.99 & 13.21 & 1.81 & 1.19 & 3.33 & 1.62 \\
0.75 & 0.99 & 13.21 & 1.82 & 1.20 & 3.32 & 1.59 \\
0.50 & 0.98 & 13.25 & 1.83 & 1.13 & 3.32 & 1.59 \\
0.25 & 0.99 & 13.22 & 1.81 & 1.18 & 3.32 & 1.59 \\
\hline\hline
$\alpha$ & r$_1^{\mathrm{HH}}$ (Å) & h$_1^{\mathrm{HH}}$ &
  r$_2^{\mathrm{HH}}$ (Å) & h$_2^{\mathrm{HH}}$ & 
  r$_3^{\mathrm{HH}}$ (Å) & h$_3^{\mathrm{HH}}$\\
\hline
1.00 & 1.57 & 1.58 & 2.40 & 1.29 & 3.86 & 1.22 \\
0.75 & 1.57 & 1.59 & 2.35 & 1.31 & 3.86 & 1.22 \\
0.50 & 1.58 & 1.60 & 2.38 & 1.24 & 3.84 & 1.21 \\
0.25 & 1.57 & 1.58 & 2.36 & 1.27 & 3.84 & 1.23 \\
\end{tabular}
\end{ruledtabular}
\end{table}

%%%%%%%%%%%%%%%%%%%%%%%%%%%%%%%%%%%%%%%%%%%%%%%%%%%%%%%%%%%%%%%%%%%%%%%%
\section{Accuracy of the MLPs trained in the SL-PIHMC process}
%%%%%%%%%%%%%%%%%%%%%%%%%%%%%%%%%%%%%%%%%%%%%%%%%%%%%%%%%%%%%%%%%%%%%%%%

To analyze the accuracy of the MLPs we need a set of  FP  data 
to compare with which are not part of the SL-PIHMC-MIX trajectory 
which is responsible for generating the MLP in question.
Conveniently the subsequent PIHMC-MIX trajectories provide such 
a data set,
as only the initial structure stems from the SL-PIHMC-MIX 
trajectories,
but the rest of the trajectory is generated independently.
We thus assume that each trial move proposed
in PIHMC-MIX represents a unique structure,
which is reasonable given that the starting
velocities are randomly initialized even
if the starting structure remains the same
when a number of HMC steps in a row are rejected.

%%%

The most frequently used quantity for determining
the quality of an MLP is the mean absolute
error (MAE) for energy per atom in the system
\begin{equation}
\sigma^{\mathrm{at}}_{\mathrm{E}}=\frac{1}{NPM}\sum_{s=1}^{P}\sum_{j=1}^{M}
\left|\mathrm{E}_{\mathrm{DFT}}^{j, s}-\mathrm{E}_{\mathrm{ML}}^{j, s}\right|
\end{equation}
where $N$ is the number of atoms in the system,
$P$ is the number of beads in the simulation,
$M$ is the number of trial moves compared, 
$\mathrm{E}_{\mathrm{DFT}}^{j,s}$ and 
$\mathrm{E}_{\mathrm{ML}}^{j,s}$
are the energies for the whole system
in bead $s$ of the $j$th trial move
calculated with DFT and MLPs respectively.
While force data has not been used to train
the MLPs, it is still useful to compare the 
forces from the DFT and MLPs.
Seeing that the force is a vector quantity,
we compare two parameters of individual
force vectors on each atom in the system.
The first being the magnitude of the forces
\begin{equation}
|F_{i,j}^{s}|=\left|\mathbf{F}_{\mathrm{DFT}}^{i,s, j}-\mathbf{F}_{\mathrm{ML}}^{i,s,j}\right|
\end{equation}
Where $F_{\mathrm{DFT}}^{i,s, j}$ and $F_{\mathrm{ML}}^{i,s,j}$
are the forces on the $i$th atom in bead $s$ of the $j$th trial move
calculated using the DFT and MLPs respectively.
The MAE for force is defined as 
\begin{equation}
\sigma_{\mathrm{F}}^{\mathrm{at}}=\frac{1}{3NPM}\sum_{j=1}^{M}
\sum_{s=1}^{P}\sum_{i=1}^{N}|F_{i,j}^{s}|.
\end{equation}
To measure the error in the direction
of the forces for each atom $i$
we use the cosine similarity of the DFT and ML force vectors
\begin{equation}
F^{i, s, j}_{\mathrm{cos}}=\frac{F_{\mathrm{DFT}}^{i, s, j}\cdot F_{\mathrm{ML}}^{i, s, j}}
{|F_{\mathrm{DFT}}^{i, s, j}||F_{\mathrm{ML}}^{i, s, j}|}
\end{equation}
The cosine similarity should be 1 if the force vectors from the DFT
and MLP are aligned,
and -1 if they are pointing in opposite directions.
We calculate the average of the cosine similarity
as a measure of the general quality of the force
vectors calculated by the MLP
\begin{equation}
  \sigma_{\theta}^{\mathrm{at}} =
  \frac{1}{NPM}\sum_{j=1}^{M}\sum_{s=1}^{P}\sum_{i=1}^{N}
  F^{i, s, j}_{\mathrm{cos}}.
\end{equation}

%%%

The results of the analysis suggested
above are given in Table SII and SIII,
for the simulations using RPBE-D3 
and using the different DFT functionals 
respectively.
Furthermore, the correlation between
$\mathrm{E_{DFT}}$ and $\mathrm{E_{ML}}$,
the 1D distributions of $|F|_{i,j}^{s}$
and $F_{cos}^{i,s,j}$,
and the correlation between $|F|_{i,j}^{s}$ and
$F_{cos}^{i,s,j}$ are plotted in Figures S2-3,
for the simulations using RPBE-D3 
and using different DFT functionals
respectively.
In analyzing the results of the 2D correlation
plots in Figures S1-2,
one should note that large discrepancies in
force vector directions, 
i.e. $F^{i, s, j}_{\mathrm{cos}}\approx -1$,
can be acceptable if $|F|_{i,j}^{s}$ is small.
The reason being that the actual forces in 
the DFT and MLPs
will be bound by $|F_{\mathrm{DFT}}^{i,s,j}|+
|F_{\mathrm{ML}}^{i,s,j}|=|F|_{i,j}^{s}$
in the extreme case $F^{i, s, j}_{\mathrm{cos}}= -1$.
The significance of these results
in relation with previous studies are discussed
in Section IV(B) of the main text.

\begin{table}[H]
\caption{Comparison of the accuracy of two MLPs trained 
using SL-HMC-MIX and SL-PIHMC-MIX both using the RPBE-D3 
DFT functional.
All results are based on taking 5000 MC steps in HMC-MIX 
or PIHMC-MIX with $\alpha=0.25$. 
Here we report the average acceptance ratios 
$\left(\left\langle A_{\mathrm{test}}\right\rangle\right)$,
the effective trajectory length $\left(t_{\mathrm{eff}}^{}\right)$ in picoseconds,
the MAE per atom $\sigma_{\mathrm{E}}^{\mathrm{at}}$ given in Eq. (S12),
the $\sigma_{\mathrm{F}}^{\mathrm{at}}$ given in Eq. (S14),
and the average of the dot products between the force
vectors from  FP  and ML potentials 
$\sigma_{\theta}^{\mathrm{at}}$ from Eq. (S16).
Lower numbers suggest a better agreement between MLP 
and DFT results,
except for $\sigma_{\theta}^{\mathrm{at}}$ which should 
be close to one.}
\begin{ruledtabular}
\begin{tabular}{ccccccc}
Method & Training &
$\left\langle A_{\mathrm{test}}\right\rangle$ & t$_{\mathrm{eff}}^{}$ & $\sigma_{\mathrm{E}}^{\mathrm{at}}$ & $\sigma_{\mathrm{F}}^{\mathrm{at}}$ & 
$\sigma_{\theta}^{\mathrm{at}}$\\
& & [\%] & [ps] & [meV/atom] & [meV/\AA]& \\
\hline
HMC-MIX & SL-HMC-MIX & 64.8 & 103.7 & 0.30 & 49.4 & 0.984\\
PIHMC-MIX & SL-HMC-MIX & 25.7 & 5.1 & 3.58 & 199.9 & 0.986\\
HMC-MIX & SL-PIHMC-MIX & 55.3 & 88.51 & 0.70 & 67.0 & 0.964\\
PIHMC-MIX & SL-PIHMC-MIX & 55.5 & 99.9 & 0.36 & 79.0 & 0.990\\
\end{tabular}
\end{ruledtabular}
\end{table}

\begin{figure}
    \centering
    \caption{ (Caption for figure on next page)
    Comparison of energies and forces from ML and  FP  potentials
    for the MLPs trained using SL-HMC-MIX (a-b) and SL-PIHMC-MIX (c-d),
    where the structures tested stem from HMC-MIX (a, c) and PIHMC-MIX (b, d)
    trajectories.
    The first row compares the energies from the DFT calculation with 
    those from the MLP, and provides the combined 
    $\sigma_{\mathrm{E}}^{\mathrm{at}}$ for all data points.
    The second row depicts the 1D distribution of 
    $\left|F_{i,j}^{s}\right|$, Eq. (S13), and 
    $F_{\mathrm{cos}}^{i,s,j}$, Eq. (S15), in blue and green, respectively.
    In the third row, we give the 2D distribution of $\left|F_{i,j}^{s}\right|$ and $F_{\mathrm{cos}}^{i,s,j}$.}
\end{figure}
\addtocounter{figure}{-1}
\begin{figure}
    \centering
    \includegraphics[width=0.99\linewidth]{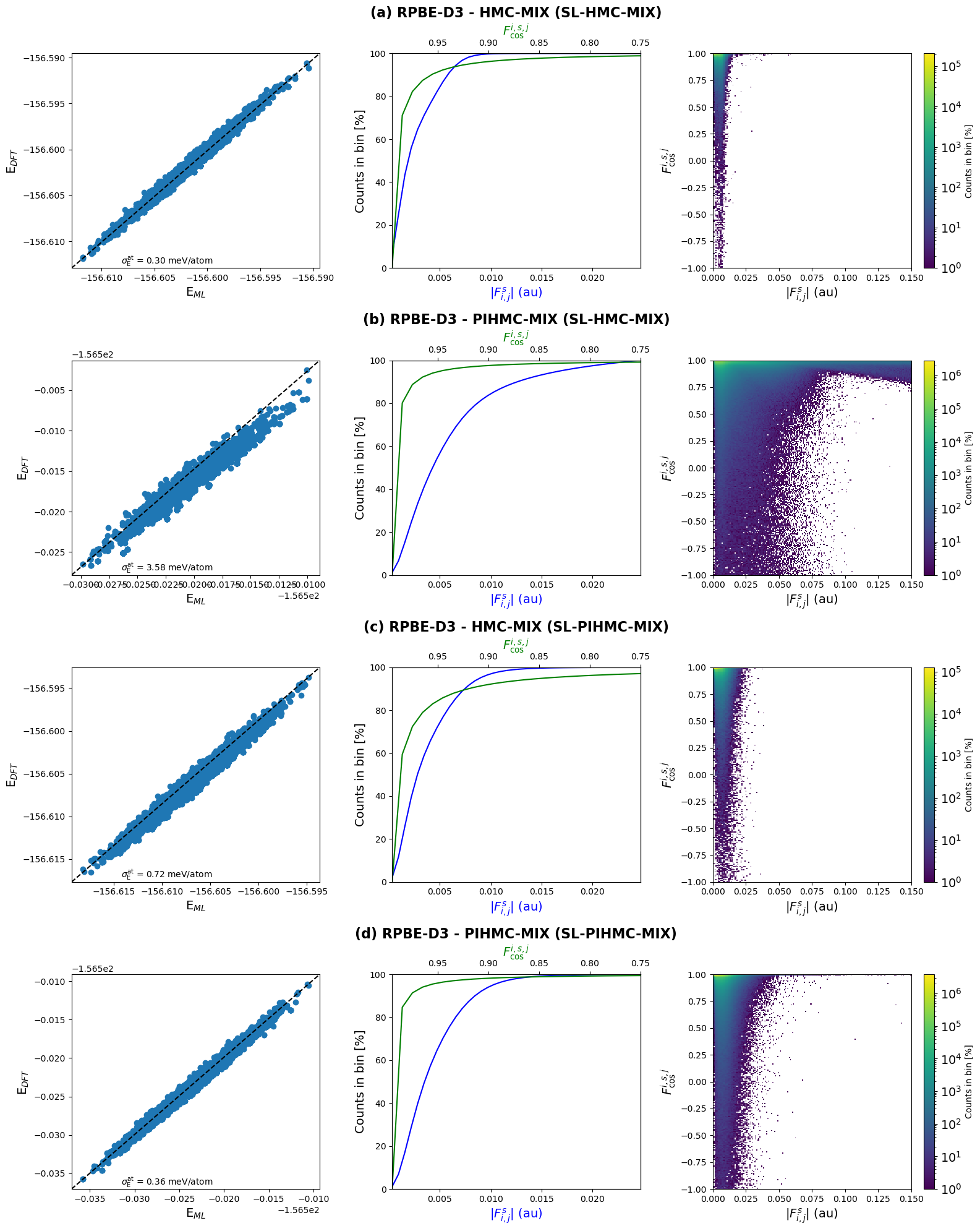}
    \caption{(See caption on previous page)}
\end{figure}

\begin{figure}
    \centering
    \includegraphics[width=0.99\linewidth]{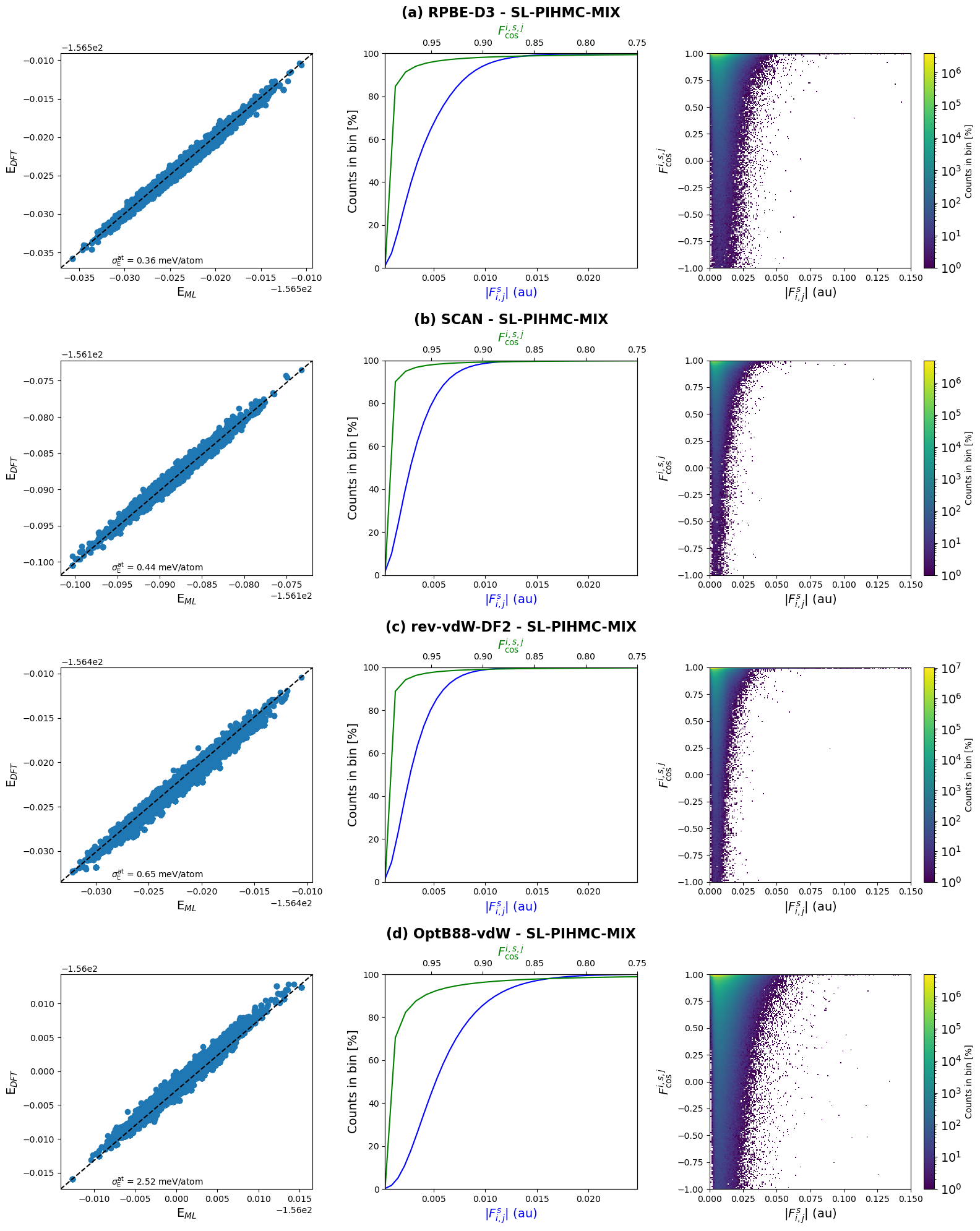}
    \caption{Comparison of energies and forces from ML and  FP  potentials
    for the MLPs trained on RPBE-D3 (a), SCAN (b), rev-vdW-DF2 (c)
    and OptB88-vdW (d).
    The rows of this figure depict the same comparisons and distributions
    as those in Figure S3.}
\end{figure}

\begin{table}[H]
\caption{Comparison of the accuracy of MLPs trained
on data from the RPBE-D3, SCAN, rev-vdW-DF2 and OptB88-vdW
functionals.
All results are based on taking 5000 MC steps in
PIHMC-MIX with $\alpha=0.25$. 
Here we report the average acceptance ratios 
$\left(\left\langle A_{\mathrm{test}}\right\rangle\right)$,
the effective trajectory length $\left(t_{\mathrm{eff}}^{}\right)$ in picoseconds,
the MAE per atom $\sigma_{\mathrm{E}}^{\mathrm{at}}$ given in Eq. (S12),
the $\sigma_{\mathrm{F}}^{\mathrm{at}}$ given in Eq. (S14),
and the average of the dot products between the force
vectors from  FP  and ML potentials 
$\sigma_{\theta}^{\mathrm{at}}$ from Eq. (S16).
Lower numbers suggest a better agreement between MLP 
and DFT results,
except for $\sigma_{\theta}^{\mathrm{at}}$ which should 
be close to one.}
 \begin{ruledtabular}
 \begin{tabular}{cccccc}
Functional & 
    $\left\langle A_{\mathrm{test}}\right\rangle$ & t$_{\mathrm{eff}}^{}$ 
    & $\sigma_{\mathrm{E}}^{\mathrm{at}}$ & $\sigma_{\mathrm{F}}^{\mathrm{at}}$ & 
$\sigma_{\theta}^{\mathrm{at}}$\\
 & [\%] & [ps] & [meV/atom] & [meV/\AA] & \\
\hline
RPBE-D3 & 55.5 & 99.9 & 0.36 & 79.0 & 0.990\\
SCAN & 54.4 & 96.8 & 0.44 & 61.3 & 0.994\\
rev-vdW-DF2 & 51.7 & 88.7 & 0.59 & 60.2 & 0.994\\
OptB88-vdW & 36.8 & 59.4 & 2.51 & 109.2 & 0.980\\
\end{tabular}
 \end{ruledtabular}
\end{table}

%%%%%%%%%%%%%%%%%%%%%%%%%%%%%%%%%%%%%%%%%%%%%%%%%%%%%%%%%%%%%%%%%%%%%%%%
\section{Comparison of RDFs from ML-MD and ML-PIMD with MLPs from SL-PIHMC-MIX and SL-HMC-MIX MLPs}
%%%%%%%%%%%%%%%%%%%%%%%%%%%%%%%%%%%%%%%%%%%%%%%%%%%%%%%%%%%%%%%%%%%%%%%%

\begin{figure}
    \centering
    \includegraphics[width=0.99\linewidth]{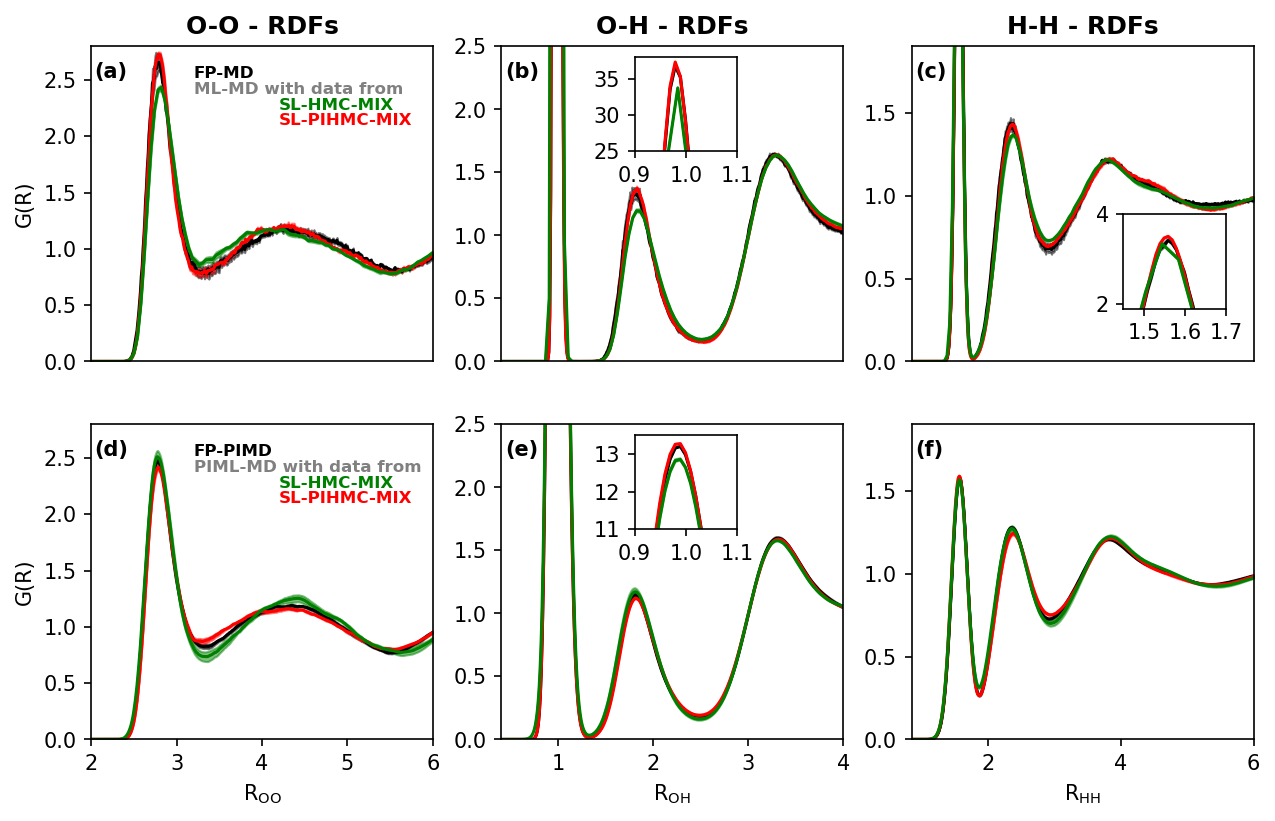}
    \caption{The O-O (a,d), O-H (b,e) and H-H (c,f) RDFs calculated using ML-MD (a-c) and 
    ML-PIMD (d-f) with MLPs based trained using SL-HMC-MIX (green) and 
    SL-PIHMC-MIX (red).
    The results of ML-MD and ML-PIMD are compared with reference 
     FP-MD (a-c) and  FP-PIMD (d-f) results in black.
    The peak heights and positions are given in Table SX.}
    \label{fig:enter-label}
\end{figure}

In the previous section we discussed the accuracy in terms of
energy and gradients when the  FP  data used to train the MLP
came from either PIHMC-MIX or HMC-MIX,
\textit{i.e.} if NQEs were included in the data set or not.
In this section, we show the performance of these trained MLPs
when used for running ML-MD and ML-PIMD for water,
to better understand the effects on the accuracy when
modelling water both with and without NQEs.
The resulting RDFs from these ML-MD and ML-PIMD simulations
are plotted in Figure S4,
where they are compared to the reference  FP-MD and  FP-PIMD
simulation results for RPBE-D3.
The peak heights and positions for the data presented in Figure
S4 are given in Table SXI.

%%%

As mentioned in the main text,
there are some minor differences between the  FP  simulations and those
using only an MLP to generate gradients for propagating the trajectory.
When comparing the MLPs trained with SL-HMC-MIX and SL-PIHMC-MIX,
we find that the former performs better for ML-MD,
while the latter performs best in the case of ML-PIMD.
This indicates that there is a limited transfer-ability of the
MLP trained on  FP  data with or without NQEs to ML-MD and ML-PIMD, respectively.

\begin{table}[H]
\caption{The positions and heights of the peaks in 
the O-O, O-H and H-H RDFs for H$_2$O presented
in Figure S4. 
All peak positions are given in Å.  
The data is denoted either with r$^{\mathrm{XY}}_i$ 
or H$^{\mathrm{XY}}_i$ referring to the peak position 
and heights respectively for the pair XY 
$\in\{\mathrm{OO}, \mathrm{OH},\mathrm{HH}\}$.
For the OO pair, we also give the position 
(r$_{\mathrm{min}}^{\mathrm{OO}}$) and height 
(h$_{\mathrm{min}}^{\mathrm{OO}}$) of the minima 
of the first interstitial region.}
\begin{ruledtabular}
\begin{tabular}{cccccccc}
 Training Data & Method & r$_1^{\mathrm{OO}}$ (Å) & h$_1^{\mathrm{OO}}$ &
 r$_{\mathrm{min}}^{\mathrm{OO}}$ (Å) & h$_{\mathrm{min}}^{\mathrm{OO}}$ &
  r$_2^{\mathrm{OO}}$ (Å) & h$_2^{\mathrm{OO}}$\\
  \hline
- &  FP-MD & 2.81 & 2.66 & 3.33 
& 0.78 & 4.22 & 1.20 \\
SL-HMC-MIX & ML-MD & 2.82 & 2.44 & 3.26 & 0.86 & 4.10 & 1.17 \\
SL-PIHMC-MIX & ML-MD & 2.79 & 2.73 & 3.28 & 0.78 & 4.31 & 1.21 \\
- &  FP-PIMD & 2.78 & 2.47 & 3.33 & 0.83 & 4.35 & 1.19 \\
SL-HMC-MIX & ML-PIMD & 2.78 & 2.51 & 3.34 & 0.74 & 4.43 & 1.25 \\
SL-PIHMC-MIX & ML-PIMD & 2.79 & 2.42 & 3.30 & 0.87 & 4.29 & 1.16 \\
 \hline\hline 
 Training Data & Method &r$_1^{\mathrm{OH}}$ (Å) & h$_1^{\mathrm{OH}}$ & 
 r$_2^{\mathrm{OH}}$ (Å) & h$_2^{\mathrm{OH}}$ & 
 r$_3^{\mathrm{OH}}$ (Å) & h$_3^{\mathrm{OH}}$\\
\hline
- &  FP-MD & 0.98 & 36.78 & 1.82 & 
1.33 & 3.28 & 1.64 \\
SL-HMC-MIX & ML-MD & 0.98 & 33.73 & 1.82 & 1.20 & 3.30 & 1.64 \\
SL-PIHMC-MIX & ML-MD & 0.98 & 37.28 & 1.83 & 1.37 & 3.29 & 1.64 \\
- &  FP-PIMD & 0.99 & 13.19 & 1.81 & 1.15 & 3.32 & 1.60 \\
SL-HMC-MIX & ML-PIMD & 0.99 & 12.85 & 1.81 & 1.17 & 3.31 & 1.58 \\
SL-PIHMC-MIX & ML-PIMD & 0.99 & 13.27 & 1.82 & 1.12 & 3.32 & 1.59 \\
\hline\hline
 Training Data & Method & r$_1^{\mathrm{HH}}$ (Å) & h$_1^{\mathrm{HH}}$ &
  r$_2^{\mathrm{HH}}$ (Å) & h$_2^{\mathrm{HH}}$ & 
  r$_3^{\mathrm{HH}}$ (Å) & h$_3^{\mathrm{HH}}$\\
\hline
- &  FP-MD & 1.56 & 3.43 & 2.34 
& 1.44 & 3.88 & 1.22 \\
SL-HMC-MIX & ML-MD & 1.54 & 3.35 & 2.38 & 1.37 & 3.78 & 1.22 \\
SL-PIHMC-MIX & ML-MD & 1.56 & 3.51 & 2.38 & 1.43 & 3.83 & 1.22 \\
- &  FP-PIMD & 1.57 & 1.57 & 2.36 & 1.28 & 3.83 & 1.21 \\
SL-HMC-MIX & ML-PIMD & 1.57 & 1.57 & 2.36 & 1.27 & 3.85 & 1.22 \\
SL-PIHMC-MIX & ML-PIMD & 1.57 & 1.59 & 2.38 & 1.24 & 3.83 & 1.21 \\
\end{tabular}
\end{ruledtabular}
\end{table}

%%%%%%%%%%%%%%%%%%%%%%%%%%%%%%%%%%%%%%%%%%%%%%%%%%%%%%%%%%%%%%%%%%%%%%%%
\section{Additional results for D$_2$O and H$_2$O}
%%%%%%%%%%%%%%%%%%%%%%%%%%%%%%%%%%%%%%%%%%%%%%%%%%%%%%%%%%%%%%%%%%%%%%%%

\begin{figure}
    \centering
    \includegraphics[width=0.99\linewidth]{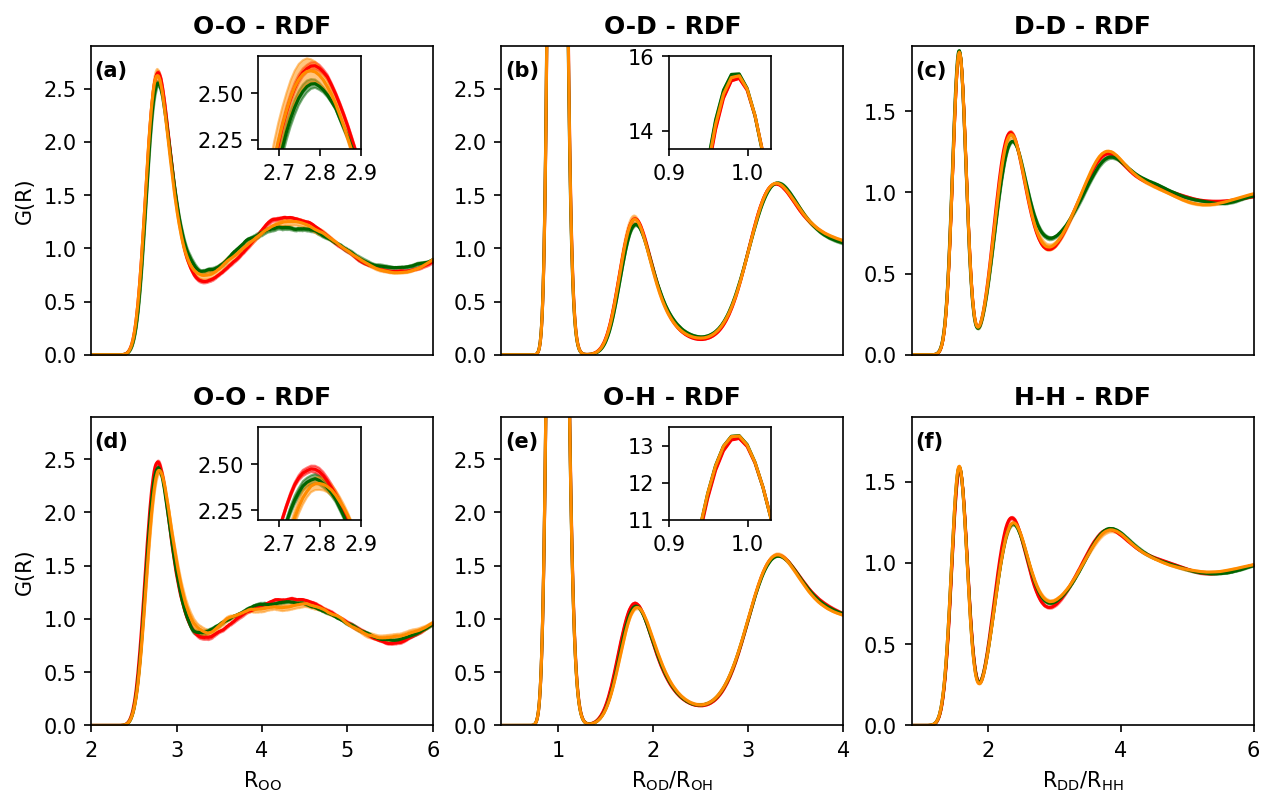}
    \caption{Comparison of the RDFs from the MLPs trained 
    using only H$_2$O structures and those using both H$_2$O and 
    D$_2$O structures. 
    In figure (a-c) we compare the O-O, O-D and D-D RDFs for
    simulations of D$_2$O using the two different MLPs, 
    the result from figure 5 in green and the potential trained on
    D$_{2}^{}$O  FP  data in orange,  
    with the reference  FP-PIMD simulation (red) using the 
    RPBE-D3 functional.
    In figure (d-f) the RDFs for the O-O, O-H and H-H pairs
    are compared for simulations of H$_2$O using the same MLPs 
    and DFT functional and color codes as parts (a-c).
    The peak positions and heights for Figure (a-c) and (d-f) are
    given in Table SXI and SXII respectively}
\end{figure}

Given the similarity of the RDFs calculated using 
ML-PIMD for D$_2$O and H$_2$O,
we have tried to improve the MLP by including data 
from a simulation of D$_2$O in the training set.
We thus ran an additional 2000 steps of
SL-PIHMC-MIX simulation for D$_2$O after
the initial 5000 steps done for H$_2$O,
$\left\langle A_\mathrm{test}^{}\right\rangle$ and $t_{\mathrm{eff}}^{}$
of this simulation are given in table SII.
We compare the RDFs for the atom pairs in
D$_2$O in Figure S5 (a-c) and for H$_2$O
in Figure S5 (d-f).
The peak heights and positions are given
in table SXI and SXII for D$_2$O and
H$_2$O respectively.
The discussion of these results are carried
out in Section IV (D) of the main text.

\begin{table}[H]
\caption{The positions and heights of the peaks 
in the O-O, O-D and D-D  RDFs for D$_2$O present in Figure S6 (a-c). 
All peak positions are given in Å.  
The data is denoted either with a r$^{\mathrm{XY}}_i$ or 
H$^{\mathrm{XY}}_i$ referring to the peak position and 
heights respectively for the pair XY 
$\in\{\mathrm{OO}, \mathrm{OD},\mathrm{DD}\}$. 
For the OO pair, we also give the position 
(r$_{\mathrm{min}}^{\mathrm{OO}}$) and height 
(h$_{\mathrm{min}}^{\mathrm{OO}}$) of the minima 
of the first interstitial region.
The experimental data stems from Ref. 
\onlinecite{soper_quantum_2008}.}
\begin{ruledtabular}
\begin{tabular}{cccccccc}
 Model & Training Set & r$_1^{\mathrm{OO}}$ (Å) & h$_1^{\mathrm{OO}}$ &
 r$_{\mathrm{min}}^{\mathrm{OO}}$ (Å) & h$_{\mathrm{min}}^{\mathrm{OO}}$ &
  r$_2^{\mathrm{OO}}$ (Å) & h$_2^{\mathrm{OO}}$\\
  \hline
 FP-PIMD & - & 2.79 & 2.65 & 3.33 & 0.69 & 4.33 & 1.29 \\
PIHMC-MIX & Only H$_2$O & 2.79 & 2.61 & 3.30 & 0.73 & 4.34 & 1.23 \\
ML-PIMD & Only H$_2$O & 2.79 & 2.55 & 3.30 & 0.78 & 4.20 & 1.20 \\
ML-PIMD & H$_2$O and D$_2$O & 2.78 & 2.62 & 3.32 & 0.75 & 4.35 & 1.25 \\
\hline
& Experiment & 2.76 & 2.62 & 3.38 & 0.79 & 4.29 & 1.15\\
 \hline\hline
 Model & Training Set &r$_1^{\mathrm{OD}}$ (Å) & h$_1^{\mathrm{OD}}$ & 
 r$_2^{\mathrm{OD}}$ (Å) & h$_2^{\mathrm{OD}}$ & 
 r$_3^{\mathrm{OD}}$ (Å) & h$_3^{\mathrm{OD}}$\\
\hline
 FP-PIMD & - & 0.99 & 15.41 & 1.81 & 1.28 & 3.29 & 1.60 \\
PIHMC-MIX & Only H$_2$O & 0.99 & 15.51 & 1.82 & 1.26 & 3.32 & 1.62 \\
ML-PIMD & Only H$_2$O & 0.99 & 15.53 & 1.82 & 1.22 & 3.32 & 1.61 \\
ML-PIMD & H$_2$O and D$_2$O & 0.99 & 15.48 & 1.80 & 1.27 & 3.30 & 1.61 \\
\hline
& Experiment & ...  & ...  & 1.77 & 1.10 & 3.20 & 1.48\\
\hline\hline
 Model & Training Set & r$_1^{\mathrm{DD}}$ (Å) & h$_1^{\mathrm{DD}}$ &
  r$_2^{\mathrm{DD}}$ (Å) & h$_2^{\mathrm{DD}}$ & 
  r$_3^{\mathrm{DD}}$ (Å) & h$_3^{\mathrm{DD}}$\\
\hline
 FP-PIMD & - & 1.57 & 1.86 & 2.34 & 1.37 & 3.81 & 1.24 \\
PIHMC-MIX & Only H$_2$O & 1.57 & 1.86 & 2.37 & 1.34 & 3.82 & 1.24 \\
ML-PIMD & Only H$_2$O & 1.57 & 1.87 & 2.37 & 1.31 & 3.86 & 1.22 \\
ML-PIMD & H$_2$O and D$_2$O & 1.57 & 1.86 & 2.35 & 1.35 & 3.82 & 1.25 \\
\hline
& Experiment & ... & ... & 2.33 & 1.41 & 3.84 & 1.21\\
\end{tabular}
\end{ruledtabular}
\end{table}

\begin{table}[H]
\caption{The positions and heights of the peaks
in the O-H RDFs for H$_2$O presented in Figure S6 (d-f). 
All peak positions are given in Å.  
The data is denoted either with r$^{\mathrm{XY}}_i$ or 
H$^{\mathrm{XY}}_i$ referring to the peak position and 
heights respectively for the pair XY 
$\in\{\mathrm{OO}, \mathrm{OH},\mathrm{HH}\}$. 
For the OO pair, we also give the position 
(r$_{\mathrm{min}}^{\mathrm{OO}}$) and height 
(h$_{\mathrm{min}}^{\mathrm{OO}}$) of the minima 
of the first interstitial region. 
The experimental reference stem from Ref. 
\onlinecite{soper_radial_2013}, 
except those marked by $^*$ which are from Ref. 
\onlinecite{soper_radial_2000}.}
\begin{ruledtabular}
\begin{tabular}{cccccccc}
 Model & Training Set & r$_1^{\mathrm{OO}}$ (Å) & h$_1^{\mathrm{OO}}$ &
 r$_{\mathrm{min}}^{\mathrm{OO}}$ (Å) & h$_{\mathrm{min}}^{\mathrm{OO}}$ &
  r$_2^{\mathrm{OO}}$ (Å) & h$_2^{\mathrm{OO}}$\\
  \hline
 FP-PIMD & - & 2.78 & 2.47 & 3.33 & 0.83 & 4.35 & 1.19 \\
PIHMC-MIX & Only H$_2$O & 2.79 & 2.53 & 3.32 & 0.78 & 4.24 & 1.22 \\
ML-PIMD & Only H$_2$O & 2.79 & 2.42 & 3.30 & 0.87 & 4.29 & 1.16 \\
ML-PIMD & H$_2$O and D$_2$O & 2.80 & 2.40 & 3.37 & 0.87 & 4.50 & 1.14 \\
\hline
 & Experiment & 2.79 & 2.50 & 3.36 & 0.78 & 4.53 & 1.12 \\
 \hline\hline
 Model & Training Set &r$_1^{\mathrm{OH}}$ (Å) & h$_1^{\mathrm{OH}}$ & 
 r$_2^{\mathrm{OH}}$ (Å) & h$_2^{\mathrm{OH}}$ & 
 r$_3^{\mathrm{OH}}$ (Å) & h$_3^{\mathrm{OH}}$\\
\hline
 FP-PIMD & - & 0.99 & 13.19 & 1.81 & 1.15 & 3.32 & 1.60 \\
PIHMC-MIX & Only H$_2$O & 0.98 & 13.24 & 1.81 & 1.18 & 3.32 & 1.59 \\
ML-PIMD & Only H$_2$O & 0.99 & 13.27 & 1.82 & 1.12 & 3.32 & 1.59 \\
ML-PIMD & H$_2$O and D$_2$O & 0.99 & 13.26 & 1.83 & 1.11 & 3.31 & 1.60 \\
\hline
 & Experiment & 0.96$^{*}_{}$ & 12.71$^{*}_{}$ & 1.86 & 1.04 & 3.27 & 1.48 \\
\hline\hline
 Model & Training Set & r$_1^{\mathrm{HH}}$ (Å) & h$_1^{\mathrm{HH}}$ &
  r$_2^{\mathrm{HH}}$ (Å) & h$_2^{\mathrm{HH}}$ & 
  r$_3^{\mathrm{HH}}$ (Å) & h$_3^{\mathrm{HH}}$\\
\hline
 FP-PIMD & - & 1.57 & 1.57 & 2.36 & 1.28 & 3.83 & 1.21 \\
PIHMC-MIX & Only H$_2$O & 1.57 & 1.59 & 2.39 & 1.27 & 3.86 & 1.23 \\
ML-PIMD & Only H$_2$O & 1.57 & 1.59 & 2.38 & 1.24 & 3.83 & 1.21 \\
ML-PIMD & H$_2$O and D$_2$O & 1.57 & 1.59 & 2.38 & 1.25 & 3.85 & 1.20 \\
\hline
  & Experiment & 1.53$^{*}_{}$ & 1.71$^{*}_{}$ & 2.43 & 1.34 & 3.84 & 1.17 \\
\end{tabular}
\end{ruledtabular}
\end{table}

\newpage

%%%%%%%%%%%%%%%%%%%%%%%%%%%%%%%%%%%%%%%%%%%%%%%%%%%%%%%%%%%%%%%%%%%%%%%%
\section{HMC-MIX Data for Various DFT Functionals}
%%%%%%%%%%%%%%%%%%%%%%%%%%%%%%%%%%%%%%%%%%%%%%%%%%%%%%%%%%%%%%%%%%%%%%%%

In this section we present the results for
HMC-MIX for the SCAN, rev-vdW-DF2 and OptB88-vdW
functionals.
In Figure S6, we show the comparisons between
PIHMC-MIX and HMC-MIX simulations using these
three functionals.
The peak positions for these RDFs are given in
Table SXIII.
These results are discussed in Section IV (E) 
of the main text.

%%%

\begin{figure}
    \centering
    \includegraphics[width=0.99\linewidth]{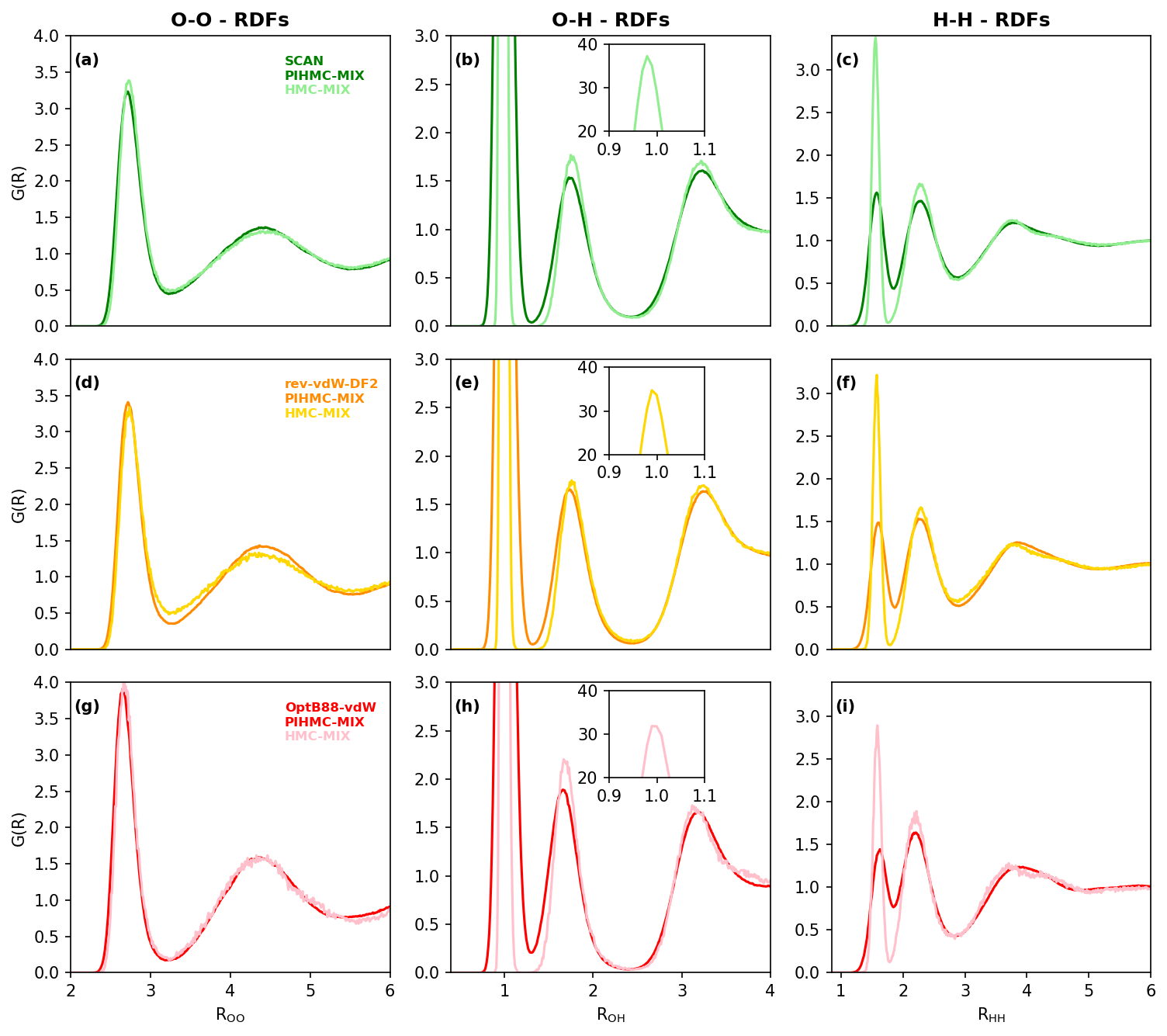}
    \caption{The RDFs for all atom pairs in H$_2$O calculated using
    PIHMC-MIX and HMC-MIX, with the HMC-MIX results given in lighter
    colors, for the (a-c) SCAN, (d-f) rev-vdW-DF2 and 
    (g-i) OptB88-vdW functionals.
    The peak heights and positions are given in Table SXIII.}
\end{figure}

\begin{table}[H]
\caption{The positions and heights of the peaks 
in the O-O, O-H and H-H RDFs for H$_2$O presented 
in Figure S7. 
All peak positions are given in Å.  
The data is denoted either with a r$^{\mathrm{XY}}_i$ 
or H$^{\mathrm{XY}}_i$ referring to the peak position 
and heights respectively for the pair XY 
$\in\{\mathrm{OO}, \mathrm{OH},\mathrm{HH}\}$.
For the OO pair, we also give the position 
(r$_{\mathrm{min}}^{\mathrm{OO}}$) and height 
(h$_{\mathrm{min}}^{\mathrm{OO}}$) of the minima of 
the first interstitial region. 
The experimental reference stem from Ref. 
\onlinecite{soper_radial_2013}, 
except those marked by $^*$ which are from Ref. 
\onlinecite{soper_radial_2000}.}
\begin{ruledtabular}
\begin{tabular}{cccccccc}
 DFT Functional & Model & r$_1^{\mathrm{OO}}$ (Å) & h$_1^{\mathrm{OO}}$ &
 r$_{\mathrm{min}}^{\mathrm{OO}}$ (Å) & h$_{\mathrm{min}}^{\mathrm{OO}}$ &
  r$_2^{\mathrm{OO}}$ (Å) & h$_2^{\mathrm{OO}}$\\
  \hline
SCAN & PIHMC-MIX & 2.72 & 3.24 & 3.23 & 0.44 & 4.36 & 1.36 \\
SCAN & HMC-MIX & 2.70 & 3.41 & 3.26 & 0.46 & 4.38 & 1.30 \\
rev-vdW-DF2 & PIHMC-MIX & 2.72 & 3.43 & 3.23 & 0.36 & 4.46 & 1.43 \\
rev-vdW-DF2 & HMC-MIX & 2.74 & 3.19 & 3.30 & 0.52 & 4.46 & 1.30 \\
optB88-vdW & PIHMC-MIX & 2.65 & 3.88 & 3.20 & 0.17 & 4.36 & 1.58 \\
optB88-vdW & HMC-MIX & 2.66 & 3.62 & 3.22 & 0.29 & 4.38 & 1.47 \\
\hline
& Experiment & 2.79 & 2.50 & 3.36 & 0.78 & 4.53 & 1.12\\
 \hline\hline
 Model & Training Set &r$_1^{\mathrm{OH}}$ (Å) & h$_1^{\mathrm{OH}}$ & 
 r$_2^{\mathrm{OH}}$ (Å) & h$_2^{\mathrm{OH}}$ & 
 r$_3^{\mathrm{OH}}$ (Å) & h$_3^{\mathrm{OH}}$\\
\hline
SCAN & PIHMC-MIX & 0.99 & 13.04 & 1.75 & 1.54 & 3.24 & 1.61 \\
SCAN & HMC-MIX & 0.98 & 33.71 & 1.74 & 1.79 & 3.22 & 1.70 \\
rev-vdW-DF2 & PIHMC-MIX & 1.00 & 12.33 & 1.74 & 1.66 & 3.26 & 1.63 \\
rev-vdW-DF2 & HMC-MIX & 0.98 & 30.85 & 1.74 & 1.68 & 3.22 & 1.68 \\
optB88-vdW & PIHMC-MIX & 1.01 & 11.82 & 1.66 & 1.89 & 3.19 & 1.66 \\
optB88-vdW & HMC-MIX & 0.98 & 28.91 & 1.70 & 2.02 & 3.18 & 1.64 \\
\hline
& Experiment & 0.96$^{*}_{}$  & 12.71$^{*}_{}$  & 1.86 & 1.04 & 3.27 & 1.48\\
\hline\hline
 Model & Training Set & r$_1^{\mathrm{HH}}$ (Å) & h$_1^{\mathrm{HH}}$ &
  r$_2^{\mathrm{HH}}$ (Å) & h$_2^{\mathrm{HH}}$ & 
  r$_3^{\mathrm{HH}}$ (Å) & h$_3^{\mathrm{HH}}$\\
\hline
SCAN & PIHMC-MIX & 1.57 & 1.56 & 2.28 & 1.47 & 3.80 & 1.21 \\
SCAN & HMC-MIX & 1.54 & 3.20 & 2.26 & 1.68 & 3.74 & 1.24 \\
rev-vdW-DF2 & PIHMC-MIX & 1.60 & 1.49 & 2.25 & 1.53 & 3.85 & 1.25 \\
rev-vdW-DF2 & HMC-MIX & 1.58 & 3.07 & 2.30 & 1.63 & 3.82 & 1.23 \\
optB88-vdW & PIHMC-MIX & 1.63 & 1.44 & 2.21 & 1.64 & 3.89 & 1.24 \\
optB88-vdW & HMC-MIX & 1.58 & 2.79 & 2.26 & 1.65 & 3.74 & 1.23 \\
\hline
& Experiment & 1.53$^{*}_{}$ & 1.71$^{*}_{}$ & 2.43 & 1.34 & 3.84 & 1.17\\
\end{tabular}
\end{ruledtabular}
\end{table}

%%%%%%%%%%%%%%%%%%%%%%%%%%%%%%%%%%%%%%%%%%%%%%%%%%%%%%%%%%%%%%%%%%%%%%%%
\section{Ice I$_\mathrm{h}^{}$ simulations with PIHMC-MIX}
%%%%%%%%%%%%%%%%%%%%%%%%%%%%%%%%%%%%%%%%%%%%%%%%%%%%%%%%%%%%%%%%%%%%%%%%

\begin{figure}
    \centering
    \includegraphics[width=0.99\linewidth]{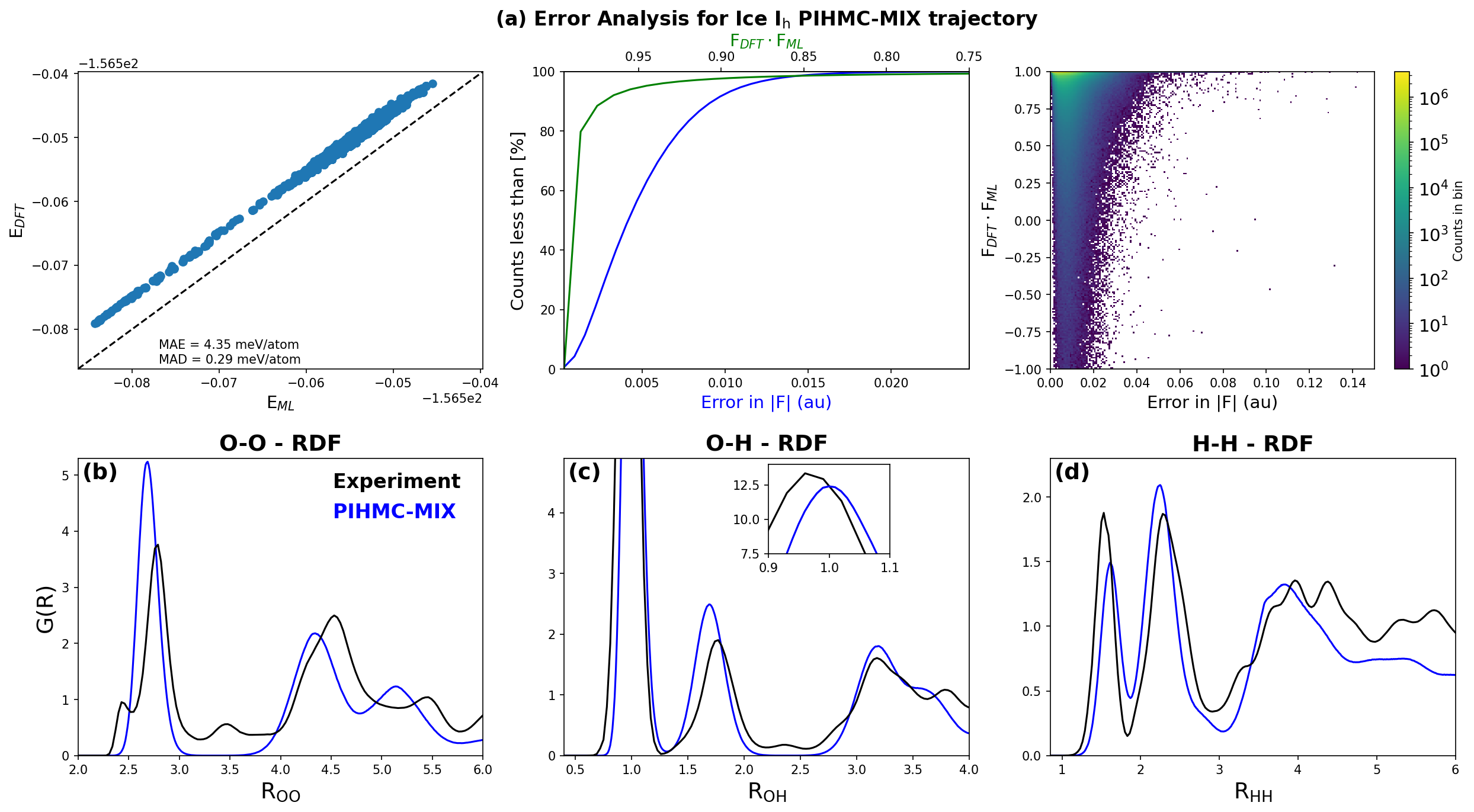}
    \caption{Comparison of energies and forces for ice I$_{\mathrm{h}}^{}$ 
    from ML and FP potentials for the MLPs trained on RPBE-D3 liquid water
    simulation.
    The rows of this figure depict the same comparisons and distributions
    as those in Figures S3 and S4.
    In figure (b-d) we compare the O-O, O-H and H-H RDFs for PIHMC-MIX simulations
    of Ice I$_{\mathrm{h}}^{}$ with the experimental reference by 
    Soper\cite{soper_radial_2000}.
    The peak positions are given in Table SXIV.}
\end{figure}

To test the transferability to other points in the Phase diagram of the PIHMC-MIX method
we have carried out a simulation for Ice I$_\mathrm{h}^{}$ at 220 K.
This simulation were initialized from the crystal structure of hexagonal ice in a parallel piped
box with side lengths $a=22.81$ \AA, $b=15.21$ \AA\ and $c=7.14$ \AA, and angles $\alpha=\beta=90^{\circ}$ and $\gamma=120^{\circ}$. 
The box contained 72 water molecules.
The simulation were run with the MLP trained from SL-PIHMC-MIX for liquid water
and the FP calculation settings being identical to those used for
the RPBE-D3 calculations described in the main text.
The simulation were carried out for 5,000 steps, with an average acceptance rate of 52.66 \%
and an effective trajectory length ($t_\mathrm{eff}^{}$) of 77.696 ps.
Which, while smaller and shorter respectively, is comparable to 
that found for the other PIHMC-MIX simulations of liquid water in this study.

%%%

The performance for the underlying MLP used to propagate the short ML-PIMD trajectories
are shown in figure S8 (a) with similar statistics as those used in Section SVII.
We find a much larger $\sigma^{\mathrm{at}}_{\mathrm{E}}$ of 4.35 meV/atom compared to
that found for the liquid water simulations.
Comparisons of the DFT and ML energies however reveal a static shift of the energy,
which can approximately be removed by considering the mean absolute 
deviation (MAD) around the absolute mean energy difference per atom in the system
\begin{equation}
\mathrm{MAD}=\frac{1}{NPM}\sum_{s=1}^{P}\sum_{j=1}^{M}
\left|\mathrm{E}_{\mathrm{DFT}}^{j, s}-\mathrm{E}_{\mathrm{ML}}^{j, s} - N\sigma^{\mathrm{at}}_{\mathrm{E}}\right|
\end{equation}
where $\sigma^{\mathrm{at}}_{\mathrm{E}}$ is defined in Eq. (S12).
This is found to be 0.29 meV/atom, which is on the same order of magnitude as the 
MAE errors found for RPBE-D3, although it should be noted that this measure might include
some fortuitous cancellations of error that makes the number smaller than an exact shifted 
$\sigma^{\mathrm{at}}_{\mathrm{E}}$.
The forces are the most important in this case, as a constant shift in the MLP energy
will not affect the acceptance criteria and thus the efficiency of the PIHMC-MIX method.
$\sigma^{\mathrm{at}}_{\mathrm{F}}$ is found to be 89.3 meV/\AA\ in this case, 
and the errors in the direction of the force vectors are also comparable to that
found for the PIHMC-MIX simulations of liquid water using RPBE-D3.
We can thus conclude that the MLP constructed from liquid water is able to reproduce the 
forces in ice I$_\mathrm{h}^{}$ with slightly less accuracy than those in liquid water,
and that the energies while shifted are reproduced with the same accuracy as in liquid water.

%%%

The O-O, O-H and H-H RDFs from the simulations are plotted in figure S8 (b), (c) and (d) respectively along with the experimental results from Soper\cite{soper_radial_2000}
and the peak positions are given in Table SXIV.
The agreement between simulation and experiment is notable worse than for liquid water,
but can be explained due to the differences in assumptions.
The current simulation were carried out using the crystal structure of ice I$^{}_{\mathrm{h}}$,
whereas the experimental study notes that there are several unstructured regions in the predicted structure, which will not be captured by the current simulation.
This is most evident in the first interstitial region in the O-O RDF,
where a non-zero RDF is reported in the experiment.
In general, we find the largest differences in the O-O RDF,
several experimental peaks are split or are broadened when compared to the RDF from our simulation of the crystalline ice I$^{}_{\mathrm{h}}$.
For the O-H and H-H RDFs the agreement between the experiment and our simulation is better,
although it is still clear that there are still extra extremes in the experiment that are not captured by simulating only the crystal.
Especially, the agreement in positions of the second O-H and H-H peaks indicate that 
the intra- and the closest inter-molecular structures are well reproduced
from our simulation.
We also note that the experimental study mentions the width of the first inter-molecular peaks,
the second O-H and H-H peaks, being an indication of the disorder of the system.
We however find similar widths due to the quantum nature of hydrogen in our simulations.
It however remains clear that the current simple simulation setup will not be able to reproduce
the experimental result.
A more thorough study is thus needed to settle this issue by targeting the neutron scattering
results directly instead of the RDFs,
a study that is beyond the scope of the current paper.

\begin{table}[H]
\caption{The positions and heights of the peaks 
in the O-O, O-H and H-H RDFs for ice I$_\mathrm{h}^{}$ (H$_2$O) presented in Figure S5 (a-c). 
All peak positions are given in Å.  
The data is denoted either with r$^{\mathrm{XY}}_i$ or 
H$^{\mathrm{XY}}_i$ referring to the peak position and 
heights respectively for the pair XY 
$\in\{\mathrm{OO}, \mathrm{OH},\mathrm{HH}\}$. 
For the OO pair we also give the position 
(r$_{\mathrm{min}}^{\mathrm{OO}}$) and height 
(h$_{\mathrm{min}}^{\mathrm{OO}}$) of the minima 
of the first interstitial region, which are a region from $\sim$3.1-3.6 \AA\ with 
h$_{}^{\mathrm{OO}}=0.0$.
For the experimental RDFs, the maxima closest resembling those found in the current 
crystal ice simulation are reported.
}
\begin{ruledtabular}
\begin{tabular}{cccccccc}
 Model & Functional & r$_1^{\mathrm{OO}}$ (Å) & h$_1^{\mathrm{OO}}$ &
 r$_{\mathrm{min}}^{\mathrm{OO}}$ (Å) & h$_{\mathrm{min}}^{\mathrm{OO}}$ &
  r$_2^{\mathrm{OO}}$ (Å) & h$_2^{\mathrm{OO}}$\\
  \hline
PIHMC-MIX & RPBE-D3 & 2.69 & 5.24 & $\sim$3.1-3.6 & 0.0 & 4.34 & 2.18 \\
Experiment & - & 2.79 & 3.76 & - & - & 4.53 & 2.50\\
 \hline\hline
 Model & Functional &r$_1^{\mathrm{OH}}$ (Å) & h$_1^{\mathrm{OH}}$ & 
 r$_2^{\mathrm{OH}}$ (Å) & h$_2^{\mathrm{OH}}$ & 
 r$_3^{\mathrm{OH}}$ (Å) & h$_3^{\mathrm{OH}}$\\
\hline
PIHMC-MIX & RPBE-D3 & 1.00 & 12.42 & 1.69 & 2.49 & 3.20 & 1.80 \\
Experiment & - & 0.96 & 13.35 & 1.77 & 1.90 & 3.18 & 1.61\\
\hline\hline
 Model & Functional & r$_1^{\mathrm{HH}}$ (Å) & h$_1^{\mathrm{HH}}$ &
  r$_2^{\mathrm{HH}}$ (Å) & h$_2^{\mathrm{HH}}$ & 
  r$_3^{\mathrm{HH}}$ (Å) & h$_3^{\mathrm{HH}}$\\
\hline
PIHMC-MIX & RPBE-D3 & 1.61 & 1.49 & 2.25 & 2.09 & 3.83 & 1.32 \\
Experiment & - & 1.53 & 1.88 & 2.28 & 1.87 & 4.02 & 1.32\\
\end{tabular}
\end{ruledtabular}
\end{table}

%%%%%%%%%%%%%%%%%%%%%%%%%%%%%%%%%%%%%%%%%%%%%%%%%%%%%%%%%%%%%%%%%%%%%%%%
\section{Hydrogen Bond Geometry}
%%%%%%%%%%%%%%%%%%%%%%%%%%%%%%%%%%%%%%%%%%%%%%%%%%%%%%%%%%%%%%%%%%%%%%%%

The hydrogen bond in liquid water is one of the key facilitators of the dynamically
distorted tetrahedral structure of the liquid.
The temperature effects on the hydrogen bonds were initially studied by Modig, Pfrommer and Halle in Ref. \onlinecite{modig_temperature-dependent_2003}, 
and later
by Yao and Kanai\cite{yao_temperature_2020}.
Here $\beta\left(\mathrm{O\cdots O-H}\right)$ is the angle between the oxygen accepting the hydrogen bond, the oxygen bound to the donating hydrogen and the donating hydrogen,
and $R^{}_{\mathrm{H\cdots O}}$ is the distance between the donating hydrogen 
and the accepting oxygen in the hydrogen bond.
In figure S9 we have plotted the distributions of these parameters for our 
HMC-MIX and PIHMC-MIX simulations compared to the experiment at 27 $^{\circ}$C,
and the reweighted averages are given in table SXV along with the values
calculated from the model interpolations in the paper by Modig, Pfrommer and Halle.
These results are discussed in Section 3 E and the conclusion of the paper.

\begin{figure}
    \centering
    \includegraphics[width=0.99\linewidth]{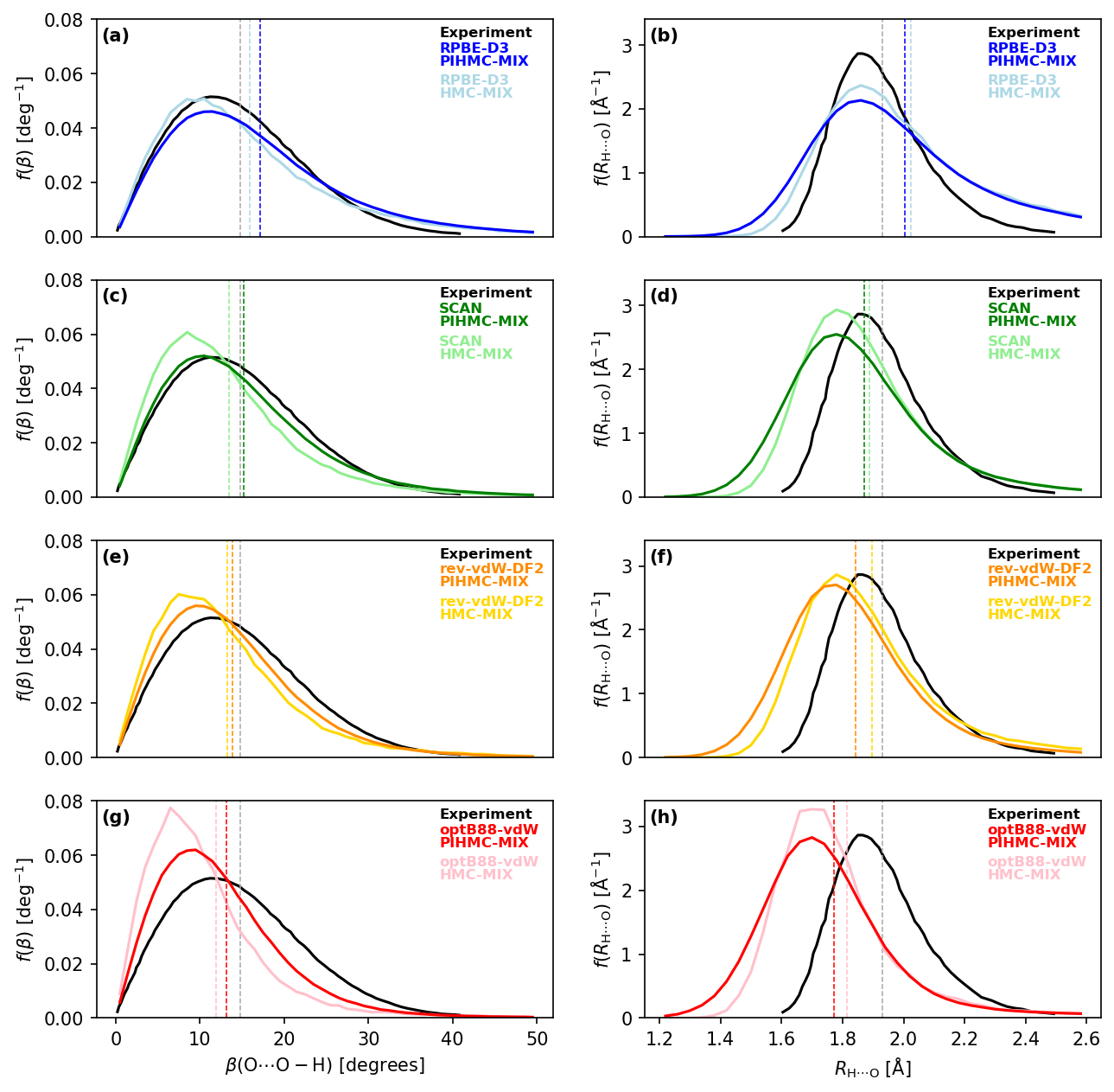}
    \caption{
    The distributions of $R_{\mathrm{H\cdots O}}^{}$, (a), (c), (e) and (g), and $\beta(\mathrm{O\cdots O-H})$, (b), (d), (f) and (h), see the text for their definition,
    for PIHMC-MIX RPBE-D3 (blue), SCAN (green), rev-vdW-DF2 (orange) and optB88-vdW (red)
    and for HMC-MIX RPBE-D3 (light blue), SCAN (light green), rev-vdW-DF2 (yellow) and optB88-vdW (pink).
    The results for RPBE-D3 given in (a) and (b), for SCAN in (c) and (d), for rev-vdW-DF2 in (e) and (f), and for optB88-vdW in (g) and (h).
    All distributions are compared with the experimental\cite{modig_temperature-dependent_2003} results at 27 $^\circ$C in black.
    The vertical dashed lines are the averages, also found in table SXV, 
    the colored being from our simulations and
    the gray being from the experimental interpolations at 298.15 K.
    }
\end{figure}

\begin{table}[H]
\caption{The averages of $R_{\mathrm{H\cdots O}}^{}$ and $\beta(\mathrm{O\cdots O-H})$ from
the HMC-MIX and PIHMC-MIX simulations of this study
compared with the interpolated values from experiment\cite{modig_temperature-dependent_2003} at 298.15 K}
\begin{ruledtabular}
\begin{tabular}{cccc}
 DFT Functional & Model & $\left\langle R_{\mathrm{H\cdots O}}^{} \right\rangle$ [\AA] & $\left\langle\beta(\mathrm{O\cdots O-H})\right\rangle$ [degrees]\\
\hline
RPBE-D3 & HMC-MIX & 2.02 & 15.93 \\
SCAN & HMC-MIX & 1.89 & 13.43 \\
rev-vdW-DF2 & HMC-MIX & 1.90 & 13.24 \\
optB88-vdW & HMC-MIX & 1.81 & 11.92 \\
\hline
RPBE-D3 & PIHMC-MIX & 2.00 & 17.15 \\
SCAN & PIHMC-MIX & 1.87 & 15.16 \\
rev-vdW-DF2 & PIHMC-MIX & 1.84 & 13.91 \\
optB88-vdW & PIHMC-MIX & 1.77 & 13.14 \\
\hline
& Ref. \onlinecite{modig_temperature-dependent_2003} & 1.93 & 14.79\\
\end{tabular}
\end{ruledtabular}
\end{table}

%%%%%%%%%%%%%%%%%%%%%%%%%%%%%%%%%%%%%%%%%%%%%%%%%%%%%%%%%%%%%%%%%%%%%%%%

%%%%%%%%%%%%%%%%%%%%%%%%%%%%%%%%%%%%%%%%%%%%%%%%%%%%%%%%%%%%%%%%%%%%%%%%
\end{document}